\documentclass[10pt,twocolumn,english,american,prx,singlecolumn,amsmath,amssymb,superscriptaddress]{revtex4-1}
\usepackage[T1]{fontenc}
\usepackage[latin9]{inputenc}
\usepackage{babel}
\usepackage{prettyref}
\usepackage{float}
\usepackage{textcomp}
\usepackage{amsmath}
\usepackage{stackrel}
\usepackage{graphicx}
\usepackage[unicode=true,pdfusetitle,
 bookmarks=true,bookmarksnumbered=false,bookmarksopen=false,
 breaklinks=true,pdfborder={0 0 1},backref=false,colorlinks=false]
 {hyperref}
\hypersetup{
 colorlinks=true,linkcolor=black,citecolor=black,urlcolor=black,filecolor=black}

\makeatletter

\newrefformat{supp}{Supplement \ref{#1}}

\usepackage{notoccite}
\usepackage{feynmp}
\newrefformat{cap}{\hyperref[#1]{Figure~\ref{#1}}}
\newrefformat{fig}{\hyperref[#1]{Figure~\ref{#1}}}
\newrefformat{tab}{\hyperref[#1]{Table ~\ref{#1}}}
\newrefformat{sec}{\hyperref[#1]{Section~\ref{#1}}}
\newrefformat{sub}{\hyperref[#1]{Section~\ref{#1}}}
\newrefformat{subsec}{\hyperref[#1]{Section~\ref{#1}}}
\newrefformat{cha}{\hyperref[#1]{Chapter~\ref{#1}}}
\newrefformat{app}{\hyperref[#1]{Appendix~\ref{#1}}}

\newcommand{\brackets}[1]{{\left( #1\right)}}

\makeatother

\begin{document}
\selectlanguage{english}%
\global\long\def\N{\mathcal{N}}%
\global\long\def\Z{\mathcal{\mathcal{Z}}}%
\global\long\def\Ndim{N_{\mathrm{dim}}}%
\global\long\def\a{\alpha}%
\global\long\def\b{\beta}%
\global\long\def\c{\gamma}%
\global\long\def\d{\delta}%
\global\long\def\.{\ .}%
\global\long\def\r{\rho}%
\global\long\def\erf{\mathrm{erf}}%

\title{A theory of data variability in Neural Network Bayesian inference}
\author{Javed Lindner}
\affiliation{Institute of Neuroscience and Medicine (INM-6) and Institute for Advanced
Simulation (IAS-6) and JARA Institute Brain Structure-Function Relationships
(INM-10), J\"ulich Research Center, J\"ulich, Germany}
\affiliation{Department of Physics, Faculty 1, RWTH Aachen University, Aachen,
Germany}
\affiliation{Institute for Theoretical Particle Physics and Cosmology, RWTH Aachen
University, Aachen, Germany}
\author{David Dahmen}
\affiliation{Institute of Neuroscience and Medicine (INM-6) and Institute for Advanced
Simulation (IAS-6) and JARA Institute Brain Structure-Function Relationships
(INM-10), J\"ulich Research Center, J\"ulich, Germany}
\author{Michael Krämer}
\affiliation{Institute for Theoretical Particle Physics and Cosmology, RWTH Aachen
University, Aachen, Germany}
\author{Moritz Helias}
\affiliation{Institute of Neuroscience and Medicine (INM-6) and Institute for Advanced
Simulation (IAS-6) and JARA Institute Brain Structure-Function Relationships
(INM-10), J\"ulich Research Center, J\"ulich, Germany}
\affiliation{Department of Physics, Faculty 1, RWTH Aachen University, Aachen,
Germany}
\date{\today}
\begin{abstract}
Bayesian inference and kernel methods are well established in machine
learning. The neural network Gaussian process in particular provides
a concept to investigate neural networks in the limit of infinitely
wide hidden layers by using kernel and inference methods. Here we
build upon this limit and provide a field-theoretic formalism which
covers the generalization properties of infinitely wide networks.
We systematically compute generalization properties of linear, non-linear,
and deep non-linear networks for kernel matrices with heterogeneous
entries. In contrast to currently employed spectral methods we derive
the generalization properties from the statistical properties of the
input, elucidating the interplay of input dimensionality, size of
the training data set, and variability of the data. We show that data
variability leads to a non-Gaussian action reminiscent of a $\varphi^{3}+\varphi^{4}$-theory.
Using our formalism on a synthetic task and on MNIST we obtain a homogeneous
kernel matrix approximation for the learning curve as well as corrections
due to data variability which allow the estimation of the generalization
properties and exact results for the bounds of the learning curves
in the case of infinitely many training data points.
\end{abstract}
\maketitle

\section{Introduction}

\begin{figure*}[t]
\begin{centering}
\includegraphics[width=7in]{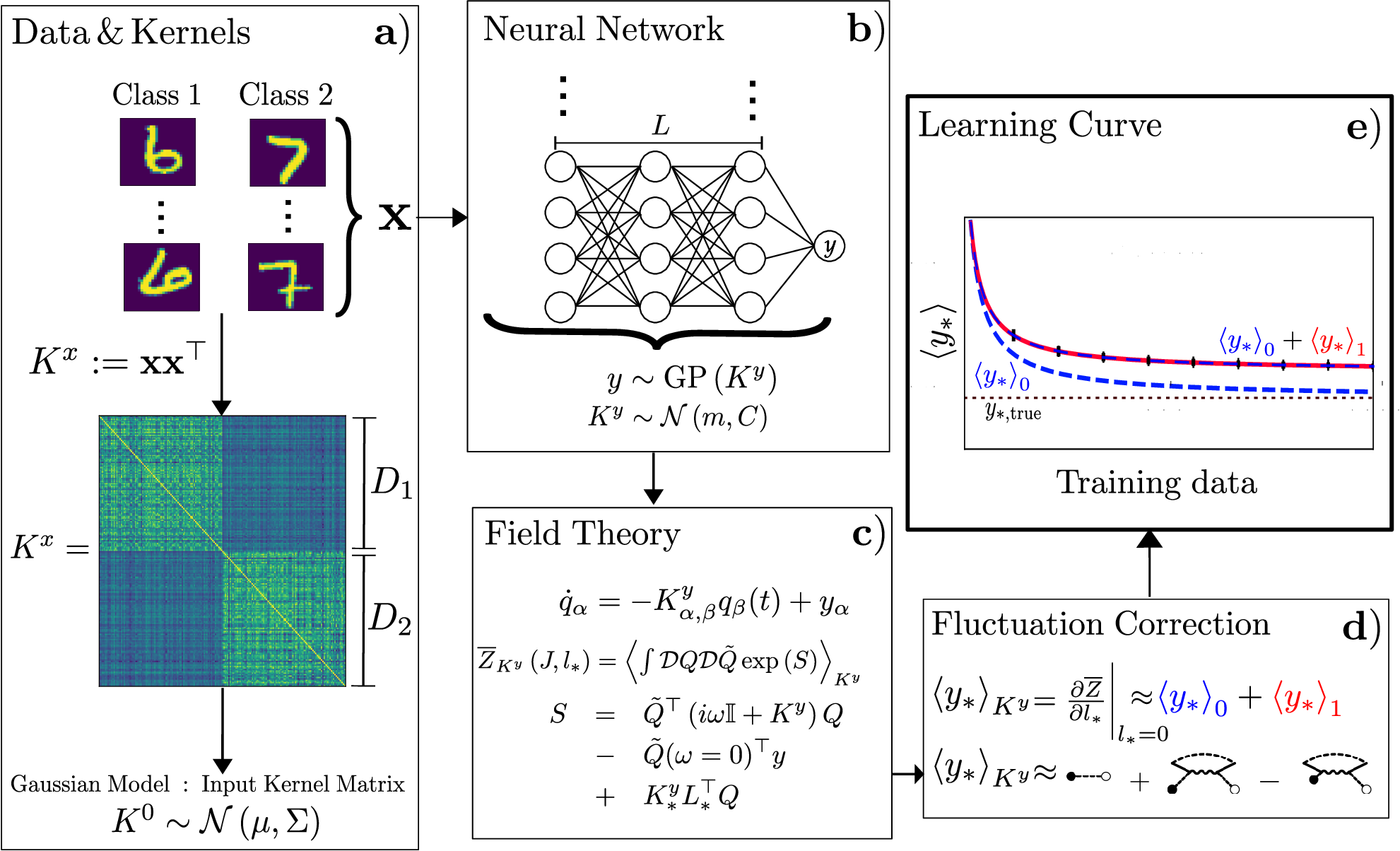}
\par\end{centering}
\caption{\textbf{Field theory of generalization in Bayesian inference. a) }A
binary classification task, such as distinguishing pairs of digits
in MNIST, can be described with the help of an overlap matrix $K^{x}$
that represents similarity across the $c=c_{1}+c_{2}$ images of the
training set of two classes, $1$ and $2$ with $D_{1}$ and $D_{2}$
samples respectively. Entries of the overlap matrix are heterogeneous.
Different drawings of $c$ example patterns each lead to different
realizations of the overlap matrix; the matrix is stochastic. We here
describe the matrix elements by a correlated multivariate Gaussian.
\textbf{b) }The data is fed through a feed-forward neural network
to produce an output $y$. In the case of infinitely wide hidden layers
and under Gaussian priors on the network weights, the output of the
network is a Gaussian process with the kernel $K^{y}$, which depends
on the network architecture and the input kernel $K^{x}$. \textbf{c)}
To obtain statistical properties of the posterior distribution, we
compute its disorder-averaged moment generating function $\overline{Z}(J,l_{*})$
diagrammatically. \textbf{d)} The leading-order contribution from
the homogeneous kernel $\langle y^{\ast}\rangle_{0}$ is corrected
by $\langle y^{\ast}\rangle_{1}$ due to the variability of the overlaps;
both follow as derivatives of $\overline{Z}(J,l_{*})$.\textbf{ e)}
Comparing the mean network output on a test point $\left\langle y^{*}\right\rangle $,
the zeroth order theory $\left\langle y_{*}\right\rangle _{0}$ (blue
dashed), the first-order approximation in the data-variability $\left\langle y_{*}\right\rangle _{0+1}$
(blue-red dashed) and empirical results (black crosses) as a function
of the amount of training data (learning curve) shows how variability
in the data set limits the network performance and validates the theory.\label{fig:graphical_abstract}}
\end{figure*}

Machine learning and in particular deep learning continues to influence
all areas of science . Employed as a scientific method, explainability,
a defining feature of any scientific method, however, is still largely
missing. This is also important to provide guarantees and to guide
educated design choices to reach a desired level of accuracy. The
reason is that the underlying principles by which artificial neural
networks reach their unprecedented performance are largely unknown.
 There is, up to date, no complete theoretical framework which fully
describes the behavior of artificial neural networks so that it would
explain the mechanisms by which neural networks operate. Such a framework
would also be useful to support architecture search and network training.

Investigating the theoretical foundations of artificial neural networks
on the basis of statistical physics dates back to the 1980s. Early
approaches to investigate neural information processing were mainly
rooted in the spin-glass literature and included the computation of
the memory capacity of the perceptron, path integral formulations
of the network dynamics \citep{Sompolinsky88_259}, and investigations
of the energy landscape of attractor network \citep{Amit85_1530,Gardner88_257,Gardner88_271}.

As in the thermodynamic limit in solid state physics, some modern
approaches deal with artificial neural networks (ANN) with an infinite
number of hidden neurons to simplify calculations. This leads to a
relation between ANNs and Bayesian inference on Gaussian processes
\citep{Neal96,Williams96_ae5e3ce4}, known as the Neural Network Gaussian
Process (NNGP) limit: The prior distribution of network outputs across
realizations of network parameters here becomes a Gaussian process
that is uniquely described by its covariance function or kernel. This
approach has been used to obtain insights into the relation of network
architecture and trainability \citep{Poole16_3360,Schoenholz17_01232,Yang19,Cui_2022,Ruben2023LearningCF}.
Other works have investigated training by gradient descent as a means
to shape the corresponding kernel \citep{Jacot18_8580}. A series
of recent studies also captures networks at finite width, including
adaptation of the kernel due to feature learning effects \citep{Cohen21_023034,Naveh21_NeurIPS,ZavatoneVeth21_NeurIPS_I,ZavatoneVeth21_NeurIPS_II,Li21_031059,Roberts22,Bordelon2022}.
Even though training networks with gradient descent is the most abundant
setup, different schemes such as Bayesian Deep Learning \citep{Murphy23_PML_Part_II}
provide an alternative perspective on training neural networks. Rather
than finding the single-best parameter realization to solve a given
task, the Bayesian approach aims to find the optimal parameter distribution.

In this work we adopt the Bayesian approach and investigate the effect
of variability in the training data on the generalization properties
of wide neural networks. We do so in the limit of infinitely wide
linear and non-linear networks. To obtain analytical insights, we
apply tools from statistical field theory to derive approximate expressions
for the predictive distribution in the NNGP limit. The remainder of
this work is structured in the following way: In \prettyref{sec:Setup}
we describe the setup of supervised learning in shallow and deep networks
in the framework of Bayesian inference and we introduce a synthetic
data set that allows us to control the degree of pattern separability,
dimensionality, and variability of the resulting overlap matrix. In
\prettyref{sec:Results} we develop the field theoretical approach
to learning curves and its application to the synthetic data set as
well as to MNIST \citep{Lecun1998}: \prettyref{sec:Field-theoretic-description-of-bayesian-inference}
presents the general formalism and shows that data variability in
general leads to a non-Gaussian process. Here we also derive perturbative
expressions to characterize the posterior distribution of the network
output. We first illustrate these ideas on the simplest but non-trivial
example of linear Bayesian regression and then generalize them first
to linear and then to non-linear deep networks. In \prettyref{sec:The-Ising-data-set}
we show results for the synthetic data set to obtain interpretable
expressions that allow us to identify how data variability affects
generalization; we then illustrate the identified mechanisms on MNIST.
In \prettyref{sec:Discussion} we summarize our findings, discuss
them in the light of the literature, and provide an outlook.

\section{Setup\label{sec:Setup}}

In this background section we outline the relation between neural
networks, Gaussian processes, and Bayesian inference. We further present
an artificial binary classification task which allows us to control
the degree of pattern separation and variability and test the predictive
power of the theoretical results for the network generalization properties.

\subsection{Neural networks, Gaussian processes and Bayesian inference}

The advent of parametric methods such as neural networks is preceded
by non-parametric approaches such as Gaussian processes. There are,
however, clear connections between the two concepts which allow us
to borrow from the theory of Gaussian processes and Bayesian inference
to describe the seemingly different neural networks. We will here
give a short recap on neural networks, Bayesian inference, Gaussian
processes, and their mutual relations.

\subsubsection{Background: Neural Networks\label{subsec:Background_Neural_Networks}}

In general a feed forward neural network maps inputs $x_{\alpha}\in\mathbb{R}^{N_{\mathrm{dim}}}$
to outputs $y_{\alpha}\in\mathbb{R}^{N_{\mathrm{out}}}$ via the transformations

\begin{align}
h_{\alpha}^{(l)} & =\mathbf{W}^{(l)}\phi^{(l)}\left(h_{\alpha}^{(l-1)}\right)\quad\mathrm{with}\quad h_{\alpha}^{0}=\mathbf{V}x_{\alpha}\,,\nonumber \\
y_{\alpha} & =\mathbf{U}\phi^{(L+1)}\left(h_{\alpha}^{(L)}\right)\,,\label{eq:Deep-FFN-Architecture}
\end{align}
where $\phi^{(l)}(x)$ are activation functions, $\mathbf{V\in}\,\mathbb{R}^{N_{h}\times N_{\mathrm{dim}}}$
are the read-in weights, $N_{\mathrm{dim}}$ is the dimension of the
input, $\mathbf{W}^{(l)}\in\mathbb{R}^{N_{h}\times N_{h}}$ are the
hidden weights, $N_{h}$ denotes the number of hidden neurons, and
$\mathbf{U\in\mathbb{R}}^{N_{\mathrm{out}}\times N_{h}}$ are the
read-out weights. Here $l$ is the layer index $1\le l\le L$ and
$L$ the number of layers of the network; we here assume layer- independent
activation functions $\phi^{(l)}=\phi$. The collection of all weights
are the model parameters\textbf{ }$\text{\ensuremath{\Theta=\{\mathbf{V},\mathbf{W}^{(1)},\ldots,\mathbf{W}^{(L)},\mathbf{U}\}}}$.
The goal of training a neural network in a supervised manner is to
find a set of parameters $\hat{\Theta}$ which reproduces the input-output
relation $\left(x_{\mathrm{tr},\a},y_{\mathrm{tr},\a}\right)_{1\le\alpha\le D}$
for a set of $D$ pairs of inputs and outputs as accurately as possible,
while also maintaining the ability to generalize. Hence one partitions
the data into a training set $\mathcal{D}_{\mathrm{tr}}\,,\,\vert\mathcal{D}_{\mathrm{tr}}\vert=D$,
and a test-set $\mathcal{D}_{\mathrm{test}}\,,\,\vert\mathcal{D}_{\mathrm{test}}\vert=D_{\mathrm{test}}$.
The training data is given in the form of the matrices $\mathbf{x}_{\mathrm{tr}}\in\mathbb{R}^{N_{\mathrm{dim}}\times N_{\mathrm{tr}}}$
and $\mathbf{y}_{\mathrm{tr}}\in\mathbb{R}^{N_{\mathrm{out}}\times N_{\mathrm{tr}}}$.
The quality of how well a neural network is able to model the relation
between inputs and outputs is quantified by a task-dependent loss
function $\mathcal{L}\left(\Theta,x_{\a},y_{\a}\right)$. Starting
with a random initialization of the parameters $\Theta$, one tries
to find an optimal set of parameters $\hat{\Theta}$ that minimizes
the loss $\sum_{\alpha=1}^{D}\mathcal{L}\left(\Theta,x_{\mathrm{tr},\a},y_{\mathrm{tr,}\a}\right)$
on the training set $\mathcal{D}_{\mathrm{tr}}$. The parameters $\hat{\Theta}$
are usually obtained through methods such as stochastic gradient descent.
The generalization properties of the network are quantified after
the training by computing the loss $\mathcal{L}\left(\hat{\Theta},x_{\alpha},y_{\alpha}\right)$
on the test set $\left(x_{\mathrm{test,}\a},y_{\mathrm{test,}\a}\right)\in\mathcal{D}_{\mathrm{test}}$,
which are data samples that have not been used during the training
process. Neural networks hence provide, by definition, a parametric
modeling approach, as the goal is to a find an optimal set of parameters
$\hat{\Theta}$.

\subsubsection{Background: Bayesian inference and Gaussian processes}

The parametric viewpoint in \prettyref{subsec:Background_Neural_Networks}
which yields a point estimate $\hat{\Theta}$ for the optimal set
of parameters can be complemented by considering a Bayesian perspective
\citep{Mackay2003,Murphy22_PML_Part_I,Murphy23_PML_Part_II}: For
each network input $x_{\alpha}$, the network equations \eqref{eq:Deep-FFN-Architecture}
yield a single output $y\left(x_{\alpha}|\Theta\right)$. One typically
considers a stochastic output $y\left(x_{\alpha}|\Theta\right)+\xi_{\alpha}$
where the $\xi_{\alpha}$ are Gaussian independently and identically
distributed (i.i.d.) with variance $\sigma_{\mathrm{reg}}^{2}$ \citep{Williams98}.
This regularization allows us to define the probability distribution
$p\left(y|x_{\alpha},\Theta\right)=\left\langle \delta\left[y_{\mathrm{\alpha}}-y(x_{\mathrm{tr},\alpha}|\Theta)-\xi_{\alpha}\right]\right\rangle _{\xi_{\alpha}}=\N\left(y_{\alpha};\,y(x_{\alpha}|\Theta),\sigma_{\mathrm{reg}}^{2}\right)$.
An alternative interpretation of $\xi_{\alpha}$ is a Gaussian noise
on the labels. Given a particular set of the network parameters $\Theta$
this implies a joint distribution $p\left(\mathbf{y}\vert\mathbf{x}_{\mathrm{tr}},\Theta\right):=\prod_{\alpha=1}^{D}\left\langle \delta\left[y_{\alpha}-y(x_{\mathrm{tr},\alpha}|\Theta)-\xi_{\alpha}\right]\right\rangle _{\{\xi_{\alpha}\}}=\prod_{\alpha=1}^{D}p(y_{\alpha}|x_{\alpha},\Theta)$
of network outputs $\{y_{\alpha}\}_{1\le\alpha\le D}$, each corresponding
to one network input $\{x_{\mathrm{tr},\alpha}\}_{1\le\alpha\le D}$.
One aims to use the training data $\mathcal{D}_{\mathrm{tr}}$ to
compute the posterior distribution for the weights $\mathbf{V},\mathbf{W}^{(1)},\ldots,\mathbf{W}^{(L)},\mathbf{U}$
by conditioning on the network outputs to agree to the desired training
values. Concretely, we here assume as a prior for the model parameters
that the parameter elements $V_{ij},W_{ij}^{(l)},U_{ij}$ are i.i.d.
according to centered Gaussian distributions $V_{ij}\sim\mathcal{N}\left(0,\sigma_{v}^{2}/N_{\mathrm{dim}}\right)$,
$W_{ij}^{(l)}\sim\mathcal{N}\left(0,\sigma_{w}^{2}/N_{h}\right)$,
and $U_{ij}\sim\mathcal{N}\left(\sigma_{u}^{2}/N_{h}\right)$.

The posterior distribution of the parameters $p\left(\Theta\vert\mathbf{x}_{\mathrm{tr}},\mathbf{y}_{\mathrm{tr}}\right)$
then follows from Bayes' theorem as

\begin{equation}
p\left(\Theta\vert\mathbf{x}_{\mathrm{tr}},\mathbf{y}_{\mathrm{tr}}\right)=\frac{p\left(\mathbf{y}_{\mathrm{tr}}\vert\mathbf{x}_{\mathrm{tr}},\Theta\right)\,p\left(\Theta\right)}{p\left(\mathbf{y}_{\mathrm{tr}}\vert\mathbf{x}_{\mathrm{tr}}\right)}\,,\label{eq:Posterior-Weight-Distribution}
\end{equation}
with the likelihood $p\left(\mathbf{y}_{\mathrm{tr}}\vert\mathbf{x}_{\mathrm{tr}},\Theta\right)$,
the weight prior $p\left(\Theta\right)$ and the model evidence $p\left(\mathbf{y}_{\mathrm{tr}}\vert\mathbf{x}_{\mathrm{tr}}\right)=\int d\Theta\,p\left(\mathbf{y}_{\mathrm{tr}}\vert\mathbf{x}_{\mathrm{tr}},\Theta\right)p\left(\Theta\right)$,
which provides the proper normalization. The posterior parameter distribution
$p\left(\Theta\vert\mathbf{x}_{\mathrm{tr}},\mathbf{y}_{\mathrm{tr}}\right)$
also determines the distribution of the network output $y_{*}$ corresponding
to a test-point $x_{*}$ by marginalizing over the parameters $\Theta$

\begin{align}
p\left(y_{*}\vert x_{*},\mathbf{x}_{\mathrm{tr}},\mathbf{y}_{\mathrm{tr}}\right) & =\int d\Theta\,p\left(y_{*}\vert x_{*},\Theta\right)p\left(\Theta\vert\mathbf{x}_{\mathrm{tr}},\mathbf{y}_{\mathrm{tr}}\right)\,,\label{eq:Inference_TestPoints_Bayes}\\
 & =\frac{p\left(y_{*},\mathbf{y}_{\mathrm{tr}}\vert x_{*},\mathbf{x}_{\mathrm{tr}}\right)}{p\left(\mathbf{y}_{\mathrm{tr}}\vert\mathbf{x}_{\mathrm{tr}}\right)}\,.\label{eq:Bayesian_Inference_Supervised_Learning}
\end{align}
One can understand this intuitively: The distribution in \eqref{eq:Posterior-Weight-Distribution}
provides a set of viable parameters $\Theta$ based on the training
data. An initial guess for the correct choice of parameters via the
prior $p\left(\Theta\right)$ is refined, based on whether the choice
of parameters accurately models the relation of the training--data,
which is encapsulated in the likelihood $p\left(\mathbf{y}_{\mathrm{tr}}\vert\mathbf{x}_{\mathrm{tr}},\Theta\right)$.
This viewpoint of Bayesian parameter selection is also equivalent
to what is known as Bayesian deep learning \citep{Murphy23_PML_Part_II}.
The distribution $p\left(y_{*},\mathbf{y}_{\mathrm{tr}}\vert x_{*},\mathbf{x}_{\mathrm{tr}}\right)$
describes the joint network outputs for all training points and the
test point. In the case of wide networks, where $N_{h}\rightarrow\infty$,
\citep{Neal96,Williams96_ae5e3ce4} showed that the distribution of
network outputs $p\left(y_{*},\mathbf{y}_{\mathrm{tr}}\vert x_{*},\mathbf{x}_{\mathrm{tr}}\right)$
approaches a Gaussian process $y\sim\N\left(0,K^{y}\right)$, where
the covariance $\langle y_{\alpha}y_{\beta}\rangle=K_{\alpha\beta}^{y}$
is also denoted as the kernel. This is beneficial, as the inference
for the network output $y_{\ast}$ for a test point $x_{\ast}$ then
also follows a Gaussian distribution with mean and covariance given
by \citep{WilliamsRasmussen06}

\begin{align}
\left\langle y_{\ast}\right\rangle  & =K_{*\alpha}^{y}\left(K^{y}\right)_{\alpha\beta}^{-1}\,y_{\mathrm{tr},\beta}\,,\label{eq:Mean-Inference-Bayesian-GP}\\
\left\langle \left(y_{\ast}-\left\langle y_{\ast}\right\rangle \right)^{2}\right\rangle  & =K_{**}^{y}-K_{*\alpha}^{y}\left(\text{\ensuremath{K^{y}}}\right)_{\alpha\beta}^{-1}\,K_{\beta*}^{y}\,,\label{eq:Variance-Inference-Bayesian-GP}
\end{align}
where summation over repeated indices is implied. There has been extensive
research in relating the outputs of wide neural networks to Gaussian
processes \citep{Neal96,Cho09,Lee18} including recent work on corrections
due to finite-width effects $N_{h}\gg1$ \citep{Cohen21_023034,Naveh21_NeurIPS,ZavatoneVeth21_NeurIPS_I,ZavatoneVeth21_NeurIPS_II,Li21_031059,Segadlo22_103401,Ariosto2022,Roberts22}.

\subsection{Our contribution}

A fundamental assumption of supervised learning is the existence of
a joint distribution $p(x_{\mathrm{tr}},y_{\mathrm{tr}})$ from which
the set of training data as well as the set of test data are drawn.
In this work we follow the Bayesian approach and investigate the effect
of variability in the training data on the generalization properties
of wide neural networks. We do so in the kernel limit of infinitely
wide linear and non-linear networks. Variability here has two meanings:
First, for each drawing of $D$ pairs of training samples $(x_{\mathrm{tr,}\alpha},y_{\mathrm{tr},\alpha})_{1\le\alpha\le D}$
one obtains a $D\times D$ kernel matrix $K^{y}$ with heterogeneous
entries; so in a single instance of Bayesian inference, the entries
of the kernel matrix vary from one entry to the next. Second, each
such drawing of $D$ training data points and one test data point
$\left(x_{*},y_{*}\right)$ leads to a different kernel $\{K_{\alpha\beta}^{y}\}_{1\le\alpha,\beta\le D+1}$,
which follows some probabilistic law $K^{y}\sim p(K^{y})$.

Our work builds upon previous results for the NNGP limit to formalize
the influence of such stochastic kernels. We here develop a field
theoretic approach to systematically investigate the influence of
the underlying kernel stochasticity on the generalization properties
of the network, namely the learning curve, the dependence of $\langle y_{\ast}\rangle$
on the number of training samples $D=|\mathcal{D}_{\mathrm{tr}}|$.
As we assume Gaussian i.i.d. priors on the network parameters, the
output kernel $K_{\alpha\beta}^{y}$ solely depends on the network
architecture and the input overlap matrix

\begin{equation}
K_{\alpha\beta}^{x}=\sum_{i=1}^{N_{\mathrm{dim}}}x_{\alpha i}x_{\beta i}\quad x_{\alpha},x_{\beta}\in\mathcal{D}_{\mathrm{tr}}\cup\mathcal{D}_{\mathrm{test}}\,,\label{eq:Definition_InputdataKernel}
\end{equation}
with $\alpha,\beta=1...D+1$. We next define a data model which allows
us to approximate the probability measure for the data variability.

\subsection{Definition of a synthetic data set\label{sec:The-Ising-data-set}}

To investigate the generalization properties in a binary classification
task, we introduce a synthetic stochastic binary classification task.
This task allows us to control the statistical properties of the data
with regard to the dimensionality of the patterns, the degree of separation
between patterns belonging to different classes, and the variability
in the kernel. Moreover, it allows us to construct training-data sets
$\mathcal{D}_{\mathrm{tr}}$ of arbitrary sizes and we will show that
the statistics of the resulting kernels is indeed representative for
more realistic data sets such as MNIST.

\begin{figure}
\begin{centering}
\includegraphics[width=3.5in]{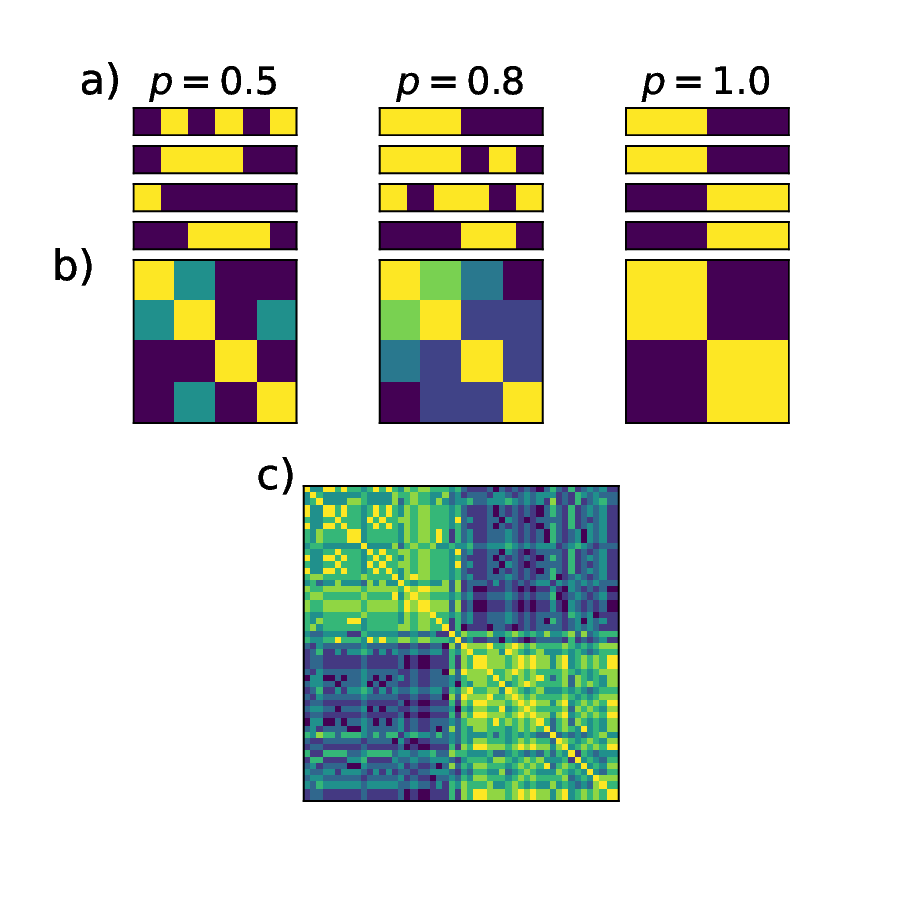}
\par\end{centering}
\caption{\label{fig:SimplePattern-DataExamples-OverlapMatrices}\textbf{Synthetic
data set.} \textbf{a)} Two sample vectors $x^{(1\le\alpha\le4)}$
for each of the two classes (upper, lower). The values for the pixels
can be $x_{i}^{(\alpha)}\in\{-1,1\}$ where yellow indicates $x_{i}^{(\alpha)}=1$
and blue $x_{i}^{(\alpha)}=-1$. The three columns correspond to three
different settings of the pixel probability $p\in\{0.5,0.8,1\}$.\textbf{
b)} Empirical overlap matrices $K^{x}$ \eqref{eq:K_zero} for $D/2=2$
patterns of class $1$ and $D/2=2$ patterns of class $2$. The entries
of the overlap matrices are $K_{\alpha\beta}^{x}\in[-1,1].$ Darker
colors indicate $K_{\alpha\beta}^{x}\approx-1$ and brighter colors
correspond to $K_{\alpha\beta}^{x}\approx1$. \textbf{c)} Empirical
overlap matrix $K^{x}$ for the same task as in a), b) with $D=50$.
Other parameters: $N_{\mathrm{dim}}=6$.}
\end{figure}
The data set consists of pattern realizations $x_{\a}\in\{-1,1\}^{N_{\mathrm{dim}}}$
with dimension $N_{\mathrm{dim}}$ even. We denote the entries $x_{\alpha i}$
of this $N_{\mathrm{dim}}$-dimensional vector for data point $\alpha$
as pixels that randomly take either of two values $x_{\a i}\in\{-1,1\}$
with respective probabilities $p(x_{\a i}=1)$ and $p(x_{\a i}=-1)$
that depend on the class $c(\alpha)\in\{1,2\}$ of the pattern realization
and whether the index $i$ is in the left half ( $i\leq\Ndim/2$)
or the right half ( $i>\Ndim/2$) of the pattern: For class $c(\alpha)=1$
each pixel $x_{\a i}\,\mathrm{with\,}1\le i\le N_{\mathrm{dim}}$
is realized independently as a binary variable as

\begin{align}
x_{\a i} & =\begin{cases}
1 & \mathrm{with}\,\:p\\
-1 & \mathrm{with}\,\:(1-p)
\end{cases}\quad\mathrm{for}\,\,i\leq\frac{N_{\mathrm{dim}}}{2}\,,\label{eq:Pixel-Value-Distribution-Class_1_Part0}\\
x_{\a i} & =\begin{cases}
1 & \mathrm{with}\,\:(1-p)\\
-1 & \mathrm{with}\,\:p
\end{cases}\quad\mathrm{for}\,\,i>\frac{N_{\mathrm{dim}}}{2}\,.\label{eq:Pixel-Value-Distribution-Class_1}
\end{align}
For a pattern $x_{\a}$ in the second class $c(\alpha)=2$ the pixel
values are distributed independently of those in the first class with
a statistics that equals the negative pixel values of the first class,
which is $P\left(x_{\alpha i}\right)=P\left(-x_{\beta i}\right)$
with $c(\beta)=1$ and $c(\alpha)=2$. There are two limiting cases
for $p$ which illustrate the construction of the patterns: In the
limit $p=1$, each pattern $x_{\a}$ in $c=1$ consists of a vector,
where the first $N_{\mathrm{dim}}/2$ pixels have the value $x_{\a i}=1$,
whereas the second half consists of pixels with the value $x_{\a,i}=-1$.
The opposite holds for patterns in the second class $c=2$. This limiting
case is shown in \prettyref{fig:SimplePattern-DataExamples-OverlapMatrices}
(right column). In the limit case $p=0.5$ each pixel assumes the
value $x_{\alpha,i}=\pm1$ with equal probability, regardless of the
pattern class-membership or the pixel position. Hence one cannot distinguish
the class membership of any of the training instances. This limiting
case is shown in \prettyref{fig:SimplePattern-DataExamples-OverlapMatrices}
(left column). If $c(\alpha)=1$ we set $y_{\mathrm{tr},\alpha}=-1$
and for $c(\alpha)=2$ we set $y_{\mathrm{tr},\alpha}=1$.We now investigate
the description of this task in the framework of Bayesian inference.
The hidden variables $h_{\alpha}^{0}$ \eqref{eq:Deep-FFN-Architecture}
in the input layer under a Gaussian prior on $V_{ij}\stackrel{\text{i.i.d.}}{\sim}\N\left(0,\sigma_{v}^{2}/N_{\mathrm{dim}}\right)$
follow a Gaussian process with kernel $K^{(0)}$ given by
\begin{align}
K_{\alpha\beta}^{0} & =\langle h_{\alpha}^{0}h_{\beta}^{0}\rangle_{V\sim\N\left(0,\frac{\sigma_{v}^{2}}{N_{\mathrm{dim}}}\right)}\,,\label{eq:K_zero}\\
 & =\frac{\sigma_{v}^{2}}{N_{\mathrm{dim}}}\,\sum_{i=1}^{N_{\mathrm{dim}}}x_{\alpha i}x_{\beta i}\,.
\end{align}
Separability of the two classes is reflected in the structure of this
input kernel $K^{0}$ as shown in \prettyref{fig:SimplePattern-DataExamples-OverlapMatrices}:
In the cases with $p=0.8$ and $p=1$ one can clearly distinguish
blocks; the diagonal blocks represent intra-class overlaps, the off-diagonal
blocks inter-class overlaps. This is not the case for $p=0.5$, where
no clear block-structure is visible. In the case of $p=0.8$ one can
further observe that the blocks are not as clear-cut as in the case
$p=1$, but rather noisy, similar to $p=0.5$. This is due to the
probabilistic realization of patterns, which induces stochasticity
in the blocks of the input kernel $K^{0}$ \eqref{eq:K_zero}. To
quantify this effect, based on the distribution of the pixel values
\eqref{eq:Pixel-Value-Distribution-Class_1} we compute the distribution
of the entries of $K^{0}$ for the binary classification task. The
mean of the overlap elements $\mu_{\alpha\beta}$ and their covariances
$\Sigma_{(\alpha\beta)(\gamma\delta)}$ are defined via
\begin{align}
\mu_{\alpha\beta} & =\left\langle K_{\alpha\beta}^{0}\right\rangle \,,\label{eq:Mean_InputOverlap_Definition}\\
\Sigma_{(\alpha\beta)(\gamma\delta)} & =\left\langle \delta K_{\alpha\beta}^{0}\,\delta K_{\gamma\delta}^{0}\right\rangle \,,\label{eq:Covariance_InputOverlap_Definition}\\
\delta K_{\alpha\beta}^{0} & =K_{\alpha\beta}^{0}-\mu_{\alpha\beta}\,,\label{eq:Deviation_Mean_Definition}
\end{align}
where the expectation value $\left\langle \cdot\right\rangle $ is
taken over drawings of $D$ training samples each. By construction
we have $\mu_{\alpha\beta}=\mu_{\beta\alpha}$. The covariance is
further invariant under the exchange of $(\alpha,\beta)\leftrightarrow(\gamma,\delta)$
and, due to the symmetry of $K_{\alpha\beta}^{0}=K_{\beta\alpha}^{0}$,
also under swapping $\alpha\leftrightarrow\beta$ and $\gamma\leftrightarrow\delta$
separately. In the artificial task-setting, the parameter $p$, the
pattern dimensionality $N_{\mathrm{dim}}$, and the variance $\sigma_{v}^{2}/N_{\mathrm{dim}}$
of each read-in weight $V_{ij}$ define the elements of $\mu_{\alpha\beta}$
and $\Sigma_{(\alpha\beta)(\gamma\delta)}$, which read (see \prettyref{supp:SupplementStatisticalProperties})

\begin{align}
\mu_{\alpha\beta} & =\sigma_{v}^{2}\,\begin{cases}
1 & \alpha=\beta\\
u & c_{\alpha}=c_{\beta}\\
-u & c_{\alpha}\neq c_{\beta}
\end{cases}\,,\nonumber \\
\Sigma_{(\alpha\beta)(\alpha\beta)} & =\frac{\sigma_{v}^{4}}{N_{\mathrm{dim}}}\kappa,\nonumber \\
\Sigma_{(\alpha\beta)(\alpha\delta)} & =\frac{\sigma_{v}^{4}}{N_{\mathrm{dim}}}\begin{cases}
\nu & \mathrm{for}\,\begin{cases}
c_{\alpha}=c_{\text{\ensuremath{\beta}}}=c_{\text{\ensuremath{\delta}}}\\
c_{\alpha}\neq c_{\beta}=c_{\delta}
\end{cases}\\
-\nu & \mathrm{for}\,\begin{cases}
c_{\alpha}=c_{\beta}\neq c_{\delta}\\
c_{\alpha}=c_{\delta}\neq c_{\beta}
\end{cases}
\end{cases}\,,\nonumber \\
\mathrm{with\quad}\kappa & :=1-u^{2}\,,\nonumber \\
\nu & :=u\,(1-u)\,,\nonumber \\
u & :=4p(p-1)+1\,.\label{eq:Statistical_Properties_SimplePattern_Setup}
\end{align}

In addition to this, the tensor elements of $\Sigma_{(\alpha\beta)(\gamma\delta)}$
are zero for the following index combinations because we fixed the
value of $K_{\alpha\alpha}^{0}$ by construction:

\begin{align}
\Sigma_{(\alpha\beta)(\gamma\delta)} & =0\quad\mathrm{with\quad\alpha\neq\beta\neq\gamma\neq\delta}\,,\nonumber \\
\Sigma_{(\alpha\alpha)(\beta\gamma)} & =0\quad\text{with\ensuremath{\quad\alpha\neq\beta\neq\gamma}}\,,\nonumber \\
\Sigma_{(\alpha\alpha)(\beta\beta)} & =0\quad\mathrm{with\quad\alpha\neq\beta}\,,\nonumber \\
\Sigma_{(\alpha\alpha)(\alpha\beta)} & =0\quad\mathrm{with\quad\alpha\neq\beta}\,,\nonumber \\
\Sigma_{(\alpha\alpha)(\alpha\alpha)} & =0\quad\mathrm{with\quad\alpha\neq\beta}\,.
\end{align}

The expressions for $\Sigma_{(\alpha\beta)(\alpha\beta)}$ and $\Sigma_{(\alpha\beta)(\alpha\delta)}$
in \eqref{eq:Statistical_Properties_SimplePattern_Setup} show that
the magnitude of the fluctuations are controlled through the parameter
$p$ and the pattern dimensionality $N_{\mathrm{dim}}$: The covariance
$\Sigma$ is suppressed by a factor of $1/N_{\mathrm{dim}}$ compared
to the mean values $\mu$. Hence we can use the pattern dimensionality
$N_{\mathrm{dim}}$ to investigate the influence of the strength of
fluctuations. As illustrated in \prettyref{fig:graphical_abstract}a,
the elements $\Sigma_{(\alpha\beta)(\alpha\beta)}$ denote the variance
of individual entries of the kernel, while $\Sigma_{(\alpha\beta)(\alpha\gamma)}$
are covariances of entries across elements of a given row $\alpha$,
visible as horizontal or vertical stripes in the color plot of the
kernel.

Equation \prettyref{eq:Statistical_Properties_SimplePattern_Setup}
implies, by construction, a Gaussian distribution of the elements
$K_{\alpha\beta}^{0}$ as it only provides the first two cumulants.
One can show that the higher-order cumulants of $K_{\alpha\beta}^{0}$
scale sub-leading in the pattern dimension and are hence suppressed
by a factor $\mathcal{O}\left(1/N_{\mathrm{dim}}\right)$ compared
to $\Sigma_{(\alpha\beta)(\gamma\delta)}$.

\section{Results\label{sec:Results}}

In this section we derive the field theoretic formalism which allows
us to compute the statistical properties of the inferred network output
in Bayesian inference with a stochastic kernel. We show that the resulting
process is non-Gaussian and reminiscent of a $\varphi^{3}+\varphi^{4}$-theory.
Specifically, we compute the mean of the predictive distribution of
this process conditioned on the training data. This is achieved by
employing systematic approximations with the help of Feynman diagrams.

Subsequently we show that our results provide an accurate bound on
the generalization capabilities of the network. We further discuss
the implications of our analytic results for neural architecture search.

\subsection{Field theoretic description of Bayesian inference\label{sec:Field-theoretic-description-of-bayesian-inference}}

\subsubsection{Bayesian inference with stochastic kernels}

In general, a network implements a map from the inputs $x_{\alpha}$
to corresponding outputs $y_{\alpha}$. In particular a model of the
form \prettyref{eq:Deep-FFN-Architecture} implements a non-linear
map $\psi:\mathbb{R}^{N_{\mathrm{dim}}}\rightarrow\mathbb{R}^{N_{h}}$
of the input $x_{\alpha}\in\mathbb{R}^{N_{\mathrm{dim}}}$ to a hidden
state $h_{\alpha}\in\mathbb{R}^{N_{h}}$. This map may also involve
multiple hidden-layers, biases and non-linear transformations. The
read-out weight $\mathbf{U}\in\mathbb{R}^{1\times N_{h}}$ links the
scalar network output $y_{\alpha}\in\mathbb{R}$ and the transformed
inputs $\psi\left(x_{\alpha}\right)$ with $1\le\alpha\le D_{\mathrm{tot}}=D+D_{\mathrm{test}}$
which yields

\begin{equation}
y_{\alpha}=\mathbf{U}\,\psi\left(x_{\alpha}\right)+\xi_{\alpha}\,,\label{eq:General_NetworkOutput}
\end{equation}
where $\xi_{\alpha}\stackrel{\text{i.i.d.}}{\sim}\N(0,\sigma_{\mathrm{reg}}^{2})$
is a regularization noise in the same spirit as in \citep{Williams98}.
 We assume that the prior on the read-out vector elements is a Gaussian
$\mathbf{U}_{i}\text{\ensuremath{\stackrel{\text{i.i.d.}}{\sim}\mathcal{N}\left(0,\sigma_{u}^{2}/N_{h}\right)}. }$The
distribution of the set of network outputs $y_{1\le\alpha\le D_{\mathrm{tot}}}$
is then in the limit $N_{h}\rightarrow\infty$ a multivariate Gaussian
\citep{Neal96}. The kernel matrix of this Gaussian is obtained by
taking the expectation value with respect to the read-out vector,
which yields

\begin{align}
\left\langle y_{\alpha}\,y_{\beta}\right\rangle _{\mathbf{U}} & =:K_{\alpha\beta}^{y}=\sigma_{u}^{2}\,K_{\alpha\beta}^{\psi}+\delta_{\alpha\beta}\,\sigma_{\mathrm{reg}}^{2}\,,\label{eq:GeneralOverlap_Definition}\\
K_{\alpha\beta}^{\psi} & =\frac{1}{N_{h}}\,\sum_{i=1}^{N_{h}}\psi_{i}\left(x_{\alpha}\right)\,\psi_{i}\left(x_{\beta}\right)\,.\label{eq:Empirical_Input_Overlap}
\end{align}
The kernel matrix $K_{\alpha\beta}^{y}$ describes the covariance
of the network's output and hence depends on the kernel matrix $K_{\alpha\beta}^{\psi}$.
The additional term $\delta_{\alpha\beta}\,\sigma_{\mathrm{reg}}^{2}$
acts as a regularization term, which is also known as a ridge regression
\citep{Hoerl00} or Tikhonov regularization \citep{Williams06}. In
the context of neural networks one can motivate the regularizer $\sigma_{\mathrm{reg}}^{2}$
by using the $L^{2}$-regularization in the readout layer. This is
also known as weight decay \citep{Goodfellow16_deepbook}. Introducing
the regularizer $\sigma_{\mathrm{reg}}^{2}$ is necessary to ensure
that one can properly invert the matrix $K_{\alpha\beta}^{y}$, ensuring
that the expressions \eqref{eq:Mean-Inference-Bayesian-GP} and \eqref{eq:Variance-Inference-Bayesian-GP}
are numerically stable.

Different drawings of sets of training data $\mathcal{D}_{\mathrm{tr}}$
lead to different realizations of kernel matrices $K^{\psi}$ and
$K^{y}$. The network output $y_{\alpha}$ hence follows a multivariate
Gaussian with a stochastic kernel matrix $K^{y}$. A more formal derivation
of the Gaussian statistics, including an argument for its validity
in deep neural networks, can be found in \citep{Lee18}. A consistent
derivation using field theoretical methods and corrections in terms
for the width of the hidden layer $N_{h}$ for deep and recurrent
networks has been presented in \citep{Segadlo22_103401}.

In general, the input kernel matrix $K^{0}$ \eqref{eq:K_zero} and
the output kernel matrix $K^{y}$ are related in a non-trivial fashion,
which depends on the specific network architecture at hand. From now
on we make an assumption on the stochasticity of $K^{0}$ and assume
that the input kernel matrix $K^{0}$ is distributed according to
a multivariate Gaussian 
\begin{align}
K^{0} & \sim\mathcal{N}\left(\mu,\:\Sigma\right)\,,\label{eq:Distribution_OverlapElements}
\end{align}
where $\mu$ and $\Sigma$ are given by \eqref{eq:Mean_InputOverlap_Definition}
and \eqref{eq:Covariance_InputOverlap_Definition}, respectively.

In the limit of large pattern dimensions $N_{\mathrm{dim}}\gg$1
this assumption is warranted for the kernel matrix $K^{0}$. This
structure further assumes, that the overlap statistics are unimodal,
which is indeed mostly the case for data such as MNIST (see \prettyref{app:Appendix_OverlapDistributions}).
Furthermore we assume that this property holds for the output kernel
matrix $K^{y}$ as well and that we can find a mapping from the mean
$\mu$ and covariance $\Sigma$ of the input kernel to the mean $m$
and covariance $C$ of the output kernel $\left(\mu_{\alpha\beta},\:\Sigma_{(\alpha\beta)(\gamma\delta)}\right)\rightarrow\left(m_{\alpha\beta},\:C_{(\alpha\beta)(\gamma\delta)}\right)$
so that $K^{y}$ is also distributed according to a multivariate Gaussian

\begin{align}
K^{y} & \sim\N(m,\:C)\,.\label{eq:Distribution_Output_KernelElements_Gaussian}
\end{align}
For each realization $K_{\alpha\beta}^{y}$, the joint distribution
of the network outputs $y_{1\le\alpha\le D_{\mathrm{tot}}}$ corresponding
to the training and test data points $\mathbf{x}$ follow a multivariate
Gaussian

\begin{equation}
p\left(\mathbf{y}\vert\mathbf{x}\right)\sim\N\big(0,K^{y}\big)\,.\label{eq:Posterior_Distribution}
\end{equation}

The kernel allows us to compute the conditional probability $p\left(y_{\ast}\vert\mathbf{x}_{\mathrm{tr}},\mathbf{y}_{\mathrm{tr}},x_{\ast}\right)$,
as defined in \prettyref{eq:Inference_TestPoints_Bayes}, for a test
point $(x_{\ast},y_{\ast})\in\mathcal{D}_{\mathrm{test}}$ conditioned
on the data from the training set $(\mathbf{x}_{\mathrm{tr}},\mathbf{y}_{\mathrm{tr}})\in\mathcal{D}_{\mathrm{tr}}$.
This distribution is Gaussian with mean and variance given by \prettyref{eq:Mean-Inference-Bayesian-GP}
and \prettyref{eq:Variance-Inference-Bayesian-GP}, respectively.
It is our goal to take into account that $K^{0}$ is a stochastic
quantity, which depends on the particular draw of the training and
test data set $(\mathbf{x}_{\mathrm{tr}},\mathbf{y}_{\mathrm{tr}})\in\mathcal{D}_{\mathrm{tr}},(x_{\ast},y_{\ast})\in\mathcal{D}_{\mathrm{test}}$.
The labels $\mathbf{y}_{\mathrm{tr}},y_{*}$ are, by construction,
deterministic and take either one of the values $\pm1$. In the following
we investigate the dependence of the mean of the predictive distribution
on the number of training samples, which we call the learning curve.
A common assumption is that this learning curve is rather insensitive
to the very realization of the chosen training points. Thus we assume
that the learning curve is self-averaging. The mean computed for a
single draw of the training data is hence expected to agree well to
the average over many such drawings. Under this assumption it is
sufficient to compute the data-averaged mean inferred network output,
which reduces to computing the disorder-average of the following quantity

\begin{align}
\left\langle y_{*}\right\rangle _{K^{y}} & =\left\langle K_{*\alpha}^{y}\left[K^{y}\right]_{\alpha\beta}^{-1}\right\rangle _{K^{y}}y_{\beta}\,.\label{eq:Def_DisorderAveraged_MeanInferredNetworkOutput}
\end{align}
To perform the disorder average and to compute perturbative corrections,
we will follow these steps
\begin{itemize}
\item construct a suitable dynamic moment-generating function $Z(K^{y})$
,
\item propagate the input stochasticity to the network output $K_{\alpha\beta}^{0}\rightarrow K_{\alpha\beta}^{y}$,
\item disorder-average the functional using the model $K_{\alpha\beta}^{y}\sim\N(m_{\alpha\beta},\:C_{(\alpha\beta)(\gamma\delta)})$,
\item and finally perform the computation of perturbative corrections using
diagrammatic techniques.
\end{itemize}

\subsubsection{Constructing the dynamic moment generating function}

Our ultimate goal is to compute learning curves. Therefore we want
to evaluate the disorder averaged mean inferred network output \eqref{eq:Def_DisorderAveraged_MeanInferredNetworkOutput}.
Both the presence of two correlated random matrices and the fact that
one of the matrices appears as an inverse complicate this process.
One alternative route is to define the moment-generating function
 
\begin{align}
Z(l_{*}) & =\int dy_{*}\exp(l_{*}y_{*})p\left(y_{*}\vert x_{*},\mathbf{x}_{\mathrm{tr}},\mathbf{y}_{\mathrm{tr}}\right)\,,\\
 & =\frac{\int dy_{*}\exp(l_{*}y_{*})p\left(y_{*},\mathbf{y}_{\mathrm{tr}}\vert x_{*},\mathbf{x}_{\mathrm{tr}}\right)}{p\left(\mathbf{y}_{\mathrm{tr}}\vert\mathbf{x}_{\mathrm{tr}}\right)}\,,\\
 & =:\frac{\mathcal{Z}(l_{*})}{\mathcal{Z}(0)}\,,
\end{align}
with joint Gaussian distributions $p\left(y_{*},\mathbf{y}_{\mathrm{tr}}\vert x_{*},\mathbf{x}_{\mathrm{tr}}\right)$
and $p\left(\mathbf{y}_{\mathrm{tr}}\vert\mathbf{x}_{\mathrm{tr}}\right)$
that each can be readily averaged over $K^{y}$. Equation \eqref{eq:Def_DisorderAveraged_MeanInferredNetworkOutput}
is then obtained as 
\begin{equation}
\left\langle y_{*}\right\rangle _{K^{y}}=\frac{\partial}{\partial l_{*}}\left\langle \frac{\mathcal{Z}(l_{*})}{\mathcal{Z}(0)}\right\rangle _{K^{y}}\bigg\rvert_{l_{*}=0}\,.
\end{equation}
A complication of this approach is that the numerator and denominator
co-fluctuate. The common route around this problem is to consider
the cumulant-generating function $W(l_{*})=\ln\Z(l_{*})$ and to obtain
$\left\langle y_{*}\right\rangle _{K^{y}}=\frac{\partial}{\partial l_{*}}\left\langle W(l_{\ast})\right\rangle _{K^{y}}$,
which, however, requires averaging the logarithm. This is commonly
done with the replica trick \citep{Fischer91,Mezard_Montanari09}.

We here follow a different route to ensure that the disorder-dependent
normalization $\mathcal{Z}(0)$ drops out and construct a dynamic
moment generating function \citep{DeDominicis78_4913}. Our goal is
hence to design a dynamic process where a time dependent observable
is related to $y_{*}$, our mean-inferred network output. We hence
define the linear process in the auxiliary variables $q_{\alpha}$

\begin{align}
\frac{\partial q_{\alpha}(t)}{\partial t} & =-K_{\alpha\beta}^{y}\,q_{\beta}(t)+y_{\alpha}\,,\label{eq:Dynamics_for_DynamicMGF}
\end{align}
for $(x_{\alpha},y_{\alpha})\in\mathcal{D}_{\mathrm{tr}}$. From this
we see directly that $q_{\alpha}(t\rightarrow\infty)=\left[K^{y}\right]_{\alpha\beta}^{-1}\,y_{\ensuremath{\beta}}$
is a fixpoint. The fact that $K_{\alpha\beta}^{y}$ is a covariance
matrix ensures that it is positive semi-definite and hence implies
the convergence to a fixpoint. We can obtain \prettyref{eq:Mean-Inference-Bayesian-GP}
$\left\langle y_{*}\right\rangle =K_{*\alpha}^{y}\left[K^{y}\right]_{\alpha\beta}^{-1}\,y_{\beta}$
from \eqref{eq:Dynamics_for_DynamicMGF} as a linear readout of $q_{\alpha}(t\rightarrow\infty)$
with the matrix $K_{*\alpha}^{y}$. Using the Martin-Siggia-Rose-deDominicis-Janssen
(MSRDJ) formalism \citep{Martin73,janssen1976_377,Stapmanns20_042124,Helias20_970}
one can express this as the first functional derivative of the moment
generating function $Z(L_{*},K^{y})$ in frequency space

\begin{align}
Z(L_{*},K^{y}) & =\int\mathcal{D}\lbrace Q,\tilde{Q}\rbrace\exp\left(S(Q,\tilde{Q},L_{*})\right)\,,\label{eq:Def_DynamicMGF_Frequency}\\
S(Q,\tilde{Q},L_{*}) & =\tilde{Q}_{\alpha}^{\top}\left(i\omega\mathbb{I}+K^{y}\right)_{\alpha\beta}Q_{\beta}\label{eq:action_S}\\
 & -\tilde{Q}_{\alpha}^{0}y_{\alpha}\nonumber \\
 & +K_{*\alpha}^{y}L_{*}^{\top}Q_{\alpha}\,,\label{eq:Action_Dynamic_MGF_Frequency}
\end{align}
where $\tilde{Q}_{\alpha}^{\top}\,(\cdots)\,Q_{\beta}=\frac{1}{2\pi}\int d\omega\tilde{Q}_{\alpha}(-\omega)\,(\cdots)\,Q_{\beta}(\text{\ensuremath{\omega)}}$
and $\tilde{Q}_{\alpha}(\omega=0)=\tilde{Q}_{\alpha}^{0}$. As $Z(L_{*},K^{y})$
is normalized such that $Z(0,K^{y})=1\quad\forall\,K^{y}$, we can
compute \eqref{eq:Def_DisorderAveraged_MeanInferredNetworkOutput}
by evaluating the functional derivative of the disorder-averaged moment-generating
function $\overline{Z}(L_{*})$ (see \prettyref{app:DynamicMGF})
at $t\rightarrow\infty$:

\begin{align}
\overline{Z}\left(L_{*}\right) & =\left\langle \int\mathcal{D}\lbrace Q,\tilde{Q}\rbrace\exp\left(S(Q,\tilde{Q},L_{*})\right)\right\rangle _{K^{y}}\,,\label{eq:Def_DisorderAveragedDynamic_MGF}\\
\left\langle y_{*}\right\rangle _{K^{y}} & =\lim_{t\rightarrow\infty}\int d\omega\exp(i\omega t)\frac{\delta\overline{Z}\left(L_{*}\right)}{\delta L_{*}(-\omega)}\bigg\rvert_{L_{*}(-\omega)=0}\,.\label{eq:Def_MeanInferredOutput_DynamicMGF}
\end{align}
By construction the distribution of the kernel matrix entries $K_{\alpha\beta}^{y}$
is a multivariate Gaussian \prettyref{eq:Distribution_OverlapElements}.
In the following we will treat the stochasticity of $K_{\alpha\beta}^{y}$
perturbatively to gain insights into the influence of input stochasticity.

\subsubsection{Perturbative treatment of the disorder averaged moment generating
function }

To compute the disorder averaged mean-inferred network output \eqref{eq:Def_DisorderAveraged_MeanInferredNetworkOutput}
we need to compute the disorder average of the dynamic moment generating
function $\overline{Z}(L_{*})=\left\langle Z(L_{*},K^{y})\right\rangle _{K^{y}}$
and its functional derivative at $L_{*}(\omega)=0$. Due to the linear
appearance of $K^{y}$ in the action \prettyref{eq:action_S} and
the Gaussian distribution for $K^{y}$ \eqref{eq:Distribution_Output_KernelElements_Gaussian}
we perform the disorder average directly and obtain the action

\begin{align}
\overline{Z}\left(L_{*}\right) & =\left\langle Z(L_{*},K^{y})\right\rangle _{K^{y}}\\
 & :=\int\mathcal{D}\lbrace Q,\tilde{Q}\rbrace\exp\left(\overline{S}(Q,\tilde{Q},L_{*})\right)\,,\\
\overline{S}(Q,\tilde{Q},L_{*}) & =\tilde{Q}_{\alpha}^{\top}\left[i\omega\mathbb{I}_{\alpha\beta}+m_{\alpha\beta}\right]Q_{\beta}\label{eq:Disorderaveraged_DynamicMGF}\\
 & -\tilde{Q}_{\alpha}^{0}y_{\alpha}\nonumber \\
 & +m_{*\alpha}L_{*}^{\top}Q_{\alpha}\nonumber \\
 & +\frac{1}{2}C_{(\alpha\beta)(\gamma\delta)}\tilde{Q}_{\alpha}^{\top}Q_{\beta}\tilde{Q}_{\gamma}^{\top}Q_{\delta}\nonumber \\
 & +C_{(*\alpha)(\beta\gamma)}L_{*}^{\top}Q_{\alpha}\tilde{Q}_{\beta}^{\top}Q_{\gamma}+\mathcal{O}\left(L_{*}^{2}\right)\,,\label{eq:Def_Action_DisorderAveragedMGF}
\end{align}
with $\tilde{Q}^{0}:=\tilde{Q}(\omega=0)$ (for details see \prettyref{app:DynamicMGF}).
As we ultimately aim to obtain corrections for the mean inferred network
output $\left\langle y_{\ast}\right\rangle _{K^{y}}$, we utilize
the action in \eqref{eq:Def_Action_DisorderAveragedMGF} and established
results from field theory to derive the leading order terms as well
as perturbative corrections diagrammatically. The presence of the
variance and covariance terms in \eqref{eq:Def_Action_DisorderAveragedMGF}
introduces corrective factors, which cannot appear in the zeroth-order
approximation, which corresponds to the homogeneous kernel that neglects
fluctuations in $K^{y}$ by setting $C_{(\alpha\beta)(\gamma\delta)}=0$.
This will provide us with the tools to derive an asymptotic bound
for the mean inferred network output $\left\langle y^{*}\right\rangle $
in the case of an infinitely large training data set. This bound is
directly controlled by the variability in the data. We provide empirical
evidence for our theoretical results for linear, non-linear, and deep-kernel-settings
and show how the results could serve as indications to aid neural
architecture search based on the statistical properties of the underlying
data set.

\subsubsection{Field theoretic elements to compute the mean inferred network output
$\left\langle y_{\ast}\right\rangle _{K^{y}}$\label{subsec:Field-theoretic-elements-diagrams}}

The field theoretic description of the inference problem in form of
an action \eqref{eq:Def_Action_DisorderAveragedMGF} allows us to
derive perturbative expressions for the statistics of the inferred
network output $\left\langle y_{*}\right\rangle _{K^{y}}$ in a diagrammatic
manner. This diagrammatic treatment for perturbative calculations
is a powerful tool and is standard practice in statistical physics
\citep{ZinnJustin96}, data analysis and signal reconstruction \citep{Ensslin2010a},
and more recently in the investigation of artificial neural networks
\citep{Dyer19_1909,Fischer22_043143}.

Comparing the action \eqref{eq:Def_Action_DisorderAveragedMGF} to
prototypical expressions from classical statistical field theory such
as the $\varphi^{3}+\varphi^{4}$ theory\citep{ZinnJustin96,Helias20_970}
one can similarly associate the elements of a field theory:
\begin{itemize}
\item $-\tilde{Q}_{\alpha}^{0}y_{\alpha}\doteq$\mbox{\centering \begin{fmffile}{SourceTerm_Monopole_Data_Small}  \begin{fmfgraph*}(10,20) \fmfbottom{i1,i2} \fmfv{decor.shape=circle,decor.filled=empty, decor.size=2thick}{i1} \fmf{plain,width=0.3mm}{i1,i2} \end{fmfgraph*}  \end{fmffile}}is
a monopole term
\item $m_{*\alpha}L_{*}^{\top}Q_{\alpha}\doteq$ \mbox{\begin{fmffile}{SourceTerm_Monopole_Inference_Small} \begin{fmfgraph*}(10,20) \fmfbottom{i1,i2} \fmfv{decor.shape=circle,decor.filled=full, decor.size=2thick}{i1} \fmf{plain,width=0.3mm}{i1,i2} \end{fmfgraph*}  \end{fmffile}}is
a source term
\item $\Delta_{\alpha\beta}:=\left(-i\omega\mathbb{I}-m\right)_{\alpha\beta}^{-1}\doteq$
\mbox{\centering \begin{fmffile}{Propagator_Small}  \begin{fmfgraph*}(20,20) \fmfbottom{i1,i2} \fmf{dashes,width=0.3mm}{i1,i2} \end{fmfgraph*} \end{fmffile}}
is a propagator that connect the fields $Q_{\alpha}(\omega),\tilde{Q}_{\beta}\left(-\omega\right)$
\item $C_{(*\alpha)(\beta\gamma)}L_{*}^{\top}Q_{\alpha}\tilde{Q}_{\beta}^{\top}Q_{\gamma}\doteq$\mbox{\centering \begin{fmffile}{ThreePointVertex_Small} \begin{fmfgraph*}(35,20) \fmfleft{i1,i2} \fmf{plain,width=0.3mm,foreground=black,tension=1.5}{i1,v1} \fmf{plain,width=0.3mm,foreground=black,tension=1.5}{i2,v1} \fmf{photon,width=0.3mm,foreground=black}{v1,v2} \fmf{plain,width=0.3mm,foreground=black,tension=1.5}{o1,v2} \fmfright{o1,o2} \fmf{plain,foreground=black,tension=1.5}{v2,o2} \fmfv{decor.shape=circle,decor.filled=full, decor.size=2thick}{i2}
\end{fmfgraph*}  \end{fmffile}}is a three-point vertex
\item $\frac{1}{2}C_{(\alpha\beta)(\gamma\delta)}\tilde{Q}_{\alpha}^{\top}Q_{\beta}\tilde{Q}_{\gamma}^{\top}Q_{\delta}\doteq$\mbox{\centering \begin{fmffile}{FourPointVertex_Small}  \begin{fmfgraph*}(35,20) \fmfleft{i1,i2} \fmf{plain,width=0.3mm,foreground=black}{i1,v1} \fmf{plain,width=0.3mm,foreground=black}{i2,v1} \fmf{photon,width=0.3mm,foreground=black}{v1,v2} \fmf{plain,width=0.3mm,foreground=black}{o1,v2} \fmfright{o1,o2} \fmf{plain,foreground=black}{o2,v2} \end{fmfgraph*}  \end{fmffile}}
is a four-point vertex.
\end{itemize}
The following rules for Feynman diagrams simplify calculations:
\begin{enumerate}
\item To obtain corrections to first order in $C\sim\mathcal{O}\left(1/N_{\mathrm{dim}}\right)$,
one has to compute all diagrams with a single vertex (three-point
or four-point) \citep{Helias20_970}. This approach assumes that the
interaction terms $C_{(\alpha\ensuremath{\beta})(\gamma\delta)}$
that stem from the variability of the data are small compared to the
mean $m_{\alpha\beta}$. In the case of strong heterogeneity one cannot
use a conventional expansion in the number of vertices $C_{(\alpha\beta)(\gamma\delta)}$
and would have to resort to other methods.
\item Vertices, source terms, and monopoles have to be connected with one
another using the propagator $\Delta_{\alpha\beta}=\left(-i\omega\mathbb{I}-m\right)_{\alpha\beta}^{-1}$
which couple $Q_{\alpha}(\omega)$ and $\tilde{Q}_{\beta}(-\omega)$
which each other.
\item We only need diagrams with a single external source term $L_{*}$
because we seek corrections to the mean-inferred network output.
\item The structure of the integrals appearing in the four-point and three-point
vertices containing $C_{(\alpha\beta)(\gamma\delta)}$ with contractions
by $\Delta_{\alpha\beta}$ or $\Delta_{\gamma\delta}$ within a pair
of indices $(\alpha\beta)$ or $(\gamma\delta)$ yield vanishing contributions;
such diagrams are known as closed response loops \citep{Helias20_970}.
This is because the propagator $\Delta_{\alpha\beta}(t-s)$ in time
domain vanishes for $t=s$, which corresponds to the integral $\int d\omega\,\Delta_{\alpha\beta}(\omega)$
over all frequencies $\omega$A detailed explanation is given in \prettyref{app:DynamicMGF}.
\item As we have frequency conservation at the vertices in the form $\frac{1}{2}\,\tilde{Q}_{\alpha}^{\top}Q_{\beta}\,C_{(\alpha\beta)(\gamma\delta)}\,\tilde{Q}_{\gamma}^{\top}Q_{\delta}$
and since by point 4. above we only need to consider contractions
by $\Delta_{\beta\gamma}$ or $\Delta_{\delta\alpha}$ by attaching
the external legs all frequencies are constrained to $\omega=0$,
so also propagators within a loop are replaced by $\Delta_{\alpha\beta}=\left(-i\omega\mathbb{I}-m\right)_{\alpha\beta}^{-1}\rightarrow-(m^{-1})_{\alpha\beta}$.
\end{enumerate}
These rules directly yield that the corrections for the disorder averaged
mean-inferred network to first order in $C_{(\alpha\beta)(\gamma\delta)}$
can only include the diagrams (see \prettyref{app:DynamicMGF})

 \begin{fmffile}{InferenceEquationFirstOrder}\begin{align} \left\langle y_*\right\rangle &\dot{=}\;\;\underbrace{\begin{gathered}\begin{fmfgraph*}(20,15)  \fmfleft{i1} \fmfright{o1}
\fmfv{decor.shape=circle,decor.filled=full, decor.size=2thick}{i1} \fmfv{decor.shape=circle,decor.filled=empty, decor.size=2thick}{o1}
\fmf{plain,foreground=black,tension=3.5}{i1,v1} \fmf{dashes,foreground=black,width=0.5mm}{v1,v2} \fmf{plain,width=0.5mm,tension=2.5,color=black}{v2,o1}
\end{fmfgraph*} \end{gathered}}_{\left\langle y_*\right\rangle_0} \; +\;\underbrace{\begin{gathered}\begin{fmfgraph*}(45,15)  \fmfleft{i1,i2} \fmfright{o1,o2}
\fmf{plain,width=0.5mm,foreground=black}{v2,o2} \fmf{plain,width=0.5mm,foreground=black}{v1,i2} \fmf{photon,width=0.5mm,foreground=black,tension=2}{v1,v2}  \fmf{dashes,width=0.5mm,foreground=black,left=0.3}{i2,o2}
\fmf{phantom,width=0.5mm,pull=7}{i1,v1}  \fmf{phantom,width=0.5mm,pull=7}{o1,v2}  \fmf{phantom,width=0.5mm,pull=7}{v1,i2}  \fmf{phantom,width=0.5mm,pull=7}{v2,o2} 
\fmffreeze \fmfv{decor.shape=circle,decor.filled=full, decor.size=2thick}{i1} \fmfv{decor.shape=circle,decor.filled=empty, decor.size=2thick}{o1}
\fmf{plain,foreground=black,tension=4}{v5,i1} \fmf{dashes,foreground=black,width=0.5mm}{v5,v9} \fmf{plain,width=0.5mm,tension=2.5,color=black}{v1,v9} 
\fmf{plain,foreground=black,tension=4}{v7,o1} \fmf{dashes,foreground=black,width=0.5mm}{v7,v8} \fmf{plain,width=0.5mm,tension=2.5,color=black}{v2,v8}
\end{fmfgraph*} \end{gathered}\;-\;\begin{gathered}\begin{fmfgraph*}(45,15)  \fmfleft{i1,i2} \fmfright{o1,o2}
\fmf{plain,width=0.5mm,foreground=black}{v2,o2} \fmf{plain,width=0.5mm,foreground=black}{v1,i2} \fmf{photon,width=0.5mm,foreground=black,tension=2}{v1,v2}  \fmf{dashes,width=0.5mm,foreground=black,left=0.3}{i2,o2}
\fmf{phantom,width=0.5mm,pull=7}{i1,v1}  \fmf{phantom,width=0.5mm,pull=7}{o1,v2}  \fmf{phantom,width=0.5mm,pull=7}{v1,i2}  \fmf{phantom,width=0.5mm,pull=7}{v2,o2} 
\fmffreeze \fmfv{decor.shape=circle,decor.filled=empty, decor.size=2thick}{o1}
\fmf{phantom,foreground=black,tension=4}{v5,i1} \fmf{phantom,foreground=black,width=0.5mm}{v5,v9} \fmf{plain,width=0.5mm,tension=1,color=black}{v1,v9}  \fmfv{decor.shape=circle,decor.filled=full, decor.size=2thick}{v9}
\fmf{plain,foreground=black,tension=4}{v7,o1} \fmf{dashes,foreground=black,width=0.5mm}{v7,v8} \fmf{plain,width=0.5mm,tension=2.5,color=black}{v2,v8}      \end{fmfgraph*}\end{gathered}}_{\left\langle y_*\right\rangle_1}\notag \\ &+\;\mathcal{O}\left(C^2\right)\end{align}\end{fmffile}which translate to our main result

\begin{align}
\left\langle y_{*}\right\rangle _{0+1} & =m_{*\alpha}m_{\alpha\beta}^{-1}\,y_{\beta}\nonumber \\
 & +m_{*\epsilon}m_{\epsilon\alpha}^{-1}C_{(\alpha\beta)(\gamma\delta)}m_{\beta\gamma}^{-1}\,m_{\delta\rho}^{-1}\,y_{\rho}\nonumber \\
 & -C_{(*\alpha)(\beta\gamma)}m_{\alpha\beta}^{-1}m_{\gamma\delta}^{-1}\,y_{\delta}+\mathcal{O}\left(C^{2}\right)\,.\label{eq:PerturbativeResult_MeanInferredNetworkOutput}
\end{align}
We here define the first line \foreignlanguage{english}{as the zeroth-order
approximation $\left\langle y_{*}\right\rangle _{0}:=m_{*\alpha}m_{\alpha\beta}^{-1}\,y_{\beta}$,
which has the same form as \eqref{eq:Mean-Inference-Bayesian-GP},
and the latter two lines as perturbative corrections $\left\langle y_{*}\right\rangle _{1}=\mathcal{O}\left(C\right)$
which are of linear order in $C$.}

\subsubsection{Evaluation of expressions for block-structured overlap matrices\label{subsec:Evaluation-of-expressions-BlockStructure}}

To evaluate the first order correction $\left\langle y_{\ast}\right\rangle _{1}$
in \eqref{eq:PerturbativeResult_MeanInferredNetworkOutput} we make
use of the fact that Bayesian inference is insensitive to the order
in which the training data are presented. We are hence free to assume
that all training samples of one class are presented en bloc. Moreover,
supervised learning assumes that all training samples are drawn from
the same distribution. As a result, the statistics is homogeneous
across blocks of indices that belong to the same class. The propagators
$-m_{\alpha\beta}^{-1}$ and interaction vertices $C_{(\alpha\beta)(\gamma\delta)}$
and $C_{(*\alpha)(\beta\gamma)}$, correspondingly, have a block structure.
To obtain an understanding how variability of the data and hence heterogeneous
kernels affect the ability to make predictions, we consider the simplest
yet non-trivial case of binary classification where we have two such
blocks.

In this section we focus on the overlap statistics given by the artificial
data set described in \prettyref{sec:The-Ising-data-set}. This data
set entails certain symmetries. Generalizing the expressions to a
less symmetric task is straightforward, but lengthy, and is deferred
to \prettyref{app:GeneralInferenceExpressions} and \prettyref{supp:Asymmetric-Overlap-Inference-1}. For the classification
task, with two classes $c_{\alpha}\in\{1,2\}$, the structure for
the mean overlaps $\mu_{\alpha\beta}$ and their covariance $\Sigma_{(\alpha\beta)(\gamma\delta)}$
at the read-in layer of the network given by \eqref{eq:Statistical_Properties_SimplePattern_Setup}
are inherited by the mean $m_{\alpha\beta}$ and the covariance $C_{(\alpha\beta)(\gamma\delta)}$
of the overlap matrix at the output of the network. In particular,
all quantities can be expressed in terms of only four parameters $a$,
$b$, $K$, $v$ whose values, however, depend on the network architecture
and will be given for linear and non-linear networks below. For four
indices $\alpha,\beta,\gamma,\delta$ that are all different

\begin{align}
m_{\alpha\alpha} & =a\,,\\
m_{\alpha\beta} & =\begin{cases}
b & c_{\alpha}=c_{\beta}\\
-b & c_{\alpha}\neq c_{\beta}
\end{cases}\,,\\
C_{(\alpha\alpha),(\gamma\delta)} & =0\,,\\
C_{(\alpha\beta)(\alpha\beta)} & =K\,,\\
C_{(\alpha\beta)(\alpha\delta)} & =\begin{cases}
v & c_{\alpha}=c_{\beta}=c_{\delta};\quad c_{\alpha}\neq c_{\beta}=c_{\delta}\\
-v & c_{\alpha}=c_{\beta}\neq c_{\delta};\quad c_{\alpha}=c_{\delta}\neq c_{\beta}
\end{cases}\,.\label{eq:Statistical_Properties_SimplePattern}
\end{align}
This symmetry further assumes that the network does not have biases
and utilizes point-symmetric activation functions $\phi(x)$ such
as $\phi(x)=\mathrm{erf(x)}$. In general, all tensors are symmetric
with regard to swapping $\alpha\leftrightarrow\beta$ as well as $\gamma\leftrightarrow\delta$
and the tensor $C_{(\alpha\beta)(\gamma\delta)}$ is invariant under
swaps of the index-pairs $(\alpha\beta)\leftrightarrow(\gamma\delta)$.
We further assume that the class label for class 1 is $y$ and that
the class label for class 2 is $-y$. In subsequent calculations and
experiments we consider the prediction for the class $y=-1$.

This setting is quite natural, as it captures the presence of differing
mean intra- and inter-class overlaps. Further $K$ and $v$ capture
two different sources of variability. Whereas $K$ is associated with
the presence of i.i.d. distributed variability on each entry of the
overlap matrix separately, $v$ corresponds to variability stemming
from correlations between different patterns. Using the properties
in \eqref{eq:Statistical_Properties_SimplePattern} one can evaluate
\eqref{eq:PerturbativeResult_MeanInferredNetworkOutput} for the inference
of test-points $*$ within class $c_{1}$ on a balanced training set
with $D$ samples explicitly to (see \prettyref{app:GeneralInferenceExpressions} and \prettyref{supp:Asymmetric-Overlap-Inference-1} )

\begin{align}
\langle y_{\ast}\rangle_{0} & =Dgy\,,\label{eq:mean_SimplePatternStatistics_MFT}\\
\langle y_{\ast}\rangle_{1} & =vg\hat{y}\left(q_{1}+3q_{2}\right)\left(D^{3}-3D^{2}+2D\right)\nonumber \\
 & +Kg\hat{y}\left(q_{1}+q_{2}\right)\left(D^{2}-D\right)\nonumber \\
 & -v\hat{y}\left(q_{1}+q_{2}\right)\left(D^{2}-D\right)\nonumber \\
 & +\mathcal{O}\left(C_{(\alpha\beta)(\gamma\delta)}^{2}\right)\quad\mathrm{for\quad}*\in c_{1}\,,\label{eq:mean_SimplePatternStatistics_FirstOrder}
\end{align}
with the additional variables

\begin{align}
g & =\frac{b}{(a-b)+bD}\,,\\
q_{2} & =-\frac{b}{(a-b)+bD}\,,\\
q_{1} & =\frac{1}{a-b}+q_{2}\,,\\
\hat{y} & =\frac{y}{(a-b)+bD}\,.\label{eq:Helper_variables_SimplePatternInferenceResult}
\end{align}
which stem from the analytic inversion of block-matrices (using the
approach from \citep{SegadloThesis}; see \prettyref{supp:AppendixElementsInverseBlockMatrix}). Carefully treating the dependencies of the parameters in \eqref{eq:Helper_variables_SimplePatternInferenceResult}
and \eqref{eq:mean_SimplePatternStatistics_FirstOrder}, one can compute
the limit $D\gg1$ and show that the $\mathcal{O}(1)$-behavior of
\eqref{eq:mean_SimplePatternStatistics_FirstOrder} for test points
$*\in c_{1}$ for the zeroth-order approximation, $\lim_{D\rightarrow\infty}\langle y_{*}\rangle_{0}:=\langle y_{*}\rangle_{0}^{(\mathrm{\infty)}}$,
and the first-order correction, $\lim_{D\rightarrow\infty}\langle y_{*}\rangle_{1}:=\langle y_{*}\rangle_{1}^{(\mathrm{\infty)}}$,
is given by

\begin{align}
\langle y_{*}\rangle_{0}^{(\mathrm{\infty)}} & =y\;,\label{eq: Limiting_Behaviour_MFT}\\
\langle y_{*}\rangle_{1}^{(\mathrm{\infty)}} & =\frac{y}{(a-b)b}\left((K-4v)-v\frac{a-b}{b}\right)\;.\label{eq:Limiting_Behaviour_FirstOrder}
\end{align}
This result implies that regardless of the amount of training data
$D$, the lowest value of the limiting behavior is controlled by the
data variability represented by $v$ and $K$. Due to the symmetric
nature of the task setting, neither the limiting behavior \eqref{eq:Limiting_Behaviour_FirstOrder}
nor the original expression \eqref{eq:mean_SimplePatternStatistics_FirstOrder}
explicitly show the dependence on the relative number of training
samples in the two respective classes $c_{1,2}$. This is due to the
fact that the task setup in \eqref{eq:Statistical_Properties_SimplePattern}
is symmetric. In the case of asymmetric statistics this behavior changes.
Moreover, the difference between variance $a$ and covariance $b$
enters the expression in a non-trivial manner

Using those results, we will investigate the implications for linear,
non-linear, and deep kernels using the artificial data set, \prettyref{sec:The-Ising-data-set},
as well as real-world data.

\subsection{Applications to linear, non-linear and deep non-linear NNGP kernels}

\subsubsection{Linear Kernel \label{subsec:Linear-Kernel}}

Before going to the non-linear case, let us investigate the implications
of \eqref{eq:mean_SimplePatternStatistics_FirstOrder} and \eqref{eq:Limiting_Behaviour_FirstOrder}
for a simple one-layer linear network. We assume that our network
consists of a read-in weight $\mathbf{V}\in\mathbb{R}^{1\times N_{\mathrm{dim}}};\mathbf{V}_{i}\sim\mathcal{N}\left(0,\sigma_{v}^{2}/N_{\mathrm{dim}}\right)$,
which maps the $N_{\mathrm{dim}}$ dimensional input vector to a one-dimensional
output space. Including a regularization noise, the output hence reads

\begin{align}
y_{\alpha} & =\mathbf{V}x_{\alpha}+\xi_{\alpha}\,.\label{eq:Definition_LinearNetwork_SingleLayer}
\end{align}
In this particular case the read-in, read-out, and hidden weights
in the general setup \eqref{eq:Deep-FFN-Architecture} coincide with
each other. Computing the average with respect to the weights $\mathbf{V}$
yields the kernel

\begin{align}
K_{\alpha\beta}^{y}=\left\langle y_{\alpha}y_{\beta}\right\rangle _{\mathbf{V}} & =K_{\alpha\beta}^{0}+\delta_{\alpha\beta}\,\sigma_{\mathrm{reg}}^{2}\,,\label{eq:LinearNetKernel}
\end{align}
where $K_{\alpha\beta}^{0}$ is given by \eqref{eq:K_zero}; it is
hence a rescaled version of the overlap of the input vectors and the
variance of the regularization noise.

We now assume that the matrix elements of the input-data overlap \eqref{eq:LinearNetKernel}
are distributed according to a multivariate Gaussian \eqref{eq:Distribution_OverlapElements}.

\begin{figure}
\begin{centering}
\includegraphics{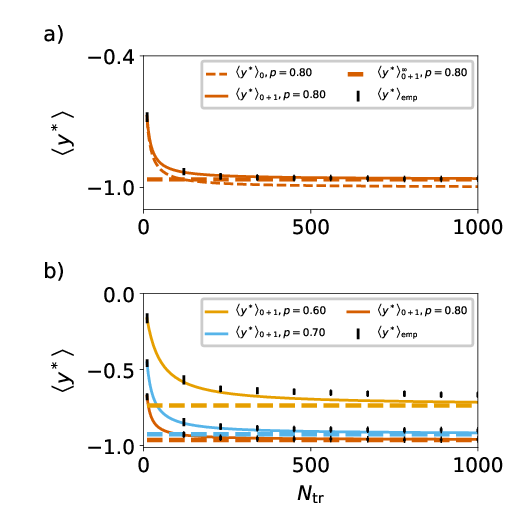}
\par\end{centering}
\caption{\label{fig:Deviations_EmpiricalTheoretical_1LayerLinearNetwork} \textbf{Predictive
mean in linear regression with heterogeneous kernel. (a)} Comparison
of empirical data (bars), zeroth-order approximation $\left\langle y^{*}\right\rangle _{0}$
(dashed), first-order corrections $\left\langle y^{*}\right\rangle _{0+1}$
(solid) and the asymptotic value $\left\langle y^{*}\right\rangle _{0+1}^{\infty}$
in the case of infinite training data (dashed horizontal line) for
a 1-layer linear network with $\sigma_{v}^{2}=1,\sigma_{\mathrm{reg}}^{2}=0.8,N_{\mathrm{dim}}=50,p=0.8$.
\textbf{(b)} Comparison of empirical inference data (bars) with first-order
results $\left\langle y^{*}\right\rangle _{0+1}$ (solid line) and
asymptotic values $\left\langle y^{*}\right\rangle _{0+1}^{\infty}$
for $p=0.6$ (orange), $p=0.7$ (blue) and $p=0.8$ (red). Empirical
results display mean and standard deviation over 50 trials with 2000
test points per trial.}
\end{figure}

As the mean and the covariance of the entries $K_{\alpha\beta}^{y}$
are given by the statistics \eqref{eq:Statistical_Properties_SimplePattern_Setup}
we evaluate \eqref{eq:mean_SimplePatternStatistics_FirstOrder} and
\eqref{eq:Limiting_Behaviour_FirstOrder} with

\begin{align}
a^{(\mathrm{Lin)}} & =\sigma_{v}^{2}+\sigma_{\mathrm{reg}}^{2}\,,\nonumber \\
b^{(\mathrm{Lin)}} & =\sigma_{v}^{2}\,u\,,\nonumber \\
K^{(\mathrm{Lin)}} & =\frac{\sigma_{v}^{4}}{N_{\mathrm{dim}}}\,(1-u^{2})\,,\nonumber \\
v^{(\mathrm{Lin)}} & =\frac{\sigma_{v}^{4}}{N_{\mathrm{dim}}}\,u\,(1-u)\,,\nonumber \\
u & :=4p(p-1)+1\,.\label{eq:eq:Statistical_Properties_1Layer_LinearKernel}
\end{align}
The asymptotic result for the first order correction, assuming that
$\sigma_{v}^{2}\neq0$, can hence be evaluated, assuming $p\neq0.5$,
as

\begin{align}
\left\langle y_{*}\right\rangle _{1}^{(\mathrm{\infty)}}= & \frac{y_{1}\sigma_{v}^{2}\frac{(1-u)}{N_{\mathrm{dim}}}}{\left(\sigma_{v}^{2}\left(1-u\right)+\sigma_{\mathrm{reg}}^{2}\right)u}\left(-2u-\frac{\sigma_{\mathrm{reg}}^{2}}{\sigma_{v}^{2}}\right)\,.\label{eq:Asymptotic_Result_Mean_1LayerLinear}
\end{align}
Using this explicit form of $\left\langle y_{*}\right\rangle {}_{1}^{(\infty)}$
one can see
\begin{itemize}
\item as $u\in[0,1]$ the corrections are always negative and hence provide
a less optimistic estimate for the generalization compared to the
zeroth-order approximation;
\item in the limit $\sigma_{v}^{2}\rightarrow\infty$ the regularizer in
\eqref{eq:Asymptotic_Result_Mean_1LayerLinear} becomes irrelevant
and the matrix inversion becomes unstable.
\item taking $\sigma_{v}^{2}\rightarrow0$ yields a setting where constructing
the limiting formula \eqref{eq:Asymptotic_Result_Mean_1LayerLinear}
is not useful, as all relevant quantities \eqref{eq:mean_SimplePatternStatistics_FirstOrder}
like $g,v,K\rightarrow0$ vanish; hence the inference yields zero
which is consistent with our intuition: $\sigma_{v}^{2}\rightarrow\text{0}$
implies that only the regularizer decides, which is unbiased with
regards to the class membership of the data. Hence the kernel cannot
make any prediction which is substantially informed by the data.
\end{itemize}
Figure \eqref{fig:Deviations_EmpiricalTheoretical_1LayerLinearNetwork}
shows that the zeroth-order approximation $\left\langle y^{*}\right\rangle _{0}$,
even though it is able to capture some dependence on the amount of
training data, is indeed too optimistic and predicts a mean-inferred
network output closer to its negative target value $y=-1$ than numerically
obtained. The first-order correction on the other hand is able to
reliably predict the results. Furthermore the limiting results $D\rightarrow\infty$
match the numerical results for different task settings $p$. These
limiting results are consistently higher than the zeroth-order approximation
$\left\langle y_{*}\right\rangle _{0}$ and depend on the level of
data variability. Deviations of the empirical results from the theory
in the case $p=0.6$ could stem from the fact that for $p=0$.5 the
fluctuations are maximal and our theory assumes small fluctuations.

\subsubsection{Non-Linear Kernel \label{subsec:Non-Linear-Kernel}}

We will now investigate how the non--linearities $\phi$ present
in typical network architectures \eqref{eq:Deep-FFN-Architecture}
influence our results for the learning curve \eqref{eq:mean_SimplePatternStatistics_FirstOrder}
and \eqref{eq:Limiting_Behaviour_FirstOrder}.

As the ansatz in \prettyref{sec:Field-theoretic-description-of-bayesian-inference}
does not make any assumption, apart from Gaussianity, on the overlap-matrix
$K^{y}$, the results presented in \prettyref{subsec:Evaluation-of-expressions-BlockStructure}
are general. One can use the knowledge of the statistics of the overlap
matrix in the read-in layer $K^{0}$ in \eqref{eq:Statistical_Properties_SimplePattern_Setup}
to extend the result \eqref{eq:mean_SimplePatternStatistics_FirstOrder}
to both non-linear and deep feed-forward neural networks.

As in \prettyref{subsec:Linear-Kernel} we start with the assumption
that the input kernel matrix is distributed according to a multivariate
Gaussian: $K_{\alpha\beta}^{0}\sim\mathcal{N}(\mu_{\alpha\ensuremath{\beta}},\Sigma_{(\alpha\beta)(\gamma\delta)})$.
In the non-linear case, we consider a read-in layer $\mathbf{V}\in\mathbb{R}^{N_{h}\times N_{\mathrm{dim}}};V_{i,j}\sim\mathcal{N}\left(0,\sigma_{v}^{2}/N_{\mathrm{dim}}\right)$,
which maps the inputs to the hidden-state space and a separate read-out
layer $\mathbf{W}\in\mathbb{R}^{1\times N_{h}};W_{i}\sim\mathcal{N}\left(0,\sigma_{w}^{2}/N_{h}\right)$,
obtaining a neural network with a single hidden layer

\begin{align}
h_{\alpha}^{(0)} & =\mathbf{V}x_{\alpha}\,,\nonumber \\
y_{\alpha} & =\mathbf{W}\phi\left(h_{\alpha}^{(0)}\right)+\xi_{\alpha}\,,\label{eq:1LayerNonlinear_Network}
\end{align}
and network kernel

\begin{align}
\left\langle y_{\alpha}y_{\beta}\right\rangle _{\mathbf{V,\mathbf{W}}} & =\frac{\sigma_{w}^{2}}{N_{h}}\text{\ensuremath{\sum_{i=1}^{N_{h}}\left\langle \phi\left(h_{\alpha i}^{(0)}\right)\phi\left(h_{\beta i}^{(0)}\right)\right\rangle _{\mathbf{V}}}}\nonumber \\
 & +\ensuremath{\delta_{\alpha\beta}\,\sigma_{\mathrm{reg}}^{2}}\,.\label{eq:OutputKernel_1LayerNonLinear}
\end{align}
As we consider the limit $N_{h}\rightarrow\infty$, one can replace
the empirical average $\frac{1}{N_{h}}\sum_{i=1}^{N_{h}}...$ with
a distributional average $\text{\ensuremath{\frac{1}{N_{h}}\sum_{i=1}^{N_{h}}...\rightarrow\langle...\rangle_{\mathbf{h}^{(0)}}}}$
\citep{Poole16_3360,Lee17_00165}. This yields the following result
for the kernel matrix $K_{\alpha\beta}^{y}$ of the multivariate Gaussian

\begin{align}
K_{\alpha\beta}^{y} & \underbrace{\rightarrow}_{N_{h}\rightarrow\infty}\sigma_{w}^{2}\text{\ensuremath{\left\langle \phi_{\alpha}^{0}\phi_{\beta}^{0}\right\rangle _{\mathbf{h}^{(0)},\mathbf{V}}}}+\ensuremath{\delta_{\alpha\beta}\,\sigma_{\mathrm{reg}}^{2}}\,.\label{eq:AverageKernel}
\end{align}
where we introduced the shorthand $\left\langle \phi\left(h_{\alpha}^{(0)}\right)\phi\left(h_{\beta}^{(0)}\right)\right\rangle :=\left\langle \phi_{\alpha}^{0}\phi_{\beta}^{0}\right\rangle $.
The expectation over the hidden states $h_{\alpha}^{(0)},h_{\beta}^{(0)}$
is with regard to the Gaussian

\begin{equation}
\left(\begin{array}{c}
h_{\alpha}^{(0)}\\
h_{\beta}^{(0)}
\end{array}\right)\sim\mathcal{N}\left(\left(\begin{array}{c}
0\\
0
\end{array}\right),\left(\begin{array}{cc}
K_{\alpha\alpha}^{0} & K_{\alpha\beta}^{0}\\
K_{\beta\alpha}^{0} & K_{\beta\beta}^{0}
\end{array}\right)\right)\,,\label{eq:Distribution_Preactivation_SingleNonLinearLayer}
\end{equation}
with the variance $K_{\alpha\alpha}^{0}$ and the covariance $K_{\alpha\beta}^{0}$
given by \eqref{eq:K_zero}. Evaluating the Gaussian integrals in
\eqref{eq:AverageKernel} is analytically possible in certain limiting
cases \citep{Williams98_1203,Cho09}. For an $\erf$-activation function,
as a prototype of a saturating activation function, this average yields

\begin{align}
\ensuremath{\left\langle \phi_{\alpha}^{0}\phi_{\alpha}^{0}\right\rangle _{\mathbf{h}^{(0)}}}= & \frac{4}{\pi}\arctan\left(\sqrt{1+4K_{\alpha\alpha}^{0}}\right)-1\,,\label{eq:Variance_SingleLayer_Erf}\\
\ensuremath{\left\langle \phi_{\alpha}^{0}\phi_{\beta}^{0}\right\rangle _{\mathbf{h}^{(0)}}}= & \frac{2}{\pi}\arcsin\left(\frac{2K_{\alpha\beta}^{0}}{1+2K_{\alpha\alpha}^{0}}\right)\,.\label{eq:Covariance_SingleLayer_Erf}
\end{align}
Equation \eqref{eq:AverageKernel} hence provides information on how
the mean overlap $m_{\alpha\ensuremath{\beta}}$ changes due to the
application of the non-linearity $\phi(\cdot)$, fixing the parameters
$a$, $b$, $K$, $v$ of the general form \prettyref{eq:Statistical_Properties_SimplePattern}
as

\begin{align}
a^{(\mathrm{Non-lin)}} & =K_{\alpha\alpha}^{y}=\sigma_{w}^{2}\text{\ensuremath{\left\langle \phi_{\alpha}^{0}\phi_{\alpha}^{0}\right\rangle _{\mathbf{h}^{(0)}}}}+\sigma_{\mathrm{reg}}^{2}\,,\label{eq:Variance_SingleLayerNonLinear}\\
b^{(\mathrm{Non-lin)}} & =K_{\alpha\beta}^{y}=\sigma_{w}^{2}\text{\ensuremath{\left\langle \phi_{\alpha}^{0}\phi_{\beta}^{0}\right\rangle _{\mathbf{h}^{(0)}}} }\,.\label{eq:Covariance_SingleLayer_NonLinear}
\end{align}
where the averages over $h^{(0)}$ are evaluated with regard to the
Gaussian \prettyref{eq:Distribution_Preactivation_SingleNonLinearLayer}
for $\phi(x)=\erf(x)$. We further require in \eqref{eq:Covariance_SingleLayer_NonLinear}
that $\alpha\neq\beta,\,c(\alpha)=c(\beta)$.

To evaluate the corrections in \eqref{eq:mean_SimplePatternStatistics_FirstOrder},
we also need to understand how the presence of the non-linearity $\phi(x)$
shapes the parameters $K,v$ that control the variability. Under the
assumption of small covariance $\Sigma_{(\alpha\beta)(\gamma\delta)}$
one can use \eqref{eq:Covariance_SingleLayer_Erf} to compute $C_{(\alpha\beta)(\gamma\delta)}$
using linear response theory. As $K_{\alpha\beta}^{0}$ is stochastic
and provided by \eqref{eq:Distribution_OverlapElements}, we decompose
$K_{\alpha\beta}^{0}$ into a deterministic kernel $\mu_{\alpha\beta}$
and a stochastic perturbation $\eta_{\alpha\beta}\sim\mathcal{N}\left(0,\Sigma_{(\alpha\beta)(\gamma\delta)}\right)$.
Linearizing \eqref{eq:Covariance_SingleLayer_NonLinear} around $\mu_{\alpha\beta}$
via Price's theorem \citep{Price58_69,Schuecker16b_arxiv}, the stochasticity
in the read-out layer yields (see \prettyref{app:DynamicMGF})

\begin{align}
C_{(\alpha\beta)(\gamma\delta)} & =\sigma_{w}^{4}\,K_{\alpha\beta}^{(\phi^{\prime})}\,K_{\gamma\delta}^{(\phi^{\prime})}\,\Sigma_{(\alpha\beta)(\gamma\delta)}\,,\\
K_{\alpha\beta}^{(\phi^{\prime})} & :=\left\langle \phi^{\prime}\left(h_{\alpha}^{(0)}\right)\phi^{\prime}\left(h_{\beta}^{(0)}\right)\right\rangle \,,
\end{align}
where $h^{(0)}$ is distributed as in \eqref{eq:Distribution_Preactivation_SingleNonLinearLayer}.
This clearly shows that the variability simply transforms with a prefactor

\begin{align}
K^{(\mathrm{Non-lin)}} & =\sigma_{w}^{4}\,K_{\alpha\beta}^{(\phi^{\prime})}\,K_{\alpha\beta}^{(\phi^{\prime})}\,\kappa\,,\\
v^{(\mathrm{Non-lin)}} & =\sigma_{w}^{4}K_{\alpha\beta}^{(\phi^{\prime})}\,K_{\alpha\delta}^{(\phi^{\prime})}\,v\,,\label{eq:K1_v1_1LayerNonLinear}
\end{align}
with $\kappa,\nu$ defined as in \eqref{eq:Statistical_Properties_SimplePattern_Setup}.
Evaluating the integral in $\left\langle \phi^{\prime}\left(h_{\alpha}^{(0)}\right)\phi^{\prime}\left(h_{\beta}^{(0)}\right)\right\rangle $
is hard in general. In fact, the integral which occurs is equivalent
to the one in \citep{molgedey92_3717} for the Lyapunov exponent and,
equivalently, in \citep{Poole16_3360,Schoenholz17_01232} for the
susceptibility in the propagation of information in deep feed-forward
neural networks. This is consistent with the assumption that our treatment
of the non-linearity follows a linear response approach as in \citep{molgedey92_3717}.
For the $\mathrm{erf}$-activation we can evaluate the kernel $K_{\alpha\beta}^{(\phi^{\prime})}$
as

\begin{align}
K_{\alpha\beta}^{(\phi^{\prime})} & =\frac{4}{\pi\left(1+2a^{(0)}\right)}\left(1-\left(\frac{2b^{(0)}}{1+2a^{(0)}}\right)^{2}\right)^{-\frac{1}{2}}\,,\label{eq:1LayerNonLinear_PhiPrime}\\
a^{(0)} & =\sigma_{v}^{2}\quad,\quad b^{(0)}=\sigma_{v}^{2}u\,,\\
u & =4p(p-1)+1\,,
\end{align}
which allows us to evaluate \eqref{eq:K1_v1_1LayerNonLinear}. Already
in the one hidden-layer setting we can see that the behavior is qualitatively
different from a linear setting: $K^{(\mathrm{Non-lin)}}$ and $v^{(\mathrm{Non-lin)}}$
scale with a linear factor which now also involves the parameter $\sigma_{v}^{2}$
in a non--linear manner.

\begin{figure}
\begin{centering}
\includegraphics[width=3.5in]{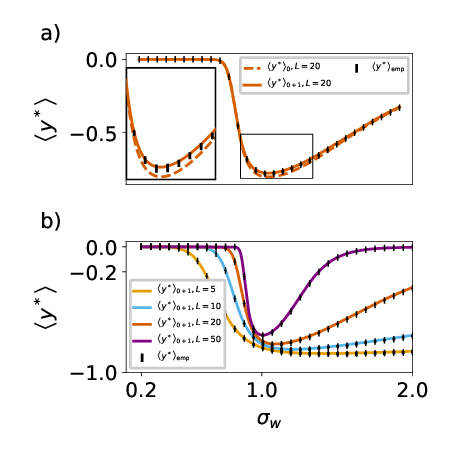}
\par\end{centering}
\caption{\label{fig:Deviations_EmpiricalTheoretical_1LayerErfNetwork} \textbf{Predictive
mean in a deep non-linear feed forward network with heterogeneous
kernel. (a)} Comparison of mean inferred network output for non-linear
network with $\phi(x)=\mathrm{erf(x)}$, five layers for different
values of the gain $\sigma_{w}$. The figure displays numerical results
(bars), zeroth-order approximation (dashed) and first-order corrections
(solid). \textbf{(b)} Similar comparison as in (a) for different network
depths $L=5,10,20,50$. In all settings we used $N_{\mathrm{dim}}=50$
for $D=100$, $p=0.8$, $\sigma_{v}^{2}=1$, $\sigma_{\mathrm{reg}}^{2}=1$.
Empirical results display mean and standard deviation over 1000 trials
with 1000 test points per trial.}
\end{figure}

\subsubsection{Multilayer-Kernel \label{subsec:Multilayer-NonLinear-Kernel}}

So far we considered single-layer networks. However, in practice the
application of multi-layer networks is often necessary. One can straightforwardly
extend the results from the non-linear case \eqref{subsec:Non-Linear-Kernel}
to the deep non-linear case. We consider the architecture introduced
in \prettyref{eq:Deep-FFN-Architecture} in \prettyref{subsec:Background_Neural_Networks}
where the variable $L$ denotes the number of hidden layers, and $1\le l\le L$
is the layer index. Similar to the computations in \prettyref{subsec:Non-Linear-Kernel}
one can derive a set of relations to obtain $K_{\alpha\beta}^{y}$
\begin{align}
K_{\alpha\beta}^{0} & =\frac{\sigma_{v}^{2}}{N_{\mathrm{dim}}}\,K_{\alpha\beta}^{x}\,,\\
K_{\alpha\beta}^{(\phi)l} & =\sigma_{w}^{2}\,\left\langle \phi\left(h_{\alpha}^{(l-1)}\right)\phi\left(h_{\beta}^{(l-1)}\right)\right\rangle \,,\\
K_{\alpha\beta}^{y} & =\sigma_{u}^{2}\,\left\langle \phi\left(h_{\alpha}^{(L)}\right)\phi\left(h_{\beta}^{(L)}\right)\right\rangle +\delta_{\alpha\beta}\,\sigma_{\mathrm{reg}}^{2}\,.\label{eq:Propagation_Kernel_DeepNonLinearNet}
\end{align}
with the setting $K^{0}\sim\mathcal{N}(\mu,\Sigma)$. As \citep{Xiao19,Poole16_3360,Schoenholz17_01232}
showed for feed-forward networks, deep non-linear networks strongly
alter both the variance and the covariance. So we expect them to influence
the generalization properties. In order to understand how the fluctuations
$\Sigma_{(\alpha\beta)(\gamma\delta)}$ transform through propagation,
one can employ the chain rule to linearize \prettyref{eq:Propagation_Kernel_DeepNonLinearNet}
and obtain

\begin{equation}
C_{(\alpha\beta)(\gamma\delta)}=\sigma_{u}^{4}\,\prod_{l=1}^{L+1}\left[K_{\alpha\beta}^{(\phi^{\prime})l}K_{\gamma\delta}^{(\phi^{\prime})l}\right]\,\Sigma{}_{(\alpha\beta)(\gamma\delta)}\,.\label{eq:Propagation_Variance_DeepNonLinear}
\end{equation}
A systematic derivation of this result as the leading order fluctuation
correction in $N_{h}^{-1}$ is found in the appendix of \citep{Segadlo22_103401}
whereas a derivation using a linear response approach is part of \prettyref{app:DynamicMGF}.

Equation \eqref{eq:Propagation_Kernel_DeepNonLinearNet} and \eqref{eq:Propagation_Variance_DeepNonLinear}
show that the kernel performance will depend on the non-linearity
$\phi$, the variances $\sigma_{v}^{2}$, $\sigma_{w}^{2}$, $\sigma_{u}^{2}$,
and the network depth $L$.

\prettyref{fig:Deviations_EmpiricalTheoretical_1LayerErfNetwork}
(a) shows the comparison of the mean inferred network output $\left\langle y^{*}\right\rangle $
for the true test label $y=-1$ between empirical results and the
first order corrections. The regime ($\sigma_{w}^{2}<1$) in which
the kernel vanishes, leads to a poor performance. The marginal regime
($\sigma_{w}^{2}\simeq1$) provides a better choice for the overall
network performance. Equation \eqref{fig:Deviations_EmpiricalTheoretical_1LayerErfNetwork}
(b) shows that the maximum absolute value for the predictive mean
is achieved slightly in the supercritical regime $\sigma_{w}^{2}>1$.
With larger number of layers, the optimum becomes more and more pronounced
and approaches the critical value $\sigma_{w}^{2}=1$ from above.
The optimum for the predictive mean to occur slightly in the supercritical
regime may be surprising with regard to the expectation that network
trainability peaks precisely at $\sigma_{w}^{2}=1$ \citep{Poole16_3360}.
In particular at shallow depths, the optimum becomes very wide and
shifts to $\sigma_{w}>1$. For few layers, even at $\sigma_{w}^{2}>1$
the increase of variance $K_{\alpha\alpha}^{y}$ per layer remains
moderate and stays within the dynamical range of the activation function.
Thus differences in covariance are faithfully transmitted by the kernel
and hence allow for accurate predictions. The theory including corrections
to linear order throughout matches the empirical results and hence
provides good estimates for choosing the kernel architecture.

\subsubsection{Experiments on Non-Symmetric Task Settings and MNIST}

In contrast to the symmetric setting in the previous subsections,
real data-sets such as MNIST exhibit asymmetric statistics so that
the different blocks in $m_{\alpha\beta}$ and $C_{(\alpha\beta)(\gamma\delta)}$
assume different values in general. All theoretical results from \prettyref{sec:Field-theoretic-description-of-bayesian-inference}
still hold. However, as the tensor elements of $m_{\alpha\beta}$
and $C_{(\alpha\beta)(\gamma\delta)}$ change, one needs to reconsider
the evaluation in \prettyref{subsec:Evaluation-of-expressions-BlockStructure}
in the most general form which yields a more general version of the
result.

\paragraph{Finite MNIST dataset}

\begin{figure}
\begin{centering}
\includegraphics{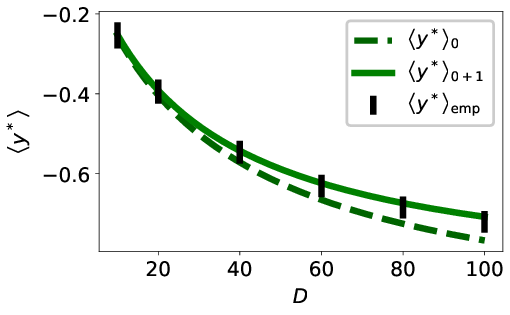}
\par\end{centering}
\caption{\textbf{\label{fig:PredictiveMean_PureMNIST}Predictive mean for a
linear network with MNIST data}: Comparison of mean inferred network
output for a linear network with 1 layer for different training set
sizes $D$. The figure displays numerical results (bars), zeroth-order
prediction (dashed) and first-order corrections (solid). Settings
$N_{\mathrm{dim}}=784$, $\sigma_{\mathrm{reg}}^{2}=2$, $D_{\mathrm{base}}=4000$.
MNIST classes $c_{1}=0$, $c_{2}=4$, $y_{c_{1}}=-1$, $y_{c_{2}}=1$;
balanced data-set in $D_{\mathrm{base}}$ and at each $D$. Empirical
results display mean and standard deviation over $1000$ trials with
$1000$ test points per trial.}
\end{figure}

First we consider a setting, where we work with the pure MNIST dataset
for two distinct labels $0$ and $4$. In this setting we estimate
the class-dependent tensor elements $m_{\alpha\beta}$ and $C_{(\alpha\beta)(\gamma\delta)}$
directly from the data. We define the data-set size per class, from
which we sample the theory as $D_{\mathrm{base}}$. The training points
are also drawn from a subset of these $D_{\mathrm{base}}$ data points.
To compare the analytical learning curve for $\langle y_{*}\rangle$
at $D$ training data-points to the empirical results, we need to
draw multiple samples of training datasets of size $D<D_{\mathrm{base}}$.
As the amount of data in MNIST is limited, these samples will generally
not be independent and therefore violate our assumption. Nevertheless
we can see in \prettyref{fig:PredictiveMean_PureMNIST} that if $D$
is sufficiently small compared to $D_{\mathrm{base}}$, the empirical
results and theoretical results match well.

\paragraph{Gaussianized MNIST dataset}

\begin{figure}
\begin{centering}
\includegraphics{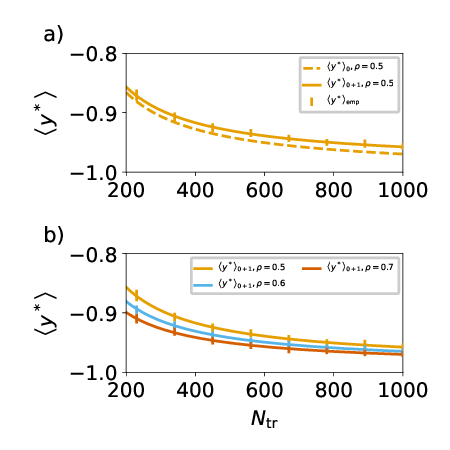}
\par\end{centering}
\caption{\textbf{\label{fig:MNISTLearningCurves}Predictive mean for a erf-network
with Gaussianized MNIST data}: \textbf{a) }Mean inferred network output
for MNIST classification with $\phi(x)=\mathrm{erf(x)}$. Figure shows
zeroth-order (dashed line), first-order (solid line), and empirical
results (bars).\textbf{ b)} Mean inferred network output in first
order approximation (solid lines) and empirical results (bars) for
MNIST classification with different ratios $\rho=D_{1}/D_{2}$ between
numbers of training samples per class $D_{1}$ and $D_{2}$, respectively;
$\rho=0.5$ (yellow), $\rho=0.6$ (blue), $\rho=0.7$ (red). Empirical
results display mean and standard deviation over 50 trials with 1000
test points per trial.}
\end{figure}

To test whether deviations in \prettyref{fig:PredictiveMean_PureMNIST}
at large $D$ stem from correlations in the samples of the dataset
we construct a generative scheme for MNIST data. This allows for the
generation of infinitely many training points and hence the assumption
that the training data is i.i.d. is fulfilled. We construct a pixel-wise
Gaussian distribution for MNIST images from the samples. We use this
model to sample as many MNIST images as necessary for the evaluation
of the empirical learning curves. Based on the class-dependent statistics
for the pixel means and the pixel covariances in the input data one
can directly compute the elements of the mean $\mu_{\alpha\beta}$
and the covariance $\Sigma_{(\alpha\beta)(\gamma\delta)}$ for the
distribution of the input kernel matrix $K_{\alpha\beta}^{0}$. We
see in \prettyref{fig:MNISTLearningCurves} that our theory describes
the results well for this data-set also for large numbers of training
samples.

Furthermore we can see that in the case of an asymmetric data-set
the learning curves depend on the balance ratio of training data $\rho=D_{1}/D_{2}$.
The bias towards class one in \prettyref{fig:MNISTLearningCurves}
b) is evident from the curves with $\rho>0.5$ predicting a lower
mean inferred network output, closer to the target label $y=-1$ of
class 1.

\section{Discussion\label{sec:Discussion}}

In this work we investigate the influence of data variability on the
performance of Bayesian inference. The probabilistic nature of the
data manifests itself in a heterogeneity of the entries in the block-structured
kernel matrix of the corresponding Gaussian process. We show that
this heterogeneity for a sufficiently large number of $D$ of data
samples can be treated as an effective non-Gaussian theory. By employing
a time-dependent formulation for the mean of the predictive distribution,
this heterogeneity can be treated as a disorder average that circumvents
the use of the replica trick; which was the basis of previous investigations
of analytical learning curves \citep{Malzahn2001}. A perturbative
treatment of the variability yields first-order corrections to the
mean in the variance of the heterogeneity that always push the mean
of the predictive distribution towards zero. In particular, we obtain
limiting expressions that accurately describe the mean in the limit
of infinite training data, qualitatively correcting the zeroth-order
approximation corresponding to homogeneous kernel matrices, is overconfident
in predicting the mean to perfectly match the training data in this
limit. This finding shows how variability fundamentally limits predictive
performance and provides not only a quantitative but also a qualitative
difference. Moreover at a finite number of training data the theory
explains the empirically observed performance accurately. We show
that our framework captures predictions in linear, non-linear shallow
and deep networks. In non-linear networks, we show that the optimal
value for the variance of the prior weight distribution is achieved
in the super-critical regime. The optimal range for this parameter
is broad in shallow networks and becomes progressively more narrow
in deep networks. These findings support that the optimal initialization
is not at the critical point where the variance is unity, as previously
thought \citep{Poole16_3360}, but that super-critical initialization
may have an advantage when considering input variability. An artificial
dataset illustrates the origin and the typical statistical structure
that arises in heterogeneous kernels, while the application of the
formalism to MNIST \citep{Lecun1998} demonstrates potential use to
predict the expected performance in real world applications.

The field theoretical formalism can be combined with approaches that
study the effect of fluctuations due to the finite width of the layers
\citep{Naveh20_01190,Yaida20,ZavatoneVeth21_NeurIPS_II,Segadlo22_103401}.
In fact, in the large $N_{h}$ limit the NNGP kernel is inert to the
training data, the so called lazy regime. At finite network width,
the kernel itself receives corrections which are commonly associated
with the adaptation of the network to the training data, thus representing
what is known as feature learning. The interplay of heterogeneity
of the kernel with such finite-size adaptations is a fruitful future
direction.

Another approach to study learning in the limit of large width is
offered by the neural tangent kernel (NTK) \citep{Jacot18_8580},
which considers the effect of gradient descent on the network output
up to linear order in the change of the weights. A combination of
the approach presented here with the NTK instead of the NNGP kernel
seems possible and would provide insights into how data heterogeneity
affects training dynamics.

The analytical results presented here are based on the assumption
that the variability of the data is small and can hence be treated
perturbatively. In the regime of large data variability, it is conceivable
to employ self-consistent methods instead, which would technically
correspond to the computation of saddle points of certain order parameters,
which typically leads to an infinite resummation of the perturbative
terms that dominate in the large $N_{h}$ limit. Such approaches may
be useful to study and predict the performance of kernel methods for
data that show little or no linear separability and are thus dominated
by variability. Another direction of extension is the computation
of the variance of the Bayesian predictor, which in principle can
be treated with the same set of methods as presented here. Finally,
since the large width limit as well as finite-size corrections, which
in particular yield the kernel response function that we employed
here, can be obtained for recurrent and deep networks in the same
formalism \citep{Segadlo22_103401} as well as for residual networks
(ResNets) \citep{Fischer2023_arxiv_07715}, the theory of generalization
presented here can straight forwardly be extended to recurrent networks
and to ResNets.\\
\\
We intend to upload the source code to produce the figures in the
manuscript to Zenodo.

\subsection*{Acknowledgments}

We thank Claudia Merger, Bastian Epping, Kai Segadlo, Alexander van
Meegen and Noah Schürholz for helpful discussions. This work was partly
supported by the German Federal Ministry for Education and Research
(BMBF Grant 01IS19077A to Jülich and BMBF Grant 01IS19077B to Aachen)
and funded by the Deutsche Forschungsgemeinschaft (DFG, German Research
Foundation) - 368482240/GRK2416, the Excellence Initiative of the
German federal and state governments (ERS PF-JARA-SDS005), and the
Helmholtz Association Initiative and Networking Fund under project
number SO-092 (Advanced Computing Architectures, ACA). Open access
publication funded by the Deutsche Forschungsgemeinschaft (DFG, German
Research Foundation) -- 491111487.\bibliographystyle{apsrev4-1_prx}
\setcounter{equation}{0} \renewcommand\theequation{A\arabic{equation}}\onecolumngrid

\appendix

\section*{Appendices}

\subsection{The dynamic moment generating function $Z(L_{*},K^{y})$\label{app:DynamicMGF}}

Ultimately we want to compute the disorder-average of

\begin{align}
y_{*}\left(K_{*\cdot},K_{\cdot\cdot}\right) & =K_{*\alpha}K_{\alpha\beta}^{-1}y_{\beta}\,,\label{eq:Appendix_MeanInferredNetworkOutputUnaveraged}\\
\left\langle y_{*}\right\rangle _{K_{*\cdot},K_{\cdot\cdot}} & =\left\langle K_{*\alpha}K_{\alpha\beta}^{-1}y_{\beta}\right\rangle _{K_{*\cdot},K_{\cdot\cdot}}\,,\label{eq:Appendix_DefMeanInferredOutput}
\end{align}
where $K_{\alpha\beta},K_{*\alpha}$ indicate the kernel matrix of
training samples and the kernel matrix between test and training samples,
respectively. As the matrices $K_{*\alpha},K_{\alpha\beta}$ are correlated
(due to the presence of the same training data in both matrices) and
$K_{\alpha\beta}$ occurs as a inverse, the average is not straightforward.
One way to evaluate \eqref{eq:Appendix_DefMeanInferredOutput} is
to compute the average of the moment generating function $Z(j)$ of
the predictive distribution and to compute its derivatives. By \prettyref{eq:Inference_TestPoints_Bayes}
the predictive distribution is given by the ratio of $p\left(y_{*},\mathbf{y}_{\mathrm{tr}}\vert x_{*},\mathbf{x}_{\mathrm{tr}}\right)$
and $p\left(\mathbf{y}_{\mathrm{tr}}\vert\mathbf{x}_{\mathrm{tr}}\right)$,
which each describe a Gaussian process with a simple dependence on
$K$ \citep{Segadlo22_103401}. Consequently, the moment generating
function of the predictive distribution is also given as a ratio

\begin{equation}
\left\langle Z(j)\right\rangle _{K}=\left\langle \frac{\mathcal{Z}(j,K)}{\mathcal{Z}(0,K)}\right\rangle _{K}\,,
\end{equation}
which requires an average over the numerator and the denominator,
which are also correlated as the normalization $\mathcal{Z}(0,K)$
also depends on the realization of $K$. We hence introduce a dynamic
approach, where the normalization $\mathcal{Z}(0,K)$ is fixed by
construction to $\mathcal{Z}(0,K)=1\:\forall K$ \citep{DeDominicis78_4913}.
We define the dynamics of an auxiliary variable $q_{\alpha}(t)$ via

\begin{equation}
\frac{\partial}{\partial t}q_{\alpha}(t)=-K_{\alpha\beta}q_{\beta}(t)+y_{\alpha}\,.\label{eq:Appendix_DefDynamics}
\end{equation}
This inhomogeneous differential equation in the long time limit converges
to

\begin{align}
q_{\alpha}(t\rightarrow\infty) & =K_{\alpha\beta}^{-1}y_{\beta}\,.\label{eq:Appendix_LimitValueDynamics}
\end{align}
We used the fact that $K_{\alpha\beta}$ is positive semi-definite,
as it is a covariance matrix. It is hence safe to assume that a fixed
point $q_{\alpha}(t\rightarrow\infty)$ exists. Further we make the
assumption, that we start the dynamics at $t\rightarrow-\infty$ with
an arbitrary initial condition. Hence we can assume that the fixed
point is achieved. A readout of $q_{\alpha}(t\rightarrow\infty)$
using the test-train kernel matrix $K_{*\alpha}$ yields the mean
inferred network output \eqref{eq:Appendix_MeanInferredNetworkOutputUnaveraged}
for a fixed realization of $K_{*\alpha},K_{\alpha\beta}$. We want
to compute the disorder average of $K_{*\alpha}q_{\alpha}(t\rightarrow\infty)$
. Using the Martin-Siggia-Rose-DeDominicis-Jannsen formalism we construct
the dynamic moment generating function $Z$ for the dynamics given
in \eqref{eq:Appendix_DefDynamics}:

\begin{align}
Z(l_{*},K) & =\int\prod_{\alpha=1}^{D}d\tilde{q}_{\alpha}\exp\left(S(q,\tilde{q},K_{*\cdot},K_{\cdot\cdot},l_{*})\right)\,,\label{eq:Appendix_DynamicMGF_TimeDomain}\\
S(q,\tilde{q},K_{*\cdot},K_{\cdot\cdot},l_{*}) & :=\int dt\tilde{q}_{\alpha}(t)\left[\frac{\partial}{\partial t}\mathbb{I}_{\alpha\beta}+K_{\alpha\beta}\right]q_{\beta}(t)-\int dt\tilde{q}_{\alpha}(t)y_{\alpha}+\int dtl_{*}(t)K_{*\alpha}q_{\alpha}(t)\.
\end{align}
We now go to Fourier space in the variables $q_{\alpha}(t),\tilde{q}_{\alpha}(t)\rightarrow Q_{\alpha}(\omega),\tilde{Q}_{\alpha}(\omega)$
using the definition

\begin{align}
q_{\alpha}(t) & =\frac{1}{2\pi}\int_{-\infty}^{\infty}d\omega\exp(i\omega t)Q_{\alpha}(\omega)\,,\label{eq:Appendix_DefFourierTrafo}\\
Q_{\alpha}(\omega) & =\int_{-\infty}^{\infty}dt\exp(-i\omega t)q_{\alpha}(t)\,.
\end{align}
Consequently products in time domain transform as

\begin{align}
\int dtq_{\alpha}(t)p_{\beta}(t) & =\frac{1}{2\pi}\int_{-\infty}^{\infty}d\omega Q_{\alpha}(\omega)P_{\beta}(-\omega):=Q_{\alpha}^{\top}P_{\beta}\label{eq:Appendix_DefinitionTimeDomainFrequencyDomainProduct}
\end{align}
where we defined the inner product in Fourier domain; we here used
that $(2\pi)^{-1}\int dt\exp\left[it\left(\omega+\omega^{\prime}\right)\right]=\delta\left(\omega+\omega^{\prime}\right)$.
Likewise the time-derivative $\partial_{t}$ transforms as

\begin{align}
\int dtq_{\alpha}(t)\frac{\partial}{\partial t}q_{\beta}(t) & =\frac{1}{2\pi}\int d\omega\,Q_{\alpha}(-\omega)i\omega Q_{\beta}(\omega)=Q_{\alpha}^{\top}(i\omega)Q_{\beta}\,.
\end{align}
Lastly the product $\int dt\tilde{q}_{\alpha}(t)y_{\alpha}$ transforms
as

\begin{align}
\int dt\tilde{q}_{\alpha}(t)y_{\alpha} & =\int dt\frac{1}{2\pi}\int d\omega\exp(i\omega t)\tilde{Q}_{\alpha}(\omega)y_{\alpha}\\
 & =\int d\omega\delta(\omega)\tilde{Q}_{\alpha}(\omega)y_{\alpha}\\
 & =\tilde{Q}_{\alpha}(\omega=0)y_{\alpha}
\end{align}
Those definitions transform the action of the dynamic moment generating
functional (MGF) \eqref{eq:Appendix_DynamicMGF_TimeDomain} to

\begin{align}
S(Q,\tilde{Q},K_{*\cdot},K_{\cdot\cdot},L_{*}) & =\frac{1}{2\pi}\int d\omega d\omega^{\prime}\delta(\omega+\omega^{\prime})\tilde{Q}_{\alpha}(\omega)\left[i\omega^{\prime}\mathbb{I}_{\alpha\beta}+K_{\alpha\beta}\right]Q_{\beta}(\omega^{\prime})-\tilde{Q}_{\alpha}(\omega=0)y_{\alpha}\nonumber \\
 & +\frac{1}{2\pi}\int d\omega\,L_{*}(\omega)K_{*\alpha}Q_{\alpha}(-\omega)\,.\label{eq:eq:Appendix_DefinitionFrequencyDomain MGF}
\end{align}
We now want to perform the disorder average. As in the main text we
assume Gaussian disorder for $K_{\alpha\beta},\,K_{*\alpha}$ according
to the statistics \eqref{eq:Distribution_Output_KernelElements_Gaussian}
and compute the disorder averaged moment generating function

\begin{equation}
\left\langle Z(L_{*},K)\right\rangle _{K}=\int\mathcal{D}\lbrace Q\tilde{Q}\rbrace\left\langle \exp\left(S(Q,\tilde{Q},K_{*\cdot},K_{\cdot\cdot},L_{*})\right)\right\rangle _{K}\,.\label{eq:Appendix_DefinitionDisorderAveragedMGF}
\end{equation}
As $K_{*\alpha},\,K_{\alpha\beta}$ appear linearly in the action
and both are Gaussian variables, the average is straightforward and
only affects the term

\begin{align}
\left\langle \exp\left(\tilde{Q}_{\alpha}^{\top}Q_{\beta}K_{\alpha\beta}+K_{*\alpha}L_{*}^{\top}Q_{\alpha}\right)\right\rangle _{K} & =\exp\left(\overline{S}\right)\,,
\end{align}
\\
with 
\begin{align}
\overline{S} & =m_{\alpha\beta}\tilde{Q}_{\alpha}^{\top}Q_{\beta}+m_{*\alpha}L_{*}^{\top}Q_{\alpha}\\
 & +\frac{1}{2}C_{(\alpha\beta)(\gamma\delta)}\tilde{Q}_{\alpha}^{\top}Q_{\beta}\tilde{Q}_{\gamma}^{\top}Q_{\delta}\nonumber \\
 & +C_{(*\alpha)(\beta\gamma)}L_{*}^{\top}Q_{\alpha}\tilde{Q}_{\beta}^{\top}Q_{\gamma}+\mathcal{O}\left(L_{*}^{2}\right)\,.\label{eq:Appendix_DisorderAveragedPartofAction}
\end{align}
In the main text and in this appendix we only consider perturbative
corrections towards the mean inferred network output. This corresponds
to computing the derivative of $\left\langle Z(L_{*},K)\right\rangle _{K}$
and subsequently setting $L_{*}=0$. Because of this we only need
to consider terms which are linear in $L_{*}$ and hence drop any
$\mathcal{O}\left(L_{*}^{2}\right)$ in \ref{eq:Appendix_DisorderAveragedPartofAction}.
Hence the full disorder averaged action reads

\begin{align}
\left\langle \exp\left(S(Q,\tilde{Q},K_{*\cdot},K_{\cdot\cdot},L_{*})\right)\right\rangle _{K} & =\exp\bigg(\tilde{Q}_{\alpha}^{\top}\left[i\omega\mathbb{I}_{\alpha\beta}+m_{\alpha\beta}\right]Q_{\beta}-\tilde{Q}_{\alpha}(\omega=0)y_{\alpha}+m_{*\alpha}L_{*}^{\top}Q_{\alpha}\nonumber \\
 & +\frac{1}{2}C_{(\alpha\beta)(\gamma\delta)}\tilde{Q}_{\alpha}^{\top}Q_{\beta}\tilde{Q}_{\gamma}^{\top}Q_{\delta}+C_{(*\alpha)(\beta\gamma)}L_{*}^{\top}Q_{\alpha}\tilde{Q}_{\beta}^{\top}Q_{\gamma}\bigg)\,.\label{eq:Appendix_FullDisorderAveragedAction}
\end{align}
We can decompose this action into a free theory with added perturbative
vertices. The free theory is simply a quadratic theory in $Q$ and
$\tilde{Q}$

\begin{align}
\left\langle \exp\left(S(Q,\tilde{Q},K_{*\cdot},K_{\cdot\cdot},L_{*})\right)\right\rangle _{K}^{(0)} & =\exp\left(\tilde{Q}_{\alpha}^{\top}\left[i\omega\mathbb{I}_{\alpha\beta}+m_{\alpha\beta}\right]Q_{\beta}-\tilde{Q}_{\alpha}(\omega=0)y_{\alpha}+m_{*\alpha}L_{*}^{\top}Q_{\alpha}\right)\\
 & =\exp\bigg(-\frac{1}{2}\left(\begin{array}{c}
\tilde{Q}_{\alpha}\\
Q_{\alpha}
\end{array}\right)^{\top}\left(\begin{array}{cc}
0 & -(i\omega\mathbb{I}_{\alpha\beta}+m_{\alpha\beta})\\
-(-i\omega\mathbb{I}_{\alpha\beta}+m_{\alpha\beta}) & 0
\end{array}\right)\left(\begin{array}{c}
\tilde{Q}_{\alpha}\\
Q_{\alpha}
\end{array}\right)\nonumber \\
 & -\tilde{Q}_{\alpha}(\omega=0)y_{\alpha}+m_{*\alpha}L_{*}^{\top}Q_{\alpha}\bigg)\,.
\end{align}
whereas the perturbative part consists of expressions reminiscent
of the $\varphi^{3}+\varphi^{4}$ theory

\begin{align}
\left\langle \exp\left(S(Q,\tilde{Q},K_{*\cdot},K_{\cdot\cdot},L_{*})\right)\right\rangle _{K}^{(1)} & =\exp\bigg(\frac{1}{2}\tilde{Q}_{\alpha}^{\top}Q_{\beta}C_{(\alpha\beta)(\gamma\delta)}\tilde{Q}_{\gamma}^{\top}Q_{\delta}\nonumber \\
 & +C_{(*\alpha)(\beta\gamma)}L_{*}^{\top}Q_{\alpha}\tilde{Q}_{\beta}^{\top}Q_{\gamma}+\mathcal{O}\left(L_{*}^{2}\right)\bigg)\,.
\end{align}
The fixed point dynamics in \eqref{eq:Appendix_DefDynamics} and \eqref{eq:Appendix_LimitValueDynamics}
can also be treated in the Fourier domain. From the inverse Fourier
transform we obtain our limit results via

\begin{align}
q_{\alpha}(t) & =\frac{1}{2\pi}\int d\omega\exp(i\omega t)Q_{\alpha}(\omega)\\
\left\langle y_{*}\right\rangle _{K} & =\lim_{t\rightarrow\infty}\left\langle \frac{1}{2\pi}\int d\omega\exp(i\omega t)K_{*\alpha}Q_{\alpha}(\omega)\right\rangle _{L}\\
\left\langle y_{*}\right\rangle _{K} & =\lim_{t\rightarrow\infty}\frac{1}{2\pi}\int d\omega\exp(i\omega t)2\pi\frac{\delta}{\delta L_{*}(-\omega)}\left[\left\langle Z(L_{*},K)\right\rangle _{K}\right]\bigg\rvert_{L_{*}(\omega)=0}
\end{align}

\subsubsection{The free case}

In the free case the action implies that we can contract the monopoles
$-\tilde{Q}_{\alpha}(\omega=0)y_{\alpha}$ and $m_{*\alpha}L_{*}^{\top}Q_{\alpha}$
with the propagator

\begin{align}
\Delta_{\alpha\beta}(\omega,\omega^{\prime}) & =\left(\begin{array}{cc}
\left\langle \tilde{Q}_{\alpha}(\omega)\tilde{Q}_{\beta}(\omega^{\prime})\right\rangle  & \left\langle Q_{\alpha}(\omega)\tilde{Q}_{\beta}(\omega^{\prime})\right\rangle \\
\left\langle \tilde{Q}_{\alpha}(\omega)Q_{\beta}(\omega^{\prime})\right\rangle  & \left\langle Q_{\alpha}(\omega)Q_{\beta}(\omega^{\prime})\right\rangle 
\end{array}\right)\\
 & =\left[\frac{1}{2\pi}\delta(\omega+\omega^{\prime})\left(\begin{array}{cc}
0 & -(i\omega\mathbb{I}+m)_{\alpha\beta}\\
-(-i\omega\mathbb{I}+m)_{\alpha\beta} & 0
\end{array}\right)\right]^{-1}\\
 & =2\pi\text{\ensuremath{\delta}(\ensuremath{\omega+\omega^{\prime})}}\left(\begin{array}{cc}
0 & -(-i\omega\mathbb{I}+m)_{\alpha\beta}^{-1}\\
-(i\omega\mathbb{I}+m)_{\alpha\beta}^{-1} & 0
\end{array}\right)\,.
\end{align}
The propagator $\Delta_{\alpha\beta}$ can hence contract $\tilde{Q}_{\alpha}(\omega)$
and $Q_{\beta}(-\omega)$ with each other. The $\delta(\omega+\omega^{\prime})$
implicitly corresponds to frequency conservation in the propagator.
From the equation above we see that effectively we get the propagator

\begin{align}
\Delta_{\alpha\beta}(\omega,\omega^{\prime}) & =\left\langle Q_{\alpha}(\omega)\tilde{Q}_{\beta}(\omega^{\prime})\right\rangle _{0}\nonumber \\
 & =-2\pi\delta(\omega+\omega^{\prime})(+i\omega\mathbb{I}+m)_{\alpha\beta}^{-1}\,.
\end{align}
Hence we can construct the diagram contracting the monopoles $-\tilde{Q}_{\alpha}(\omega=0)y_{\alpha}$
and $m_{*\alpha}L_{*}^{\top}Q_{\alpha}$ using $\Delta_{\alpha,\beta}$

\begin{align}
\frac{\delta}{\delta L_{*}(-\omega^{\prime})}\left[\left\langle Z(l_{*},K)\right\rangle _{0,K}\right]\bigg\rvert_{L_{*}(\omega)=0} & =\frac{\delta}{\delta L_{*}(-\omega^{\prime})}\left[-m_{*\alpha}\frac{1}{2\pi}\int d\omega L_{*}(-\omega)\left\langle Q_{\alpha}(\omega)\tilde{Q}_{\beta}(\omega=0)\right\rangle _{0}y_{\beta}\right]\bigg\rvert_{L_{*}(\omega)=0}\\
 & =\frac{\delta}{\delta L_{*}(-\omega^{\prime})}\left[m_{*\alpha}\frac{1}{2\pi}2\pi\int d\omega L_{*}(-\omega)(i\omega\mathbb{I}+m)_{\alpha\beta}^{-1}\delta(\omega)y_{\beta}\right]\bigg\rvert_{L_{*}(\omega)=0}\\
 & =\delta(\omega^{\prime})m_{*\alpha}(i\omega^{\prime}\mathbb{I}+m)_{\alpha\beta}^{-1}y_{\beta}\,.
\end{align}
This result yields the zeroth-order approximation for the mean inferred
network output as

\begin{align}
\left\langle y_{*}\right\rangle _{0,K} & =\lim_{t\rightarrow\infty}\frac{1}{2\pi}\int d\omega\exp(i\omega t)2\pi\frac{\delta}{\delta L_{*}(-\omega)}\left[\left\langle Z(L_{*},K)\right\rangle _{0,K}\right]\bigg\rvert_{L_{*}(\omega)=0}\\
 & =\lim_{t\rightarrow\infty}\int d\omega\exp(i\omega t)\delta(\omega^{\prime})m_{*\alpha}(+i\omega^{\prime}\mathbb{I}_{\alpha\beta}+m_{\alpha\beta})^{-1}y_{\beta}\\
 & =m_{*\alpha}m_{\alpha\beta}^{-1}y_{\beta}\,.
\end{align}
which is the well known result of Gaussian process inference, as it
has to be. Diagrammatically this corresponds to evaluating the contraction
of a monopole with a source term via the propagator leading to the
zeroth-order contribution

\begin{fmffile}{LowestOrder}\begin{align}\left\langle y_*\right\rangle_0 &\dot{=} \hspace{1.3cm} \vspace{-0.1cm}\begin{gathered}\vspace{-0.1cm}\begin{fmfgraph*}(60,30)  \fmfleft{i1} \fmfright{o1} \fmfv{decor.shape=circle,decor.filled=full, decor.size=5thick}{i1} \fmfv{decor.shape=circle,decor.filled=empty, decor.size=5thick}{o1} \fmf{plain,foreground=black,tension=3.5,width=0.6mm}{i1,v1} \fmf{dashes,foreground=black,width=0.6mm,label=$\Delta_{\alpha\beta}(\omega=0)$}{v1,v2} \fmf{plain,width=0.6mm,tension=2.5,color=black}{v2,o1} \fmflabel{$\frac{1}{2\pi}m_{*\alpha}$}{i1} \fmflabel{$y_{\beta}$}{o1} \end{fmfgraph*} \end{gathered}\hspace{0.5cm}=m_{*\alpha}m^{-1}_{\alpha\beta}y_\beta \quad .\end{align}\end{fmffile} 

\subsubsection{Vanishing response loop contributions \label{subsec:Vanishing-response-loop}}

The perturbative expressions, which contribute to the first-order
approximation of the mean inferred network output $\left\langle y_{+}\right\rangle _{0+1}$,
can also be translated into Feynman diagrams. In this subsection we
elaborate on vanishing response loops which limit the set of admissable
Feynman diagrams. This is related to the functional formulation of
of the starting point for our dynamical approach \eqref{eq:Appendix_DefDynamics}.

In order to go from \eqref{eq:Appendix_DefDynamics}. to the functional
formulation in \eqref{eq:Appendix_DynamicMGF_TimeDomain} we follow
along the lines of \citep{Helias20_970} and start by discretizing
the differential equation in the Ito convention. Assuming that $t\in[0,T]$
and discretizing the interval in $N$ steps of width $h=T/N$ we get
in the Ito-convention

\begin{align}
q_{\alpha,t+1}-q_{\a,t} & =-K_{\alpha\beta}q_{\beta,t}h+y_{\alpha}h\,,\label{eq:AppendixVanishingLoops_DiscretizedODE}\\
\mathrm{with} & \quad h=\frac{T}{N};\,t=ih,i\in\{0,1...N\}\,.
\end{align}
We enforce the relation between $q_{\alpha,t+1},q_{\alpha,t}$ explicitly
using the Fourier transform of the Dirac delta distribution

\begin{equation}
\delta(q)=\frac{1}{2\pi i}\int_{-i\infty}^{i\infty}d\tilde{q}\,\exp\left(\tilde{q}q\right)
\end{equation}
yielding the action

\begin{align}
Z(\lbrace j_{\alpha,t},\tilde{j}_{\alpha,t}\rbrace) & =\int\mathcal{D}q\mathcal{D}\tilde{q}\exp(S)\\
\int\mathcal{D}q\mathcal{D}\tilde{q} & =\prod_{\alpha,t}\int_{-\infty}^{\infty}dq_{\alpha,t}\prod_{\alpha,t}\int_{-i\infty}^{i\infty}\frac{d\tilde{q}_{\alpha,t}}{2\pi i}\\
S & =\sum_{\alpha,t}\tilde{q}_{\alpha,t+1}\left(q_{\alpha,t+1}-q_{\alpha,t}+K_{\alpha\beta}q_{\beta,t}h\right)-\sum_{\alpha,t}\tilde{q}_{\alpha,t+1}y_{\alpha}h+\sum_{\alpha,t}j_{\alpha,t}q_{\alpha,t}
\end{align}
where we introduced source terms for $q_{\alpha,t}$ as $j_{\alpha,t}$.
A natural way to introduce source terms for the auxiliary variables
$\tilde{q}_{\alpha,t}$ is to insert a source $\tilde{j}_{\alpha,t}$
on the right hand side of \eqref{eq:AppendixVanishingLoops_DiscretizedODE}
as $q_{\alpha,t+1}-q_{\a,t}=-K_{\alpha\beta}q_{\beta,t}h+y_{\alpha}h+\tilde{j}_{\alpha,t}$.
Hence the action with both sources reads

\begin{align}
S & =\sum_{\alpha,t}\tilde{q}_{\alpha,t+1}\left(q_{\alpha,t+1}-q_{\alpha,t}+K_{\alpha\beta}q_{\beta,t}h\right)-\sum_{\alpha,t}\tilde{q}_{\alpha,t+1}y_{\alpha}h+\sum_{\alpha,t}j_{\alpha,t}q_{\alpha,t}-\sum_{\alpha,t}\tilde{q}_{\alpha,t+1}\tilde{j}_{\alpha,t}\,.
\end{align}
This shows that one can interpret the correlator between $q_{\alpha,t}$
and $\tilde{q}_{\alpha,s}$ as the linear response of an infinitesimal
perturbation at time $s$ on the state at time $t$. Assuming the
free theory we simply replace $K_{\alpha\beta}\rightarrow m_{\alpha\beta}$.
The correlator reads

\begin{equation}
\frac{\partial Z}{\partial j_{\alpha,t}\partial\tilde{j}_{\beta,s}}\bigg\rvert_{j=0,\tilde{j}=0}=\left\langle q_{\alpha,t}\tilde{q}_{\beta,s+1}\right\rangle _{0}\bigg\rvert_{j=0,\tilde{j}=0}
\end{equation}
Going back to a continuous time setting, this corresponds to introducing
a small regulator $\epsilon\rightarrow0$ into the correlator between
the variable $q_{\alpha}(t)$ and the response field $\tilde{q}_{\beta}(s)$

\begin{align}
\frac{\delta Z}{\delta j_{\alpha}(t)\delta\tilde{j}_{\beta}(s)}\bigg\rvert_{j=0,\tilde{j}=0} & =\lim_{\epsilon\searrow0}\left\langle q_{\alpha}(t)\tilde{q}_{\alpha}(s+\epsilon)\right\rangle _{0}\,.\label{eq:AppendixVanishingLoops_DefRegulatedPropagator}
\end{align}
On a technical level the presence of this regulating term $\epsilon$
enforces causality in the response function. It further leads to vanishing
equal time responses

\begin{align}
\frac{\delta Z}{\delta j_{\alpha}(t)\delta\tilde{j}_{\beta}(t)}\bigg\rvert_{j=0,\tilde{j}=0} & =\lim_{\epsilon\searrow0}\left\langle q_{\alpha}(t)\tilde{q}_{\alpha}(t+\epsilon)\right\rangle _{0}=0
\end{align}
The reason for this is that one can show, that the response propagator
is $\left\langle q_{\alpha}(t)\tilde{q}_{\alpha}(s)\right\rangle _{0}\propto H(t-s)$,
where $H(t-s)$ is the Heaviside function \citep{Helias20_970}. This
has the direct consequence that integrals of the form

\begin{equation}
\int dt\left\langle q_{\alpha}(t)\tilde{q}_{\beta}(t)\right\rangle _{0}=0
\end{equation}
vanish trivially. Further, because of \eqref{eq:AppendixVanishingLoops_EigenvalueFrequencyIntegral},
the propagator in the frequency domain now includes an additional
factor $\exp(-i\omega\epsilon)\,,\epsilon>0$. Hence integrals of
the form

\begin{equation}
\frac{1}{2\pi}\int d\omega\frac{\delta Z}{\delta J_{\alpha}(-\omega)\delta\tilde{J}_{\beta}(\omega)}\bigg\rvert_{J=0,\tilde{J}=0}
\end{equation}
vanish as well. One can see this by starting from the definition

\begin{align}
\frac{1}{2\pi}\int d\omega\frac{\delta Z}{\delta J_{\alpha}(-\omega)\delta\tilde{J}_{\beta}(\omega)}\bigg\rvert_{J=0,\tilde{J}=0} & =\frac{1}{2\pi}\int d\omega\left\langle Q_{\alpha}(\omega)\tilde{Q}_{\beta}(-\omega)\right\rangle e^{-i\omega\epsilon}\nonumber \\
 & =-\frac{1}{2\pi}\int_{-\infty}^{\infty}d\omega\,(i\omega\mathbb{I}+m)_{\alpha\beta}^{-1}e^{-i\omega\epsilon}\nonumber \\
 & =-\frac{1}{2\pi}\int_{-\infty}^{\infty}d\omega\sum_{\gamma}(i\omega\mathbb{I}+m)_{\alpha\gamma}^{-1}\mathbb{I}_{\gamma\beta}e^{-i\omega\epsilon}\nonumber \\
 & =-\frac{1}{2\pi}\int_{-\infty}^{\infty}d\omega\sum_{\gamma,n}(i\omega\mathbb{I}+m)_{\alpha\gamma}^{-1}v_{\gamma}^{(n)}v_{\beta}^{(n)}e^{-i\omega\epsilon}\nonumber \\
 & =-\frac{1}{2\pi}\int_{-\infty}^{\infty}d\omega\sum_{n}(i\omega\mathbb{I}+\lambda^{(n)})^{-1}v_{\alpha}^{(n)}v_{\beta}^{(n)}e^{-i\omega\epsilon}\label{eq:AppendixVanishingLoops_EigenvalueFrequencyIntegral}
\end{align}
where we utilized the fact the matrix $m$ is symmetric and hence
the eigenvectors $v^{(n)}$ corresponding to the eigenvalues $\lambda^{(n)}$
of $m$ form a complete and orthonormal basis set. Further we know
that $m$ is a covariance matrix and hence positive semi-definite
and therefore $\lambda^{(n)}\in\mathbb{R}^{+}$. We treat the terms
of $\sum_{n}$ in \eqref{eq:AppendixVanishingLoops_EigenvalueFrequencyIntegral}
individually. Integrals of the form

\begin{equation}
\frac{1}{2\pi}\int_{-\infty}^{\infty}d\omega(i\omega\mathbb{I}+\lambda^{(n)})^{-1}e^{-i\omega\epsilon}=0
\end{equation}
vanish due to the regulator by the residue theorem: The term $(i\omega\mathbb{I}+\lambda^{(n)})^{-1}$
creates a pole in the upper complex half plane of $\omega$, whereas
the regulator $\exp(-i\omega\epsilon)$ requires one to close the
contour integration in the lower half-plane, where no pole is present.
Hence the integral vanishes and response loops vanish in frequency
space as well. The consequence for the set of Feynman diagrams is
that diagrams, which contract a pair of legs of the same vertex will
vanish; this is sometimes referred to as closed response loops. This
greatly reduces the number of allowed Feynman diagrams as we will
see in the next section.

\subsubsection{The interacting case: First-order corrections for the mean inferred
network output $\left\langle y_{*}\right\rangle _{0+1}$}

Taking into account perturbations, we now need to deal with contractions
of the higher order terms

 \begin{fmffile}{Test_ExplainingVertices} \begin{align} \mathrm{3-point\,Vertex:}&C_{(*\alpha)(\beta\gamma)}L_{*}^{\top}Q_{\alpha}\tilde{Q}_{\beta}^{\top}Q_{\gamma}=\hspace{0.3cm} \begin{gathered}\begin{fmfgraph*}(45,15)  \fmfleft{i1,i2} \fmfright{o1,o2} \fmf{plain,width=0.3mm,foreground=black}{v2,o2} \fmf{plain,width=0.3mm,foreground=black}{v2,o1} \fmf{plain,width=0.3mm,foreground=black}{v1,i2} \fmf{photon,width=0.3mm,foreground=black,tension=2}{v1,v2}  \fmf{phantom,width=0.3mm,pull=7}{i1,v1}  \fmf{phantom,width=0.3mm,pull=7}{o1,v2}  \fmf{phantom,width=0.3mm,pull=7}{v1,i2}  \fmf{phantom,width=0.3mm,pull=7}{v2,o2}  \fmffreeze \fmf{phantom,foreground=black,tension=4}{v5,i1} \fmf{phantom,foreground=black,width=0.3mm}{v5,v9} \fmf{plain,width=0.3mm,tension=1,color=black}{v1,v9}  \fmfv{decor.shape=circle,decor.filled=full, decor.size=2thick,label=$L_*$,l.d=0.01h+0.08w}{v9} \fmfv{decor.shape=none,decor.filled=empty,label=$Q_{\alpha}$,l.d=0.01h+0.01w}{i2} \fmfv{decor.shape=none,decor.filled=empty,label=$\tilde{Q}_\beta$,l.d=0.01h+0.01w}{o2} \fmfv{decor.shape=none,decor.filled=empty,label=$Q_{\gamma}$,l.d=0.01h+0.01w}{o1} \end{fmfgraph*}\end{gathered}\\ \nonumber \\ 4-\mathrm{point\,Vertex:}&\frac{1}{2}\tilde{Q}_{\alpha}^{\top}Q_{\beta}C_{(\alpha\beta)(\gamma\delta)}\tilde{Q}_{\gamma}^{\top}Q_{\delta}= \hspace{0.3cm}\vspace{-1cm} \begin{gathered}\begin{fmfgraph*}(45,15)  \fmfleft{i1,i2} \fmfright{o1,o2} \fmf{plain,width=0.3mm,foreground=black}{v2,o2} \fmf{plain,width=0.3mm,foreground=black}{v1,i2} \fmf{plain,width=0.3mm,foreground=black}{v2,o1} \fmf{plain,width=0.3mm,foreground=black}{v1,i1} \fmf{photon,width=0.3mm,foreground=black,tension=2}{v1,v2}  \fmf{phantom,width=0.3mm,pull=7}{i1,v1}  \fmf{phantom,width=0.3mm,pull=7}{o1,v2}  \fmf{phantom,width=0.3mm,pull=7}{v1,i2}  \fmf{phantom,width=0.3mm,pull=7}{v2,o2}  \fmffreeze \fmfv{decor.shape=none,decor.filled=empty, decor.size=2thick,label=$\tilde{Q}_{\alpha}$,l.d=0.01h+0.01w}{i2} \fmfv{decor.shape=none,decor.filled=empty, decor.size=2thick,label=$Q_{\beta}$,l.d=0.01h+0.01w}{i1} \fmfv{decor.shape=none,decor.filled=empty, decor.size=2thick,label=$\tilde{Q}_{\gamma}$,l.d=0.01h+0.01w}{o2} \fmfv{decor.shape=none,decor.filled=empty, decor.size=2thick,label=$Q_{\delta}$,l.d=0.01h+0.01w}{o1}\end{fmfgraph*}\end{gathered}  \end{align} \end{fmffile}Taking
the $3-\mathrm{point}$ vertex as an example: The propagator $\Delta_{\alpha\beta}$
can only contract auxiliary fields $\tilde{Q}_{\alpha}$ and fields
$Q_{\beta}$ with each other. We are not able to contract two auxiliary
fields or two real fields with each other. In the case of the 3-point
vertex one could hence either contract $Q_{\alpha},\tilde{Q}_{\beta}$
or $\tilde{Q}_{\beta},Q_{\gamma}$ which corresponds to

\begin{fmffile}{V3_Vanishing_and_NonVanishing_Diagram} \begin{align}C_{(*\alpha)(\beta\gamma)}\int d\omega d\omega^{\prime}L_{*}(-\omega)Q_{\gamma}(\omega^{\prime})\Delta_{\alpha\beta}(\omega;\omega^{\prime})&=\; \begin{gathered}\begin{fmfgraph*}(45,15)  \fmfleft{i1,i2} \fmfright{o1,o2} \fmf{plain,width=0.3mm,foreground=black}{v2,o2} \fmf{plain,width=0.3mm,foreground=black}{v1,i2} \fmf{photon,width=0.3mm,foreground=black,tension=2}{v1,v2}  \fmf{dashes,width=0.3mm,foreground=black,left=0.3,label=$\Delta_{\alpha\beta}(\omega;\omega^{\prime})$,l.d=0.2h}{i2,o2} \fmf{phantom,width=0.3mm,pull=7}{i1,v1}  \fmf{phantom,width=0.3mm,pull=7}{o1,v2}  \fmf{phantom,width=0.3mm,pull=7}{v1,i2}  \fmf{phantom,width=0.3mm,pull=7}{v2,o2}  \fmffreeze \fmf{phantom,foreground=black,tension=4}{v5,i1} \fmf{phantom,foreground=black,width=0.3mm}{v5,v9} \fmf{plain,width=0.3mm,tension=1,color=black}{v1,v9}  \fmfv{decor.shape=circle,decor.filled=full, decor.size=2thick}{v9} \fmf{plain,foreground=black,tension=4}{v7,o1} \fmf{plain,foreground=black,width=0.3mm}{v7,v8} \fmfv{decor.shape=none,decor.filled=empty,label=$Q_{\gamma}$,l.side=right,l.d=-0.1h+0.1w}{v7} \fmfv{decor.shape=none,decor.filled=full,label=$L_*$,l.side=right,l.d=0.04w-0.01h,decor.size=2thick}{v9} \fmf{plain,width=0.3mm,tension=2.5,color=black}{v2,v8}     \end{fmfgraph*}\end{gathered}\hspace{0.1cm}\quad ,\\[10pt]
\vspace{0.1cm}C_{(*\alpha)(\beta\gamma)}\int d\omega L_{*}(-\omega)Q_{\alpha}(\omega)\int d\omega^{\prime}\Delta_{\alpha\beta}(\omega^{\prime},\omega^{\prime})&=\; \begin{gathered}\hspace{0.2cm}\begin{fmfgraph*}(45,15)  \fmfleft{i1,i2} \fmfright{o1,o2} \fmf{plain,width=0.3mm,foreground=black}{v2,o2} \fmf{plain,width=0.3mm,foreground=black}{v2,o1} \fmf{plain,width=0.3mm,foreground=black}{v1,i2} \fmf{photon,width=0.3mm,foreground=black,tension=2}{v1,v2}  \fmf{dashes,width=0.3mm,foreground=black,right=0.3,label=$\Delta_{\alpha\beta}(\omega^{\prime};\omega^{\prime})$}{o1,o2} \fmf{phantom,width=0.3mm,pull=7}{i1,v1}  \fmf{phantom,width=0.3mm,pull=7}{o1,v2}  \fmf{phantom,width=0.3mm,pull=7}{v1,i2}  \fmf{phantom,width=0.3mm,pull=7}{v2,o2}  \fmffreeze \fmf{phantom,foreground=black,tension=4}{v5,i1} \fmf{phantom,foreground=black,width=0.3mm}{v5,v9} \fmf{plain,width=0.3mm,tension=1,color=black}{v1,v9}  \fmfv{decor.shape=circle,decor.filled=full, decor.size=2thick}{v9} \fmfv{decor.shape=none,decor.filled=empty,label=$Q_{\alpha}$,l.d=0.01h+0.01w}{i2} \fmf{plain,foreground=black}{v7,o1} \fmf{plain,foreground=black,width=0.3mm}{v7,v8} \fmfv{decor.shape=none,decor.filled=full,label=$L_*$,l.side=right,l.d=0.04w-0.01h, decor.size=2thick}{v9} \end{fmfgraph*}\end{gathered}\hspace{1.8cm}=0\,.\end{align} \end{fmffile}

where the second expression vanishes, as we close a response loop.
Similarly the 4-point vertex could be within a pair ($\tilde{Q}_{\alpha}^{\top}Q_{\beta}$
or $\tilde{Q}_{\gamma}^{\top}Q_{\delta}$) or between pairs ($\tilde{Q}_{\alpha}^{\top}Q_{\delta}$
or $Q_{\beta}\tilde{Q}_{\gamma}^{\top}$). Again we see that the first
option yields $0$, as we close a response loop. Diagrammatically
this reads

\begin{fmffile}{V4_Combined_Expressions} \begin{align}\frac{1}{2}C_{(\alpha\beta)(\gamma\delta)}\int d\omega d\omega^{\prime}\Delta_{\beta\gamma}(\omega;\omega^{\prime})\tilde{Q}_{\alpha}(-\omega)Q_{\delta}(\omega^{\prime})&=\; \begin{gathered} \hspace{0.1cm}\;\begin{fmfgraph*}(45,15)  \fmfleft{i1,i2} \fmfright{o1,o2} \fmf{plain,width=0.3mm,foreground=black}{v2,o2} \fmf{plain,width=0.3mm,foreground=black}{v1,i2} \fmf{plain,width=0.3mm,foreground=black}{v2,o1} \fmf{plain,width=0.3mm,foreground=black}{v1,i1} \fmf{photon,width=0.3mm,foreground=black,tension=2}{v1,v2}  \fmf{phantom,width=0.3mm,pull=7}{i1,v1}  \fmf{phantom,width=0.3mm,pull=7}{o1,v2}  \fmf{phantom,width=0.3mm,pull=7}{v1,i2}  \fmf{phantom,width=0.3mm,pull=7}{v2,o2}  \fmffreeze \fmfv{decor.shape=none,decor.filled=empty, decor.size=2thick,label=$\tilde{Q}_{\alpha}$,l.d=0.01h+0.01w}{i2} \fmfv{decor.shape=none,decor.filled=empty, decor.size=2thick,label=$Q_{\delta}$,l.d=0.01h+0.01w}{o2} \fmf{dashes,width=0.3mm,right=0.3,label=$\Delta_{\beta\gamma}(\omega ;\omega^{\prime})$,l.d=0.2h}{i1,o1} \end{fmfgraph*}\end{gathered}\hspace{0.5cm}\;,\\[32pt] \frac{1}{2}C_{(\alpha\beta)(\gamma\delta)}\int d\omega d\omega^{\prime}\Delta_{\alpha\delta}(\omega;\omega^{\prime})Q_{\beta}(\omega)\tilde{Q}_{\gamma}(-\omega^{\prime})&=\; \begin{gathered} \hspace{0.1cm}\;\begin{fmfgraph*}(45,15)  \fmfleft{i1,i2} \fmfright{o1,o2} \fmf{plain,width=0.3mm,foreground=black}{v2,o2} \fmf{plain,width=0.3mm,foreground=black}{v1,i2} \fmf{photon,width=0.3mm,foreground=black,tension=2}{v1,v2}  \fmf{dashes,width=0.3mm,left=0.3,label=$\Delta_{\alpha\delta}(\omega ;\omega^{\prime})$,l.d=0.2h}{i2,o2} \fmf{phantom,width=0.3mm,pull=7}{i1,v1}  \fmf{phantom,width=0.3mm,pull=7}{o1,v2}  \fmf{phantom,width=0.3mm,pull=7}{v1,i2}  \fmf{phantom,width=0.3mm,pull=7}{v2,o2}  \fmffreeze \fmf{plain,foreground=black,tension=4}{v5,i1} \fmf{plain,foreground=black,width=0.3mm}{v5,v9} \fmf{plain,width=0.3mm,tension=2.5,color=black}{v1,v9}  \fmf{plain,foreground=black,tension=4}{v7,o1} \fmf{plain,foreground=black,width=0.3mm}{v7,v8} \fmf{plain,width=0.3mm,tension=2.5,color=black}{v2,v8} \fmfv{decor.shape=none,decor.filled=empty,label=$Q_{\beta}$,l.side=right,l.d=-0.1h+0.1w}{v5} \fmfv{decor.shape=none,decor.filled=empty,label=$\tilde{Q}_{\gamma}$,l.side=right,l.d=-0.1h+0.1w}{v7} \end{fmfgraph*}\end{gathered}\hspace{0.5cm}\;, \\[10pt] \frac{1}{2}C_{(\alpha\beta)(\gamma\delta)}\int d\omega\Delta_{\gamma\delta}(\omega,\omega)\int d\omega^{\prime}\tilde{Q}_{\alpha}(-\omega^{\prime})Q_{\beta}(\omega^{\prime})&=\; \begin{gathered} \hspace{0.2cm}\;\begin{fmfgraph*}(45,15)  \fmfleft{i1,i2} \fmfright{o1,o2} \fmf{plain,width=0.3mm,foreground=black}{v2,o2} \fmf{plain,width=0.3mm,foreground=black}{v1,i2} \fmf{plain,width=0.3mm,foreground=black}{v2,o1} \fmf{plain,width=0.3mm,foreground=black}{v1,i1} \fmf{photon,width=0.3mm,foreground=black,tension=2}{v1,v2}  \fmf{dashes,width=0.3mm,right=0.3,label=$\Delta_{\gamma\delta}(\omega;\omega)$,l.d=0.2h}{o1,o2} \fmf{phantom,width=0.3mm,pull=7}{i1,v1}  \fmf{phantom,width=0.3mm,pull=7}{o1,v2}  \fmf{phantom,width=0.3mm,pull=7}{v1,i2}  \fmf{phantom,width=0.3mm,pull=7}{v2,o2}  \fmffreeze \fmfv{decor.shape=none,decor.filled=empty,label=$Q_{\beta}$,l.side=right,l.d=-0.3h+0.1w}{i1} \fmfv{decor.shape=none,decor.filled=empty,label=$\tilde{Q}_{\alpha}$,l.side=right,l.d=-0.2h+0.1w}{i2} \end{fmfgraph*} \end{gathered}\hspace{1.5cm}=0\;, \\[5pt] \frac{1}{2}C_{(\alpha\beta)(\gamma\delta)}\int d\omega\Delta_{\alpha\beta}(\omega,\omega)\int d\omega^{\prime}\tilde{Q}_{\gamma}(-\omega^{\prime})Q_{\delta}(\omega^{\prime})&=\; \begin{gathered} \vspace{-0.1cm}\hspace{1.3cm}\;\begin{fmfgraph*}(45,15)  \fmfleft{i1,i2} \fmfright{o1,o2} \fmf{plain,width=0.3mm,foreground=black}{v2,o2} \fmf{plain,width=0.3mm,foreground=black}{v1,i2} \fmf{plain,width=0.3mm,foreground=black}{v2,o1} \fmf{plain,width=0.3mm,foreground=black}{v1,i1} \fmf{photon,width=0.3mm,foreground=black,tension=2}{v1,v2}  \fmf{dashes,width=0.3mm,left=0.3,label=$\Delta_{\alpha\beta}(\omega;\omega)$,l.d=0.1h}{i1,i2} \fmf{phantom,width=0.3mm,pull=7}{i1,v1}  \fmf{phantom,width=0.3mm,pull=7}{o1,v2}  \fmf{phantom,width=0.3mm,pull=7}{v1,i2}  \fmf{phantom,width=0.3mm,pull=7}{v2,o2}  \fmffreeze \fmfv{decor.shape=none,decor.filled=empty,label=$Q_{\delta}$,l.side=right,l.d=-0.2h+0.05w}{o1} \fmfv{decor.shape=none,decor.filled=empty,label=$\tilde{Q}_{\gamma}$,l.side=right,l.d=-0.2h+0.05w}{o2} \end{fmfgraph*} \end{gathered}\hspace{0.1cm}=0\;. \end{align} \end{fmffile}

\subsection{Diagrams for the mean of inferred network output to first order in
$C_{(*\beta)(\gamma\delta)},C_{(\alpha\beta)(\gamma\delta)}$\label{subsec:Appendix_all_Diagrams}}

As we elaborated in the main text in \prettyref{subsec:Field-theoretic-elements-diagrams},
we use a field theoretic framework to evaluate the perturbative corrections
to the mean inferred network output, based on the averaged moment
generating function, which reads

\begin{align}
\left\langle Z(L_{*},K)\right\rangle _{K} & =\int\mathcal{D}Q\mathcal{D}\tilde{Q}\left\langle \exp\left(S(Q,\tilde{Q},K_{*\cdot},K_{\cdot\cdot},L_{*})\right)\right\rangle _{K}:=\int\mathcal{D}Q\mathcal{D}\tilde{Q}\exp\left(\overline{S}(Q,\tilde{Q},L_{*})\right)\,,\nonumber \\
\overline{S}(Q,\tilde{Q},L_{*}) & =\tilde{Q}_{\alpha}^{\top}\left[i\omega\mathbb{I}_{\alpha\beta}+m_{\alpha\beta}\right]Q_{\beta}\nonumber \\
 & -\tilde{Q}_{\alpha}(\omega=0)y_{\alpha}\nonumber \\
 & +m_{*\alpha}L_{*}^{\top}Q_{\alpha}\nonumber \\
 & +\frac{1}{2}C_{(\alpha\beta)(\gamma\delta)}\tilde{Q}_{\alpha}^{\top}Q_{\beta}\tilde{Q}_{\gamma}^{\top}Q_{\delta}\nonumber \\
 & +C_{(*\alpha)(\beta\gamma)}L_{*}^{\top}Q_{\alpha}\tilde{Q}_{\beta}^{\top}Q_{\gamma}\,,\label{eq:Appendix4_Action_DisorderAverageMGF}
\end{align}
where we introduced the notation $1/(2\pi)\int d\omega\,\tilde{Q}(-\omega)Q(\omega):=\tilde{Q}^{\top}Q$.
We use the Einstein convention for the Greek indices $\alpha,\beta,\gamma,\delta$
which run over the training-data exclusively and $*$ denotes the
test-point which is by definition not part of the training set. From
the way that we constructed our theory, we can obtain the disorder
averaged mean inferred network output by computing derivatives w.r.t
to $L_{*}$

\begin{equation}
\left\langle y_{*}\right\rangle _{K}=\lim_{t\rightarrow\infty}\frac{1}{2\pi}\int d\omega\exp(i\omega t)2\pi\frac{\delta}{\delta L_{*}(-\omega)}\left[\left\langle Z(L_{*},K)\right\rangle _{K}\right]\bigg\rvert_{L_{*}(\omega)=0}\,.
\end{equation}
As we are not able to solve the integral occurring in \eqref{eq:Appendix4_Action_DisorderAverageMGF} in
a closed form, we treat the third and fourth order terms in \eqref{eq:Appendix4_Action_DisorderAverageMGF}
perturbatively. We want to study the perturbative terms up to linear
order in the expression for $\langle y_{*}\rangle$. A systematic
way to do this is to introduce Feynman diagrams. We there introduce
the diagrammatic notation shown in \prettyref{subsec:Field-theoretic-elements-diagrams}.

\subsection{The mean and covariance in the synthetic data set\label{app:The-statistical-properties}}

As presented in the main text, we need to consider the statistical
properties of the overlaps between patterns. In the case of the synthetic
task settings one can evaluate those properties directly. The patterns
in the synthetic task have length $N_{\mathrm{dim}}$, where $\Ndim$
is even. Each of the $N_{\mathrm{dim}}$ pixels can take the value
$\pm1$ . The value of each pixel is drawn independently. The probability
to be $\pm1$ is given by the parameter $p\in[0,1]$ , the class membership
of the pattern and the relative position of the pixel in the pattern
according to

\begin{align}
x_{\alpha i} & =\begin{cases}
1 & \mathrm{with}\,\:p\\
-1 & \mathrm{with}\,\:(1-p)
\end{cases}\quad\mathrm{for}\,\,i\leq\frac{N_{\mathrm{dim}}}{2}\ ,\label{eq:Pixel-Value-Distribution-Class_1_Part0-1}\\
x_{\a i} & =\begin{cases}
1 & \mathrm{with}\,\:(1-p)\\
-1 & \mathrm{with}\,\:p.
\end{cases}\quad\mathrm{for}\,\,i>\frac{N_{\mathrm{dim}}}{2}\.\label{eq:Pixel-Value-Distribution-Class_1-1}
\end{align}
For a pattern $x^{(\alpha)}$ in the second class $c=2$ the pixel
values are distributed according to

\begin{align}
x_{\a i} & =\begin{cases}
-1 & \mathrm{with}\,\:p\\
1 & \mathrm{with}\,\:(1-p)
\end{cases}\quad\mathrm{for}\,\,i\leq\frac{N_{\mathrm{dim}}}{2}\ ,\label{eq:Pixel-Value-Distribution-Class_2_Part0-1}\\
x_{\a i} & =\begin{cases}
-1 & \mathrm{with}\,\:(1-p)\\
1 & \mathrm{with}\,\:p
\end{cases}\quad\mathrm{for}\,\,i>\frac{N_{\mathrm{dim}}}{2}\.\label{eq:Pixel-Value-Distribution-Class_2-1}
\end{align}
We aim to compute the mean $\mu_{\alpha\beta}$ and the covariances
$\Sigma_{(\alpha\beta)(\gamma\delta)}$ of the overlaps

\begin{equation}
K_{xx}(\alpha,\beta)=\sum_{i=1}^{N_{\mathrm{dim}}}x_{\a i}x_{\text{\ensuremath{\b}}i}.\label{eq:Input_Overlap_Appendix}
\end{equation}
Following the calculations in \prettyref{supp:SupplementStatisticalProperties} this yields

\begin{align}
\mu_{\alpha\beta} & =N_{\mathrm{dim}}\begin{cases}
1 & \alpha=\beta\ ,\\
\hat{\mu} & c_{\alpha}=c_{\beta}\ ,\,\\
\brackets{-\hat{\mu}} & c_{\alpha}\neq c_{\beta}\.
\end{cases}\\
\Sigma_{(\alpha\beta)(\alpha\delta)} & =\Ndim\begin{cases}
\hat{\mu}(1-\hat{\mu}) & \mathrm{for\;}\begin{cases}
c_{\alpha}=c_{\beta}=c_{\delta}\\
c_{\alpha}\ne c_{\beta}=c_{\delta}
\end{cases}\\
-\hat{\mu}(1-\hat{\mu}) & \mathrm{for\;}\begin{cases}
c_{\alpha}=c_{\beta}\neq c_{\delta}\\
c_{\alpha}=c_{\delta}\neq c_{\beta}
\end{cases}
\end{cases}\,.
\end{align}
with $\hat{\mu}:=4p(p-1)+1$.

\subsection{Distribution of Overlap-Matrix elements for MNIST and FashionMNIST\label{app:Appendix_OverlapDistributions}}

\begin{figure}[H]
\begin{centering}
\includegraphics[scale=0.8]{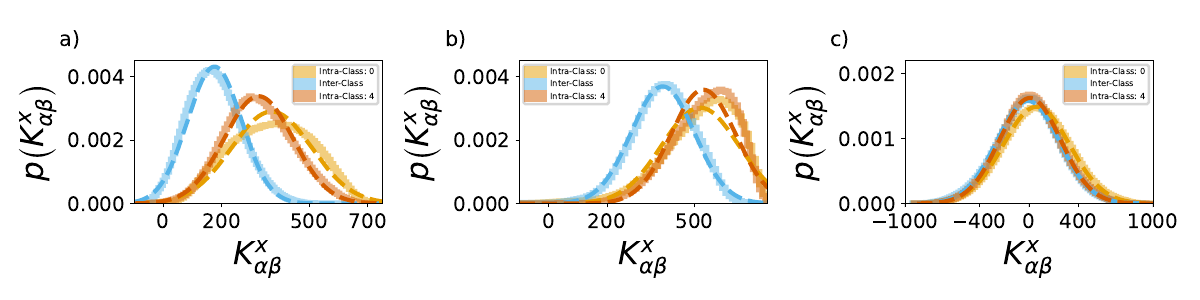}
\par\end{centering}
\caption{\textbf{\label{fig:Appendix_OverlapDistribution_MNIST_FMNIST_CIFAR10}Distribution
of pixel values in input kernels for MNIST, FashionMNIST, and CIFAR-10.
}Normalized inter-class (blue) and intra-class (red, yellow) distributions
of the pixel values in the input-kernel $K_{\alpha\beta}^{x}$. Solid
lines indicate histograms, dashed lines indicate Gaussian approximations.
Results are displayed for \textbf{a)} MNIST,\textbf{ b)} FashionMNIST
and\textbf{ c) }CIFAR-10. Results are displayed for $N_{\mathrm{dat}}=2000$
training samples with a balance ratio of $0.5$. Class labels $c_{1}=0,c_{2}=4$.}
\end{figure}

As stated in the main text, we make the assumption that the elements
of the input kernel $K_{\alpha\beta}^{x}=x_{\alpha}x_{\beta}^{\top}$
are distributed according to multivariate Gaussian distributions.
It is, a priori, not clear, whether this is the case for real data-sets
such as MNIST, FashionMNIST or CIFAR-10 (see \prettyref{fig:Appendix_OverlapDistribution_MNIST_FMNIST_CIFAR10})
. Upon investigating the distribution of the pixel values for the
input kernel $K_{\alpha\beta}^{x}$ we can see, that a Gaussian approximation
provides a good description. One can however also see, that the distribution
of CIFAR-10 values is centered around 0 indicating, that the overlap
between to patterns $x_{\alpha i},x_{\beta i}$ is close to orthogonal
and hence a simple dot-product kernel might not be informative for
this task setting.

\subsection{General expressions for the inference formula in asymmetric block
structured task settings\label{app:GeneralInferenceExpressions}}

\selectlanguage{english}%
We here treat the most general case for the statistics of overlaps
as they occur in real world data sets, such as MNIST. As in the main
text we assume that the elements of $K_{*\alpha}^{y},K_{\alpha\beta}^{y}$
are distributed according to a multivariate Gaussian distribution

\begin{equation}
K_{\alpha\beta}^{y}\sim\mathcal{N}\left(m_{\alpha\beta},C_{(\alpha\beta)(\gamma\delta)}\right)\,.
\end{equation}
where $a,b,c,d$ can be either train or test points. The most general
choice is to assume that the statistics only depends on the class
membership of $a,b,c,d$. We assume a binary classification task with
the classes $c_{1},c_{2}$. Hence the mean of the kernel matrices
read

\begin{equation}
m_{\alpha\beta}=\begin{cases}
a & \text{\ensuremath{\alpha=\beta}}\\
b & \alpha\neq\beta,c_{\alpha}=c_{\beta}=c_{1}\\
d & \alpha\neq\beta,c_{\alpha}=c_{\beta}=c_{2}\\
c & c_{\alpha}\neq c_{\beta}
\end{cases}\,,\label{eq:Mean_Blocks}
\end{equation}
which is a block matrix. Both the variance $C_{(\alpha\beta)(\alpha\beta)}$
and the covariance $C_{(\alpha\beta)(\gamma\delta)}$ inherit the
block structure as well

\begin{align}
C_{(\alpha\beta)(\alpha\beta)} & =\begin{cases}
K_{1} & c_{\alpha}=c_{\beta}=c_{1}\\
\overline{K}_{1} & c_{\alpha}=c_{\beta}=c_{2}\\
K_{2}=\overline{K}_{2} & c_{\alpha}\neq c_{\beta}
\end{cases}\,,\label{eq:Variance_Blocks}\\
C_{(\alpha\beta)(\alpha\delta)} & =\begin{cases}
v_{1} & c_{\alpha}=c_{1},c_{\beta}=c_{1},c_{\delta}=c_{1}\\
v_{2} & c_{\alpha}=c_{1},c_{\beta}=c_{2},c_{\delta}=c_{2}\\
v_{3} & c_{\alpha}=c_{1},c_{\beta}=c_{1},c_{\delta}=c_{2}\\
v_{4}=v_{3} & c_{\alpha}=c_{1},c_{\beta}=c_{2},c_{\delta}=c_{1}\\
\overline{v}_{1} & c_{\alpha}=c_{2},c_{\beta}=c_{2},c_{\delta}=c_{2}\\
\overline{v}_{2} & c_{\alpha}=c_{2},c_{\beta}=c_{1},c_{\delta}=c_{1}\\
\overline{v}_{3} & c_{\alpha}=c_{2},c_{\beta}=c_{2},c_{\delta}=c_{1}\\
\overline{v}_{4}=\overline{v}_{3} & c_{\alpha}=c_{2},c_{\beta}=c_{1},c_{\delta}=c_{2}
\end{cases}\,.\label{eq:Covariance_Blocks}
\end{align}
\foreignlanguage{american}{Further the tensor elements $C_{(\alpha\beta)(\gamma\delta)}$
for the case where all indices are different $\alpha\neq\beta\neq\gamma\neq\delta$
yields zero. This follows directly from}

\selectlanguage{american}%
\begin{align}
C_{(\alpha\beta)(\gamma\delta)} & =\left\langle \left(K_{\alpha\beta}^{y}-\left\langle K_{\gamma\delta}^{y}\right\rangle \right)\left(K_{\gamma\delta}^{y}-\left\langle K_{\gamma\delta}^{y}\right\rangle \right)\right\rangle \nonumber \\
 & =\left\langle K_{\alpha\beta}^{y}K_{\gamma\delta}^{y}\right\rangle -\left\langle K_{\alpha\beta}^{y}\right\rangle \left\langle K_{\gamma\delta}^{y}\right\rangle \,,\\
\mathrm{i.i.d\,samples\,}\alpha,\beta,\gamma,\delta: & =\left\langle K_{\alpha\beta}^{y}\right\rangle \left\langle K_{\gamma\delta}^{y}\right\rangle -\left\langle K_{\alpha\beta}^{y}\right\rangle \left\langle K_{\gamma\delta}^{y}\right\rangle =0\,.
\end{align}
\foreignlanguage{english}{Hence the tensor $C_{(\alpha\beta)(\gamma\delta)}$
is sparse by construction due to the assumption of independent and
identically distributed training data samples. One additional assumption
we make in the main text is that the diagonal elements of the kernel
matrix $K_{\alpha\alpha}^{y}$ are deterministic $K_{\alpha\alpha}^{y}=\left\langle K_{\alpha\alpha}^{y}\right\rangle =a$.
From this follows directly}

\begin{align}
C_{(\alpha\alpha)(\beta\gamma)} & =\left\langle \left(K_{\alpha\alpha}^{y}-\left\langle K_{\alpha\alpha}^{y}\right\rangle \right)\left(K_{\beta\gamma}^{y}-\left\langle K_{\beta\gamma}^{y}\right\rangle \right)\right\rangle \nonumber \\
 & =\left\langle \left(a-a\right)\left(K_{\beta\gamma}^{y}-\left\langle K_{\beta\gamma}^{y}\right\rangle \right)\right\rangle =0\,,\\
\rightarrow & C_{(\alpha\alpha)(\alpha\beta)}=C_{(\alpha\alpha)(\alpha\alpha)}=0\,.
\end{align}
\foreignlanguage{english}{As $m_{\alpha\beta}^{-1}$ corresponds to
the propagator in the expressions in our main text, we state the matrix
elements of the inverse of a bipartite block structured matrix \prettyref{eq:Mean_Blocks}
in }\prettyref{supp:AppendixElementsInverseBlockMatrix} \foreignlanguage{english}{.
Based on the action }\eqref{eq:Appendix_FullDisorderAveragedAction}\foreignlanguage{english}{
we now want to compute the corrections at first order to the mean-inferred
network output which is given by}

\selectlanguage{english}%
\begin{equation}
\langle y_{*}\rangle_{0+1}=g_{*\alpha}y_{\alpha}+g_{*\alpha}C_{(\alpha\beta)(\gamma\delta)}m_{\beta\gamma}^{-1}m_{\delta\delta^{\prime}}^{-1}y_{\delta^{\prime}}-C_{(*\alpha)(\beta\gamma)}m_{\alpha\beta}^{-1}m_{\gamma\gamma^{\prime}}^{-1}y_{\gamma^{\prime}}\,.\label{eq:Appendix_FullExpressionMeanInferredNetworkOutput}
\end{equation}
\foreignlanguage{american}{This can be rewritten as}

\begin{align}
\langle y^{*}\rangle_{0+1} & =g_{*\alpha}y_{\alpha}+V_{4}(*)-V_{3}(*)\,,\\
V_{3}(*) & =\sum_{\alpha,\beta,\gamma}C_{(*\alpha)(\beta\gamma)}m_{\alpha\beta}^{-1}\hat{y}_{\gamma}\,,\\
V_{4}(*) & =\sum_{\alpha,\beta,\gamma,\delta}g_{*\alpha}C_{(\alpha\beta)(\gamma\delta)}m_{\beta\gamma}^{-1}\hat{y}_{\delta}\,,\\
g_{*,\alpha}= & \sum_{\alpha^{\prime}}m_{*\alpha^{\prime}}m_{\alpha^{\prime}\alpha}^{-1}\,,\\
\hat{y}_{\gamma} & =\sum_{\gamma^{\prime}}m_{\gamma\gamma^{\prime}}^{-1}y_{\gamma^{\prime}}\,,
\end{align}
where we introduced the shorthand notation $g_{*\alpha},\hat{y},V_{3}(*),V_{4}(*)$
to simplify subsequent calculations in this part of the appendix.
Evaluating the expression \eqref{eq:Appendix_FullExpressionMeanInferredNetworkOutput}
is numerically expensive, because it requires computing contractions
over $D$ training examples in up to $5$ indices and hence scales
as $\mathcal{O}(D^{5})$. However, we know from the definition of
our problem setting that the tensors $C_{(\alpha\beta)(\gamma\delta)},C_{(*\alpha)(\beta\gamma)}$
are sparse and have block structure. We can therefore simplify the
problem and compute the contractions analytically. We will do so by
computing $V_{3}(*)$ and $V_{4}(*)$ individually and combine the
results later. It is important to note that, even though the approach
below is general, it is practically restricted to a binary classification
problem. The reason for this is that in our calculation we exploit
the block structure in the tensors by splitting the contractions over
the indices $\alpha,\beta...$ into two parts, assuming that the data
is presented in an ordered fashion

\begin{equation}
\sum_{\alpha}=\sum_{c_{\alpha}=c_{1}}+\sum_{c_{\beta}=c_{2}}\,.
\end{equation}
\foreignlanguage{american}{In principle, an extension to more classes
is possible in an analogous way, but it requires the inversion of
block matrices with more than two blocks. In the case of $V_{3}(*)$
the contractions over the three indices $\alpha,\beta,\gamma$ hence
produces eight terms}

\begin{align}
\left(\sum_{\alpha,c(\alpha)=1}+\sum_{\beta,c(\alpha)=2}\right)\left(\sum_{\beta,c(\beta)=1}+\sum_{\beta,c(\beta)=2}\right)\left(\sum_{\gamma,c(\gamma)=1}+\sum_{\gamma,c(\gamma)=2}\right) & =\sum_{c(\alpha)=c(\beta)=c(\gamma)=1}\nonumber \\
 & +\sum_{c(\alpha)=c(\beta)=1,c(\gamma)=2}+....\sum_{c(\alpha)=c(\beta)=c(\gamma)=2}\,,
\end{align}
whereas $V_{4}(*)$ requires the evaluation of 16 individual terms.
As the number of terms grows exponentially with the number of classes
we restrict ourselves to binary classification. In the remainder of
this section we denote $D_{1}$ as the number of training samples
in class 1 and $D_{2}$ as the number of training samples in class
2.

\subsubsection{Three Point term $V_{3}(*)$}

We evaluate $V_{3}(*)$ by splitting the contractions in eight terms.
However, due to the structure of $C_{(*\alpha)(\beta,\gamma)}$ two
of those terms vanish\foreignlanguage{american}{. The reason for this
is that these terms contain tensor elements $C_{(*\cdot)(\cdot,\cdot)}$
which vanish by construction. Take}

\selectlanguage{american}%
\begin{equation}
\sum_{c_{\alpha}=1,c_{\beta}=2,c_{\gamma}=2}C_{(*,\alpha)(\beta,\gamma)}m_{\alpha,\beta}^{-1}\hat{y}_{\gamma}
\end{equation}
for example. We know, that $C_{(*\alpha)(\beta\gamma)}=0$ if neither
$\alpha=\beta$ nor $\alpha=\gamma$ and because the test point $*\notin\{\alpha,\beta,\gamma\}$.
However, as $c_{\alpha}=1$ and both $c_{\beta}=c_{\gamma}=2$ and
the set of the two patterns are distinct, there is no index combination
of $\alpha\beta\gamma$ which yields $C_{(*\alpha)(\beta\gamma)}\neq0$.
Hence this term vanishes. Due to symmet\foreignlanguage{english}{ry
the equivalent term with $c_{\alpha}=2,c_{\beta}=c_{\gamma}=1$ vanishes
as well. In this appendix we will present the calculation for the
first term. The calculations for the remaining seven terms can be
found in }\prettyref{supp:Asymmetric-Overlap-Inference-1} \foreignlanguage{english}{.\newline}

\selectlanguage{english}%

\paragraph{\noindent Example calculation for case 1: $c_{\alpha}=1,c_{\beta}=1,c_{\gamma}=1$\newline}

\selectlanguage{american}%
\noindent \newline For the first term we will spell out the calculation
steps explicitly. The calculations for the subsequent terms follow
along similar lines. We start by introducing the notation

\begin{equation}
C_{(\alpha\beta)(\gamma\delta)}^{c_{\alpha},c_{\beta},c_{\gamma},c_{\delta}}\,\mathrm{with}\,c_{\alpha},c_{\beta},c_{\gamma},c_{\delta}\in\{1,2\}\,.
\end{equation}
From \prettyref{supp:AppendixElementsInverseBlockMatrix} \foreignlanguage{english}{we know that the expressions
$\hat{y}_{\gamma}$ have block structure as well and $\hat{y}_{\gamma}=\hat{y}_{1}\forall\gamma\,\mathrm{with\,}c_{\gamma}=1$.
To evaluate the first sum}

\selectlanguage{english}%
\begin{equation}
\sum_{\alpha,\beta,\gamma}^{c_{\alpha}=c_{\beta}=c_{\gamma}=c_{1}}C_{(*,\alpha)(\beta,\gamma)}m_{\alpha,\beta}^{-1}\hat{y}_{\gamma}\,.
\end{equation}
We can start by recognizing that the expression $C_{(*\alpha)(\beta\gamma)}$
is only unequal to zero if either $\alpha=\beta$ or $\alpha=\gamma$.
Hence we can split the sum

\begin{equation}
\sum_{\alpha,\beta,\gamma}^{c_{\alpha}=c_{\beta}=c_{\gamma}=c_{1}}C_{(*,\alpha)(\beta,\gamma)}m_{\alpha,\beta}^{-1}\hat{y}_{\gamma}=\sum_{\alpha=\beta,\gamma}^{c_{\alpha}=c_{\beta}=c_{\gamma}=c_{1}}C_{(*,\alpha)(\alpha,\gamma)}m_{\alpha,\alpha}^{-1}\hat{y}_{\gamma}+\sum_{\alpha=\gamma,\beta}^{c_{\alpha}=c_{\beta}=c_{\gamma}=c_{1}}C_{(*,\alpha)(\beta,\alpha)}m_{\alpha,\beta}^{-1}\hat{y}_{\gamma}\,.
\end{equation}
From this, and exploiting the block structure in $C_{(*\alpha)(\beta\gamma)},\hat{y}_{\gamma},m_{\alpha\beta}^{-1}$
we get

\begin{align}
\sum_{\alpha,\beta,\gamma}^{c_{\alpha}=c_{\beta}=c_{\gamma}=c_{1}}C_{(*,\alpha)(\beta,\gamma)}m_{\alpha,\beta}^{-1}\hat{y}_{\gamma} & =\sum_{\alpha=\beta,\gamma}^{c_{\alpha}=c_{\beta}=c_{\gamma}=c_{1}}C_{(*,\alpha)(\alpha,\gamma)}q_{1}\hat{y}_{\gamma}\nonumber \\
 & +\sum_{\alpha=\gamma,\beta}^{c_{\alpha}=c_{\beta}=c_{\gamma}=c_{1}}C_{(*,\alpha)(\beta,\alpha)}q_{2}\hat{y}_{\gamma}\,,\nonumber \\
 & =C_{(*\alpha)(\alpha,\gamma)}^{*,1,1,1}D_{1}(D_{1}-1)\hat{y}_{1}\left(q_{1}+q_{2}\right)\,.
\end{align}
As the subsequent calculations are analogous we will simply state
the results for the sake of brevity if no additional considerations
need to be taken care of.
\selectlanguage{american}%

\paragraph{\newline \foreignlanguage{english}{\noindent Total expression for
$V_{3}(*)$\newline}}

\noindent \newline \foreignlanguage{english}{From the symmetries
of $C$ and and by combining the results from the eight terms we can
write down the full expressions for $V_{3}(*)$ as}

\selectlanguage{english}%
\begin{align}
V_{3}(*) & =C_{(*\alpha)(\alpha,\gamma)}^{*,1,1,1}N_{1}(N_{1}-1)\hat{y}_{1}\left(q_{1}+q_{2}\right)\nonumber \\
 & +C_{(*\alpha)(\alpha,\gamma)}^{*,1,1,2}N_{1}N_{2}q_{1}\hat{y}_{2}\nonumber \\
 & +C_{(*\alpha)(\beta,\alpha)}^{*,1,2,1}N_{1}N_{2}\hat{y}_{1}r\nonumber \\
 & +C_{(*\alpha,)(\beta,\alpha)}^{*,2,1,2}N_{1}N_{2}r\hat{y}_{2}\nonumber \\
 & +C_{(*\alpha)(\alpha,\gamma)}^{*,2,2,1}N_{1}N_{2}q_{1}^{\prime}\hat{y}_{1}\nonumber \\
 & +N_{2}(N_{2}-1)\hat{y}_{2}C_{(*\alpha)(\alpha,\gamma)}^{*,2,2,2}\left(q_{1}^{\prime}+q_{2}^{\prime}\right)\,,\label{eq:Appendix_V3_Raw}
\end{align}
which reduces after some calculations (see \foreignlanguage{american}{\prettyref{supp:Asymmetric-Overlap-Inference-1}})
to

\begin{align}
V_{3}(*\in c_{1}) & =v_{1}N_{1}(N_{1}-1)\left(q_{1}\hat{y}_{1}+q_{2}\hat{y}_{1}\right)\nonumber \\
 & +v_{3}N_{1}N_{2}\left(q_{1}\hat{y}_{2}+\hat{y}_{1}r\right)\nonumber \\
 & +\overline{v}_{2}N_{1}N_{2}\left(r\hat{y}_{2}+q_{1}^{\prime}\hat{y}_{1}\right)\nonumber \\
 & +\overline{v}_{3}N_{2}(N_{2}-1)\left(q_{1}^{\prime}\hat{y}_{2}+q_{2}^{\prime}\hat{y}_{2}\right)\,,\\
V_{3}(*\in c_{2}) & =v_{3}N_{1}(N_{1}-1)\left(q_{1}\hat{y}_{1}+q_{2}\hat{y}_{1}\right)\nonumber \\
 & +v_{2}N_{1}N_{2}\left(q_{1}\hat{y}_{2}+\hat{y}_{1}r\right)\nonumber \\
 & +\overline{v}_{3}N_{1}N_{2}\left(r\hat{y}_{2}+q_{1}^{\prime}\hat{y}_{1}\right)\nonumber \\
 & +\overline{v}_{1}N_{2}(N_{2}-1)\left(q_{1}^{\prime}\hat{y}_{2}+q_{2}^{\prime}\hat{y}_{2}\right)\,.
\end{align}

\subsubsection{Four point expression $V_{4}(*)$}

The calculations in this subsection are analogous to the calculations
for $V_{3}(*)$, the difference being that instead of eight terms
we now have four contractions over the training data indices $\alpha,\beta,\gamma,\delta$
and hence we need to consider sixteen terms. Here we present the calculation
for the first term in detail. The remaining 15 terms are part of  \foreignlanguage{american}{\prettyref{supp:Asymmetric-Overlap-Inference-1}}.

\selectlanguage{american}%

\paragraph{\newline \foreignlanguage{english}{\noindent Example calculation
for term 1: $c_{\alpha}=1,c_{\beta}=1,c_{\gamma}=1,c_{\delta}=1$\newline}}

\noindent \newline For the first term we will spell out the calculation
steps explicitly. The calculations for the subsequent terms follow
along similar lines. We use the same notation as in the previous subsection
for tensor elements

\begin{equation}
C_{(\alpha\beta)(\gamma\delta)}^{c_{\alpha},c_{\beta},c_{\gamma},c_{\delta}}\,\mathrm{with\,}c_{\alpha},c_{\beta},c_{\gamma},c_{\delta}\in\{1,2\}\,.
\end{equation}
For the first term we get:

\selectlanguage{english}%
\begin{align}
\sum_{c_{\alpha}=1,c_{\beta}=1,c_{\gamma}=1,c_{\delta}=1}g_{*,\alpha}C_{(\alpha,\beta)(\gamma,\delta)}m_{\beta,\gamma}^{-1}\hat{y}_{\delta} & =g_{*1}\hat{y}_{1}D_{1}(D_{1}-1)C_{(\alpha,\beta)(\alpha,\beta)}^{1,1,1,1}(q_{1}+q_{2})\nonumber \\
 & +g_{*1}\hat{y}_{1}D_{1}(D_{1}-1)(D_{1}-2))C_{(\alpha,\beta)(\alpha,\delta)}^{1,1,1,1}\left(q_{1}+3q_{2}\right)\,.
\end{align}
The occurring terms have different origins. First we can state that
$g_{*\alpha}=g_{*1}$ and $\hat{y}_{\delta}=\hat{y}_{1}$ because
of the block structure. Next we can decompose the sum into relevant
expressions. One of them covers the cases where either $\alpha=\gamma,\beta=\delta$
and another $\alpha=\delta,\beta=\gamma$ which produce the following
sums

\begin{equation}
\sum_{\alpha=\gamma,\beta=\delta}^{c_{\alpha}=c_{\beta}=c_{1}}g_{*,\alpha}C_{(\alpha,\beta)(\alpha,\beta)}m_{\beta,\alpha}^{-1}\hat{y}_{\beta}+\sum_{\alpha=\delta,\beta=\gamma}^{c_{\alpha}=c_{\beta}=c_{1}}g_{*,\alpha}C_{(\alpha,\beta)(\beta,\alpha)}m_{\beta,\beta}^{-1}\hat{y}_{\alpha}\,.\label{eq:Appendix_V4_Term_1_Sums1}
\end{equation}
In both sums the tensor element yields $C_{(\alpha\beta)(\alpha\beta)}^{1,1,1,1}$
because of the exchange symmetry in the tensor $C$. Both sums simplify
to

\begin{equation}
g_{*1}C_{(\alpha\beta)(\alpha\beta)}^{1,1,1,1}\hat{y}_{1}\left(\sum_{\alpha=\gamma,\beta=\delta}^{c_{\alpha}=c_{\beta}=c_{1}}m_{\beta,\alpha}^{-1}+\sum_{\alpha=\delta,\beta=\gamma}^{c_{\alpha}=c_{\beta}=c_{1}}m_{\beta,\beta}^{-1}\right)\,.
\end{equation}
\foreignlanguage{american}{Because of the block structure in the propagator
$m_{\alpha\beta}^{-1}$ we can also replace $m_{\alpha\beta}^{-1}=q_{2}$
and $m{{}^-}_{\beta\beta}^{1}=q_{1}$ according to our definitions
in \prettyref{supp:AppendixElementsInverseBlockMatrix}. The sums evaluate to}

\selectlanguage{american}%
\begin{align}
\sum_{\alpha=\gamma,\beta=\delta}^{c_{\alpha}=c_{\beta}=c_{1}}m_{\beta,\alpha}^{-1}+\sum_{\alpha=\delta,\beta=\gamma}^{c_{\alpha}=c_{\beta}=c_{1}}m_{\beta,\beta}^{-1} & =\sum_{\alpha=\gamma,\beta=\delta}^{c_{\alpha}=c_{\beta}=c_{1}}q_{2}+\sum_{\alpha=\delta,\beta=\gamma}^{c_{\alpha}=c_{\beta}=c_{1}}q_{1}\nonumber \\
 & =D_{1}\left(D_{1}-1\right)\left(q_{2}+q_{1}\right)\,.
\end{align}
Here we counted the terms in the sums in the following way: We know
that we need to enforce $\alpha\neq\beta$ because otherwise the tensor
would yield $C_{(\alpha\alpha)(\alpha\alpha)}=0$ and hence the terms
would vanish. We can choose any of the $D_{1}$ training examples
for the index $\alpha$. Because $\alpha\neq\beta$ we are left with
$D_{1}-1$ choices for $\beta$ which produces the corresponding prefactor.
In addition to the sums \eqref{eq:Appendix_V4_Term_1_Sums1} one also
needs to consider the sub cases that only two indices in $C_{(\alpha,\beta)(\gamma,\delta)}$
are equal. This produces the sums

\begin{align}
\sum_{c_{\alpha}=1,c_{\beta}=1,c_{\gamma}=1,c_{\delta}=1}g_{*,\alpha}C_{(\alpha,\beta)(\gamma,\delta)}m_{\beta,\gamma}^{-1}\hat{y}_{\delta} & =\sum_{c_{\alpha}=1,c_{\beta}=1,c_{\delta}=1}g_{*,\alpha}C_{(\alpha,\beta)(\beta,\delta)}m_{\beta,\beta}^{-1}\hat{y}_{\delta}\nonumber \\
 & +\sum_{c_{\alpha}=1,c_{\beta}=1,c_{\delta}=1}g_{*,\alpha}C_{(\alpha,\beta)(\alpha,\delta)}m_{\beta,\alpha}^{-1}\hat{y}_{\delta}\nonumber \\
 & +\sum_{c_{\alpha}=1,c_{\beta}=1,c_{\gamma}=1}g_{*,\alpha}C_{(\alpha,\beta)(\gamma,\alpha)}m_{\beta,\gamma}^{-1}\hat{y}_{\alpha}\nonumber \\
 & +\sum_{c_{\alpha}=1,c_{\beta}=1,c_{\gamma}=1}g_{*,\alpha}C_{(\alpha,\beta)(\gamma,\beta)}m_{\beta,\gamma}^{-1}\hat{y}_{\beta}\,.
\end{align}
As above we can extract the terms $g_{*1},C,\hat{y}_{1}$ due to the
block structure in the tensors and we can insert the explicit expressions
for the propagator elements $m_{\beta\beta}^{-1},m_{\alpha\beta}^{-1}$:

\begin{align}
\sum_{c_{\alpha}=1,c_{\beta}=1,c_{\gamma}=1,c_{\delta}=1}g_{*,\alpha}C_{(\alpha,\beta)(\gamma,\delta)}m_{\beta,\gamma}^{-1}\hat{y}_{\delta} & =g_{*1}\hat{y}_{1}C_{(\alpha\beta)(\alpha\delta)}^{1,1,1,1}\bigg(\sum_{c_{\alpha}=1,c_{\beta}=1,c_{\delta}=1}q_{1}+\sum_{c_{\alpha}=1,c_{\beta}=1,c_{\delta}=1}q_{2}\nonumber \\
 & +\sum_{c_{\alpha}=1,c_{\beta}=1,c_{\gamma}=1}q_{2}+\sum_{c_{\alpha}=1,c_{\beta}=1,c_{\gamma}=1}q_{2}\bigg)\text{\,.}
\end{align}
Again we need to consider the admissible elements for the sums: We
assume that the index $\alpha$ can take any of the $D_{1}$ values.
The index $\beta$ can hence take $D_{1}-1$ values as we require
$\alpha\neq\beta$. Likewise $\text{\ensuremath{\gamma}}$ can take
$D_{1}-2$ values as we have $\alpha\neq\gamma,\beta\neq\gamma$.
If this were not the case we would over count elements of the sum
as cases where $\gamma=\beta$ are already covered in \eqref{eq:Appendix_V4_Term_1_Sums1}.
Combining all of these results we get the first term

\selectlanguage{english}%
\begin{align}
\sum_{c_{\alpha}=1,c_{\beta}=1,c_{\gamma}=1,c_{\delta}=1}g_{*,\alpha}C_{(\alpha,\beta)(\gamma,\delta)}m_{\beta,\gamma}^{-1}\hat{y}_{\delta} & =g_{*1}\hat{y}_{1}D_{1}(D_{1}-1)C_{(\alpha,\beta)(\alpha,\beta)}^{1,1,1,1}(q_{1}+q_{2})\nonumber \\
 & =+g_{*1}\hat{y}_{1}D_{1}(D_{1}-1)(D_{1}-2))C_{(\alpha,\beta)(\alpha,\delta)}^{1,1,1,1}\left(q_{1}+3q_{2}\right)\,.
\end{align}

\selectlanguage{american}%

\paragraph{\newline \foreignlanguage{english}{\noindent} Total expression
for $V_{4}(*)$\foreignlanguage{english}{\newline}}

\noindent \newline \foreignlanguage{english}{Considering the symmetries
and the definition of $C$, see }\prettyref{eq:Variance_Blocks} \foreignlanguage{english}{ and \prettyref{eq:Covariance_Blocks}, 
and by combining all 16 terms, the total expression for $V_{4}(*)$
hence reads:}

\selectlanguage{english}%
\begin{align}
V_{4}(*) & =g_{*1}\hat{y}_{1}D_{1}(D_{1}-1)*\left(K_{1}(q_{1}+q_{2})+v_{1}(D_{1}-2)\left(q_{1}+3q_{2}\right)\right)\nonumber \\
 & +g_{*2}\hat{y}_{2}D_{2}(D_{2}-1)*\left(\overline{K}_{1}(q_{1}^{\prime}+q_{2}^{\prime})+\overline{v}_{1}(D_{2}-2)\left(q_{1}^{\prime}+3q_{2}^{\prime}\right)\right)\nonumber \\
 & +D_{1}D_{2}(D_{1}-1)(g_{*1}(q_{1}+q_{2})\hat{y}_{2}+g_{*1}4r\hat{y}_{1}+g_{*2}\hat{y}_{1}(q_{1}+q_{2}))v_{3}\nonumber \\
 & +D_{2}(D_{2}-1)D_{1}(g_{*2}(q_{2}^{\prime}+q_{1}^{\prime})\hat{y}_{1}+g_{*2}4r\hat{y}_{2}+g_{*1}\hat{y}_{2}(q_{1}^{\prime}+q_{2}^{\prime}))\overline{v}_{3}\nonumber \\
 & +\overline{v}_{2}D_{1}D_{2}(D_{1}-1)\left(r*g_{*1}\hat{y}_{2}+g_{*1}\hat{y}_{1}q_{1}^{\prime}+g_{*2}\hat{y}_{2}q_{2}+g_{*2}\hat{y}_{1}r\right)\nonumber \\
 & +v_{2}D_{1}D_{2}(D_{2}-1)\left(g_{*1}\hat{y}_{2}r+g_{*1}\hat{y}_{1}q_{2}^{\prime}+g_{*2}\hat{y}_{2}q_{1}+g_{*2}\hat{y}_{1}r\right)\nonumber \\
\mathrm{} & +K_{2}D_{1}D_{2}\left(g_{*1}\hat{y}_{2}r+g_{*2}\hat{y}_{1}r+g_{*1}\hat{y}_{1}q_{1}^{\prime}+g_{*2}\hat{y}_{2}q_{1}\right)\,.
\end{align}

\subsubsection{Full result}

Taking the full results from $V_{3}(*)$ and $V_{4}(*)$ we get:

\begin{align}
\left\langle y_{*}\right\rangle _{\text{0+1}} & =g_{*1}D_{1}y_{1}+g_{*2}D_{2}y_{2}\nonumber \\
 & +K_{1}(q_{1}+q_{2})g_{*1}\hat{y}_{1}D_{1}(D_{1}-1)+\overline{K}_{1}(q_{1}^{\prime}+q_{2}^{\prime})g_{*2}\hat{y}_{2}D_{2}(D_{2}-1)\nonumber \\
 & +v_{1}\hat{y}_{1}D_{1}(D_{1}-1)\left[(D_{1}-2)\left(q_{1}+3q_{2}\right)*g_{*1}-\left(q_{1}+q_{2}\right)\right]\nonumber \\
 & +\overline{v}_{1}(D_{2}-2)\left(q_{1}^{\prime}+3q_{2}^{\prime}\right)*g_{*2}\hat{y}_{2}D_{2}(D_{2}-1)\nonumber \\
 & +v_{3}D_{1}D_{2}\left[(D_{1}-1)(g_{*1}(q_{1}+q_{2})\hat{y}_{2}+g_{*1}4r\hat{y}_{1}+g_{*2}\hat{y}_{1}(q_{1}+q_{2}))-\left(q_{1}\hat{y}_{2}+\hat{y}_{1}r\right)\right]\nonumber \\
 & +\overline{v}_{3}D_{2}(D_{2}-1)\left[D_{1}(g_{*2}(q_{2}^{\prime}+q_{1}^{\prime})\hat{y}_{1}+g_{*2}4r\hat{y}_{2}+g_{*1}\hat{y}_{2}(q_{1}^{\prime}+q_{2}^{\prime}))-\left(q_{1}^{\prime}\hat{y}_{2}+q_{2}^{\prime}\hat{y}_{2}\right)\right]\nonumber \\
 & +\overline{v}_{2}D_{1}D_{2}\left[(D_{1}-1)\left(r*g_{*1}\hat{y}_{2}+g_{*1}\hat{y}_{1}q_{1}^{\prime}+g_{*2}\hat{y}_{2}q_{2}+g_{*2}\hat{y}_{1}r\right)-\left(r\hat{y}_{2}+q_{1}^{\prime}\hat{y}_{1}\right)\right]\nonumber \\
 & +v_{2}D_{1}D_{2}(D_{2}-1)\left(g_{*1}\hat{y}_{2}r+g_{*1}\hat{y}_{1}q_{2}^{\prime}+g_{*2}\hat{y}_{2}q_{1}+g_{*2}\hat{y}_{1}r\right)\nonumber \\
 & +K_{2}D_{1}D_{2}\left(g_{*1}\hat{y}_{2}r+g_{*2}\hat{y}_{1}r+g_{*1}\hat{y}_{1}q_{1}^{\prime}+g_{*2}\hat{y}_{2}q_{1}\right)\,.\label{eq:Appendix_FullAsymmetricExpression}
\end{align}

\subsubsection{Symmetric Case}

The expression in \eqref{eq:Appendix_FullAsymmetricExpression} can
be greatly simplified if one considers a symmetric task setting such
as in the Ising task of the main text. Following the calculations
in the Supplement \foreignlanguage{american}{\prettyref{supp:Asymmetric-Overlap-Inference-1}} and exploiting
that in the symmetric case we have

\begin{align}
K_{1},\overline{K}_{1},K_{2} & \rightarrow K\,,\\
v_{1},v_{2},\overline{v}_{1},\overline{v}_{2} & \rightarrow v\,,\\
v_{3},v_{4},\overline{v}_{3},\overline{v}_{4} & \rightarrow-v\,,
\end{align}
\foreignlanguage{american}{the expression reduces to the statement
in the main text} \prettyref{eq:mean_SimplePatternStatistics_MFT}
and \prettyref{eq:mean_SimplePatternStatistics_FirstOrder}.
\selectlanguage{american}%

\subsection{Limiting value for $D\rightarrow\infty$ of $\left\langle y_{*}\right\rangle _{0+1}$
in the symmetric setting}

\selectlanguage{english}%
We can now compute expansions of the mean of the predictive distribution
in $1/D\ll1$ in order to obtain the leading order expressions for
large number of training samples and hence get the asymptotic value
for $\lim_{D\rightarrow\infty}\left\langle y_{*}\right\rangle _{0+1}$.
Starting from the full expression \prettyref{eq:mean_SimplePatternStatistics_MFT}
and \prettyref{eq:mean_SimplePatternStatistics_FirstOrder} we compute
the limiting value for $D\rightarrow\infty$. Following along the
lines of the Supplement \foreignlanguage{american}{\prettyref{supp:LimitingValue} },
we recover in $\mathcal{O}(1)$ the result from the main text

\begin{equation}
\langle y\rangle_{0+1}(D\rightarrow\infty)=y_{1}+\frac{y_{1}}{b}\left(\frac{1}{a-b}(K-4v)-\frac{v}{b}\right)\,.
\end{equation}
\section*{Supplemental Material}
\setcounter{subsection}{0}
\setcounter{equation}{0} \renewcommand\theequation{S\arabic{equation}}\onecolumngrid
\subsection{The statistical properties of the data set\label{supp:SupplementStatisticalProperties}}

\subsubsection{The mean and covariance in the synthetic data set}

As presented in the main text, we need to consider the statistical
properties of the overlaps between patterns \prettyref{eq:Definition_InputdataKernel}.
In the case of the synthetic task settings one can evaluate those
properties directly. The patterns in the synthetic task have length
$N_{\mathrm{dim}}$, where $\Ndim$ is even. Each of the $N_{\mathrm{dim}}$
pixels can take the value $\pm1$. The value of each pixel is drawn
independently. This is a simplification compared to real data, where
correlations exist between pixels. Still, this model serves us to
discover the dominant contributions to the variability of the overlap
matrices. The probability to be $\pm1$ is given by the parameter
$p\in[0,1]$, the class membership of the pattern and the relative
position of the pixel in the pattern according to

\begin{align}
x_{\alpha i} & =\begin{cases}
1 & \mathrm{with}\,\:p\\
-1 & \mathrm{with}\,\:(1-p)
\end{cases}\quad\mathrm{for}\,\,i\leq\frac{N_{\mathrm{dim}}}{2}\ ,\label{eq:Pixel-Value-Distribution-Class_1_Part0-1-1}\\
x_{\a i} & =\begin{cases}
1 & \mathrm{with}\,\:(1-p)\\
-1 & \mathrm{with}\,\:p.
\end{cases}\quad\mathrm{for}\,\,i>\frac{N_{\mathrm{dim}}}{2}\.\label{eq:Pixel-Value-Distribution-Class_1-1-1}
\end{align}
For a pattern $x_{\alpha}$ in the second class $c=2$ the pixel values
are distributed according to

\begin{align}
x_{\a i} & =\begin{cases}
-1 & \mathrm{with}\,\:p\\
1 & \mathrm{with}\,\:(1-p)
\end{cases}\quad\mathrm{for}\,\,i\leq\frac{N_{\mathrm{dim}}}{2}\ ,\label{eq:Pixel-Value-Distribution-Class_2_Part0-1-1}\\
x_{\a i} & =\begin{cases}
-1 & \mathrm{with}\,\:(1-p)\\
1 & \mathrm{with}\,\:p
\end{cases}\quad\mathrm{for}\,\,i>\frac{N_{\mathrm{dim}}}{2}\.\label{eq:Pixel-Value-Distribution-Class_2-1-1}
\end{align}
We aim to compute the mean $\mu_{\alpha\beta}$ and the covariances
$\Sigma_{(\alpha\beta)(\gamma\delta)}$ of the overlaps

\begin{equation}
K_{\alpha\beta}^{x}=\sum_{i=1}^{N_{\mathrm{dim}}}x_{\a i}x_{\text{\ensuremath{\b}}i}.\label{eq:Input_Overlap_Appendix-1}
\end{equation}
To this end we define

\begin{align}
\mu_{\alpha\beta} & =\left\langle K_{\alpha\beta}^{x}\right\rangle \ ,\label{eq:Def_Mean_Patterns_Appendix}\\
\Sigma_{(\alpha\beta)(\gamma\delta)} & =\left\langle \delta K_{\alpha\beta}^{x}\delta K_{\gamma\delta}^{x}\right\rangle \ ,\label{eq:Def_Covariance_Patterns_Appendix}\\
 & =\left\langle K_{\alpha\beta}^{x}K_{\gamma\delta}^{x}\right\rangle -\mu_{\alpha\beta}\mu_{\gamma\delta}\,,\\
\delta K_{\alpha\beta}^{x} & =K_{\alpha\beta}^{x}-\mu_{\alpha\beta}\,.
\end{align}

As each pixel value is drawn independently, we can compute $\mu_{\alpha\beta}$
as

\begin{align}
\mu_{\alpha,\beta} & =\left\langle K_{\alpha\beta}^{x}\right\rangle \nonumber \\
 & =\left\langle \sum_{i=1}^{N_{\mathrm{dim}}}x_{\a i}x_{\text{\ensuremath{\b}}i}\right\rangle =\sum_{i=1}^{N_{\mathrm{dim}}}\left\langle x_{\a i}x_{\text{\ensuremath{\b}}i}\right\rangle \.\label{eq:Appendix_ComputeMean_IntermediateStep}
\end{align}

We can obtain the value $\left\langle x_{\a i}x_{\text{\ensuremath{\b}}i}\right\rangle $
based on the class memberships $c_{\alpha},c_{\beta}\in\{1,2\}$ of
$\alpha,\beta$
\begin{enumerate}
\item If $\alpha=\beta$, the average reads $\left\langle x_{\alpha i}^{2}\right\rangle =1\quad\forall i$.
Hence $\mu_{\alpha\alpha}=N_{\mathrm{dim}}$.
\item If $c_{\alpha}=c_{\beta}$, the product $x_{\a i}x_{\text{\ensuremath{\b}}i}=1$
if either both $x_{\a i}=x_{\b i}=1$ or $x_{\a i}=x_{\b i}=-1$.
This happens, according to \eqref{eq:Pixel-Value-Distribution-Class_1_Part0-1-1},
\eqref{eq:Pixel-Value-Distribution-Class_2_Part0-1-1} with the probability
$p^{2}+(1-p)^{2}$. On the other hand, the product $x_{\a i}x_{\text{\ensuremath{\b}}i}=-1$
if one of the pixel values is positive and the other negative. This
event has the probability $2p(1-p)$. Combining those results we obtain
$\mu_{\alpha\beta}=N_{\mathrm{dim}}\left(4p(p-1)+1\right)$.
\item If $c_{\alpha}\neq c_{\beta}$ we obtain a similar result as in 2.)
except for the fact, that the sign is changed. Hence $\mu_{\alpha\beta}=-N_{\mathrm{dim}}\left(4p(p-1)+1\right)$.
\end{enumerate}
Hence we get for the mean with $\hat{\mu}:=4p(p-1)+1$ we get

\begin{equation}
\mu_{\alpha\beta}=N_{\mathrm{dim}}\begin{cases}
1 & \alpha=\beta\ ,\\
\hat{\mu} & c_{\alpha}=c_{\beta}\ ,\\
\brackets{-\hat{\mu}} & c_{\alpha}\neq c_{\beta}\.
\end{cases}\label{eq:Means_SimplePattern_Appendix}
\end{equation}

Similarly one can proceed for the covariance $\Sigma_{(\alpha\beta)(\gamma\delta)}$
which depends on the class memberships of $\alpha,\beta,\gamma,\delta$.
\begin{enumerate}
\item If either $\alpha=\beta$ or $\gamma=\delta$, we get, by definition
\eqref{eq:Def_Mean_Patterns_Appendix}, $K_{\alpha\alpha}^{x}=N_{\mathrm{dim}}$
which is a deterministic value. This directly implies via \eqref{eq:Def_Covariance_Patterns_Appendix}
that $\Sigma_{(\alpha\alpha)(\beta\gamma)}=0$. By extension any covariance
term where more than two indices coincide, such as $\Sigma_{(\alpha\alpha)(\alpha\beta)},\,\Sigma_{(\alpha\alpha)(\alpha\alpha)}$
will also vanish.
\item If all indices differ $\alpha\neq\beta\neq\gamma\neq\delta$ one gets
by definition \eqref{eq:Def_Covariance_Patterns_Appendix} and independence
\begin{align}
\Sigma_{(\alpha\beta)(\gamma\delta)} & =\left\langle K_{\alpha\beta}^{x}K_{\gamma\delta}^{x}\right\rangle -\mu_{\alpha\beta}\mu_{\gamma\delta}\nonumber \\
 & =\left\langle K_{\alpha\beta}^{x}\right\rangle \left\langle K_{\gamma\delta}^{x}\right\rangle -\mu_{\alpha\beta}\mu_{\gamma\delta}=0\.\label{eq:Appendix_ComputeCovariance_DifferentIndices}
\end{align}
\item The element $\Sigma_{(\alpha\beta)(\alpha\beta)}$ for $\alpha\neq\beta$
is the variance of $\sum_{i}x_{\alpha i}x_{\beta i}$, which is additive
as each term is i.i.d. Hence one can write 
\begin{align}
\Sigma_{(\alpha\beta)(\alpha\beta)} & =\sum_{i=1}^{N_{\mathrm{dim}}}\mathrm{Var}\left(x_{\alpha i}x_{\beta i}\right)\nonumber \\
 & \overset{\mathrm{i.i.d.}}{=}N_{\mathrm{dim}}\left(\left\langle \left(\underbrace{x_{\alpha i}x_{\beta i}}_{=\pm1}\right)^{2}\right\rangle -\left\langle x_{\alpha i}x_{\beta i}\right\rangle ^{2}\right)\nonumber \\
 & \overset{\eqref{eq:Means_SimplePattern_Appendix}}{=}N_{\mathrm{dim}}\left(1-\hat{\mu}^{2}\right)\.\label{eq:Appendix_Compute_Variance_TwoIndices}
\end{align}
\item If $\Sigma_{(\alpha\beta)(\alpha\delta)}$ one needs to carefully
consider the class membership of $\alpha,\beta,\delta$. In general
we get:
\begin{align}
\Sigma_{(\alpha\beta)(\gamma\delta)} & =\left\langle \sum_{i,j=1}^{N_{\mathrm{dim}}}x_{\alpha i}x_{\beta i}x_{\alpha j}x_{\delta j}\right\rangle \nonumber \\
 & -\mu_{\alpha\beta}\mu_{\gamma\delta}\nonumber \\
 & =\left\langle \sum_{i=j=1}^{N_{\mathrm{dim}}}\underbrace{\left(x_{\alpha i}\right)^{2}}_{=1}x_{\beta i}x_{\delta i}\right\rangle \nonumber \\
 & +\left\langle \sum_{i=1}^{N_{\mathrm{dim}}}x_{\alpha i}x_{\beta i}\sum_{j=1,i\neq j}^{N_{\mathrm{dim}}}x_{\alpha j}x_{\delta j}\right\rangle \nonumber \\
 & -\mu_{\alpha\beta}\mu_{\alpha\delta}\.\label{eq:Appendix_Compute_Covariance_ThreeIndices_Def}
\end{align}
As the indices $i,j$ in the sum are independent, one can reduce the
average to
\begin{align}
\left\langle \sum_{i,j=1,i\neq j}^{N_{\mathrm{dim}}}x_{\alpha i}x_{\beta i}x_{\alpha j}x_{\delta j}\right\rangle  & =\sum_{i=1}^{N_{\mathrm{dim}}}\left\langle x_{\alpha i}x_{\beta i}\right\rangle \nonumber \\
 & \times\sum_{j=1,j\neq i}^{N_{\mathrm{dim}}}\left\langle x_{\alpha j}x_{\delta j}\right\rangle \nonumber \\
 & =\Ndim\left(\Ndim-1\right)\mu_{\alpha\beta}\mu_{\alpha\delta}\.\label{eq:Appendix_Compute_Covariance_ThreeIndices_Step_1}
\end{align}
Hence the covariance reads
\begin{align}
\Sigma_{(\alpha\beta)(\alpha\delta)} & =\mu_{\beta\delta}+\frac{1}{N_{\mathrm{dim}}}\left(\Ndim-1\right)\mu_{\alpha\beta}\mu_{\alpha\delta}\nonumber \\
 & -\mu_{\alpha\beta}\mu_{\alpha\delta}\nonumber \\
 & =\mu_{\beta\delta}-\frac{1}{\Ndim}\mu_{\alpha\beta}\mu_{\alpha\delta}\.\label{eq:Appendix_Compute_Covariance_ThreeIndices_Step_2}
\end{align}
As we know, $\mu_{\alpha\beta}$ yields either $\Ndim(4p(p-1)+1)$
or $-\Ndim(4p(p-1)+1)$, depending on whether the classes are equal
or different. Therefore we can see, that if $c_{\alpha}=c_{\beta}=c_{\delta}$
or $c_{\beta}=c_{\delta}\neq c_{\alpha}$, we will obtain the same
result for $\Sigma_{(\alpha\beta)(\alpha\delta)}$. In this case $\mu_{\beta\delta}=\Ndim(4p(p-1)+1)$
and $\mu_{\alpha\beta},\mu_{\alpha\delta}$ both yield $\Ndim(4p(p-1)+1)$
or $-\Ndim(4p(p-1)+1).$ This culminates in
\begin{align}
\Sigma_{(\alpha\beta)(\alpha\delta)} & =\Ndim\left(\hat{\mu}-\hat{\mu}^{2}\right)\nonumber \\
 & =\Ndim\hat{\mu}\left(1-\hat{\mu}\right)\.\label{eq:Appendix_Compute_Covariance_ThreeIndices_Step_3}
\end{align}
The same reasoning goes for $c_{\alpha}=c_{\beta}\neq c_{\delta}$
or $c_{\alpha}=c_{\delta}\neq c_{\beta}$. In this case $\mu_{\beta\delta}=-\Ndim(4p(p-1)+1)$
and one of the terms $\mu_{\alpha\beta},\mu_{\alpha\delta}$ yields
$-\Ndim(4p(p-1)+1)$. This yields
\begin{align}
\Sigma_{(\alpha\beta)(\alpha\delta)} & =\Ndim\left(-\hat{\mu}+\hat{\mu}^{2}\right)\nonumber \\
 & =-\Ndim\hat{\mu}\left(1-\hat{\mu}\right)\.\label{eq:Appendix_Compute_Covariance_ThreeIndices_Step_4}
\end{align}
\linebreak{}
Hence we can summarize the properties for $\Sigma_{(\alpha\beta)(\alpha\delta)}$
with $\hat{\mu}=4p(p-1)+1$
\begin{equation}
\Sigma_{(\alpha\beta)(\alpha\delta)}=\Ndim\begin{cases}
\hat{\mu}(1-\hat{\mu}) & \mathrm{for\;}\begin{cases}
c_{\alpha}=c_{\beta}=c_{\delta}\\
c_{\alpha}\ne c_{\beta}=c_{\delta}
\end{cases}\,,\\
-\hat{\mu}(1-\hat{\mu}) & \mathrm{for\;}\begin{cases}
c_{\alpha}=c_{\beta}\neq c_{\delta}\\
c_{\alpha}=c_{\delta}\neq c_{\beta}
\end{cases}\,.
\end{cases}\label{eq:Appendix_Summary_CovarianceTerms_ThreeIndices}
\end{equation}
\end{enumerate}
Finally, the assumption that patterns are drawn independently from
the described distribution corresponds to the fundamental assumption
of supervised learning, which assumes that the observed data stems
from an unknown distribution and that different samples are drawn
independent of one another. This corresponds to the case of an infinite
number of samples to draw the finite training set from. A finite set
to start with would induce correlations between patterns.

\subsubsection{Sub--leading corrections to Gaussian approximations of overlaps
\label{app:Sub=002013leading-corrections-OverlapDistribution}}

We assumed in the main text, that the overlap distribution is Gaussian
\prettyref{eq:Distribution_OverlapElements}. We can show, that this
is approximately the case and that higher order cumulants are suppressed
by the term $\mathcal{O}\left(1/N_{\mathrm{dim}}\right)$.

We define

\begin{equation}
y:=\frac{1}{N_{\mathrm{dim}}}\sum_{i=1}^{N_{\mathrm{dim}}}x_{\alpha i}x_{\beta i}\.\label{eq:Empirical_Overlap_SimplePattern}
\end{equation}
We start by constructing the cumulant generating function $W(j)$
of $y$ :

\begin{align}
W(j) & =\ln\left\langle \exp\left(jy\right)\right\rangle _{y}\\
 & =\ln\left\langle \exp\left(j\frac{1}{N_{\mathrm{dim}}}\sum_{i=1}^{N_{\mathrm{dim}}}x_{\alpha i}x_{\beta i}\right)\right\rangle _{x_{\alpha},x_{\beta}}\,.
\end{align}
As each of the terms in the sum of \prettyref{eq:Empirical_Overlap_SimplePattern}
is distributed independently, the cumulant generating function $W(j)$
decomposes into the sum of the cumulant generating functions for each
of the terms in the sum

\begin{align}
W(j) & =\ln\left\langle \exp\left(j\frac{1}{N_{\mathrm{dim}}}\sum_{i=1}^{N_{\mathrm{dim}}}x_{\alpha i}x_{\beta i}\right)\right\rangle _{x_{\alpha},x_{\beta}}\\
 & =\sum_{i=1}^{N_{\mathrm{dim}}}\ln\left\langle \exp\left(\frac{j}{N_{\mathrm{dim}}}x_{\alpha i}x_{\beta i}\right)\right\rangle _{x_{\alpha i},x_{\beta i}}\\
 & =N_{\mathrm{dim}}W_{1}\left(\frac{j}{N_{\mathrm{dim}}}\right),
\end{align}
where we used that each term in the sum $x_{\alpha i}x_{\beta i}$
is distributed identically in the step from line two to line three
and defined $W_{1}(j)=\ln\left\langle \exp\left(jx_{\alpha1}x_{\beta1}\right)\right\rangle _{x_{\alpha1},x_{\beta1}}$.
From the distribution of $x_{\alpha i}x_{\beta i}$ we know that

\begin{align}
W_{1}(j) & =\ln\left(qe^{-j}+(1-q)e^{j}\right)\,,\\
\kappa_{1}^{(n)} & =\frac{\partial W_{1}(j)}{\partial j}\bigg\rvert_{j=0}\,,
\end{align}
with $q=p$ or $q=1-p$ (depending on whether the class membership
of $x_{\alpha}$ is different from $x_{\beta}$ or not). The $n$-th
cumulant of $x_{\alpha i}x_{\beta i}$ hence scales as $\kappa_{1}^{(n)}\propto\mathcal{O}(1)$.
As the $n$-th cumulant of $y$ is given by the derivatives of $W(j)$,
the scaling follows from $\kappa_{1}^{(n)}$, the definition of $W(j)$
and by the chain rule

\begin{align}
\kappa^{(n)} & =\frac{\partial^{n}W(j)}{\partial j^{n}}\bigg\rvert_{j=0}\\
 & =\left(\frac{1}{N_{\mathrm{dim}}}\right)^{n-1}\kappa_{1}^{(n)}\,.
\end{align}
Non-Gaussian terms, such as cumulants higher than $\kappa^{(3)}$
are hence suppressed at least by a factor $1/N_{\mathrm{dim}}$ compared
to the first two cumulants $\kappa^{(1)}\propto\mathcal{O}(1),\kappa^{(2)}\propto\mathcal{O}\left(1/N_{\mathrm{dim}}\right)$.

\begin{figure}
\begin{centering}
\includegraphics[scale=0.6]{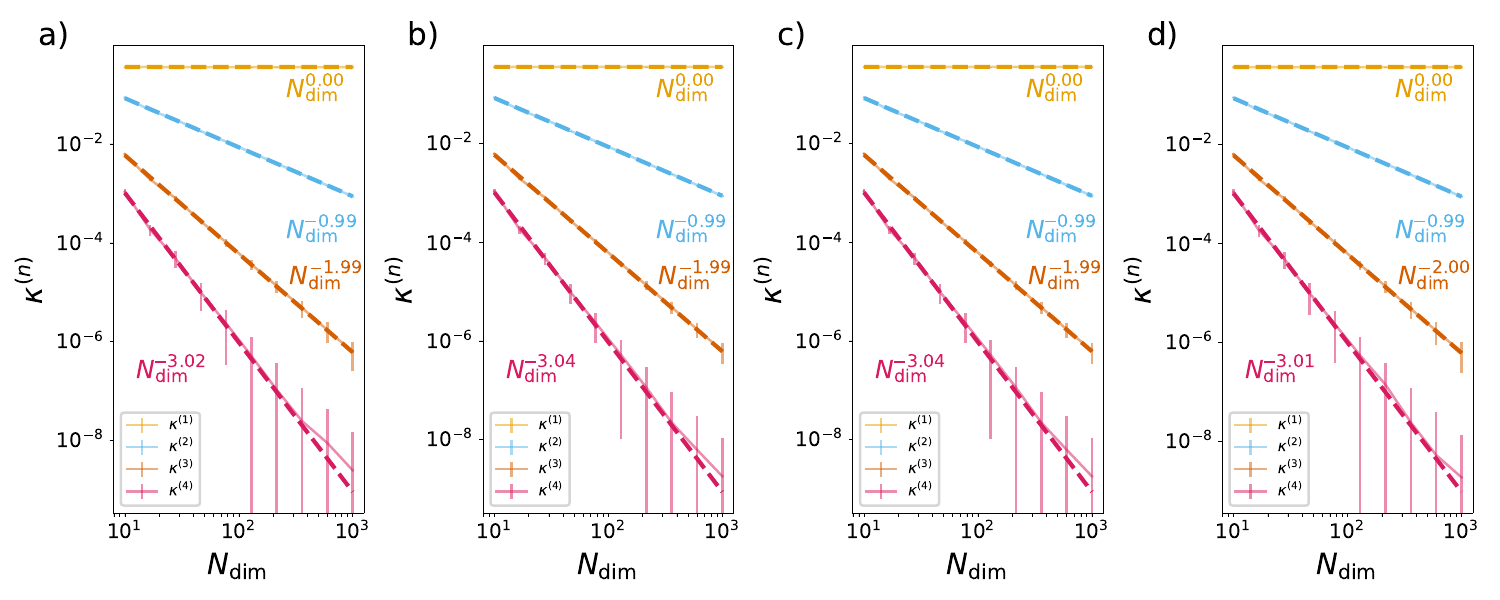}
\par\end{centering}
\caption{\textbf{\label{fig:AppendixScalingNonGaussianContributions_OverlapDistribution}Scaling
of cumulants with pattern dimensionality }Scaling of the first (yellow),
second (blue), third (brown) and fourth (purple) cumulant of the input
overlap distribution with the pattern dimensionality $N_{\mathrm{dim}}$.
Results are shown for the overlap statistics of a) the intra-class
overlap distribution of class 1, b) the intra-class overlap distribution
of class 2 and c,d) the inter-class overlap distribution between patterns
of class 1 and class 2. Dashed results display linear fit in log-log
plot. Settings: $p=0.8,$ $N_{\mathrm{dim}}=10,\ldots,10^{3},$$N_{\mathrm{trial}}=400$,
$N_{\mathrm{dat}}=2000$.}
\end{figure}

This comes with a notable consequence: If one would want to use the
moment generating function in the main text to compute corrections
up to second order in the vertex $C_{(\alpha\beta)(\gamma\delta)}\sim\mathcal{O}\left(1/N_{\mathrm{dim}}\right)$,
one has to perturbatively consider the third cumulant $\kappa_{(\alpha\beta)(\gamma\delta)(\mu\rho)}^{(3)}\sim\mathcal{O}\left(1/N_{\mathrm{dim}}^{2}\right)$
(formally a six-point vertex) as well to treat the perturbations consistently.
This scaling is visualized in \prettyref{fig:AppendixScalingNonGaussianContributions_OverlapDistribution},
which shows that empirical results for the cumulants $\kappa^{(1)}...\kappa^{(4)}$
of the overlap distributions match the scaling $\kappa^{(n)}\sim\left(1/N_{\mathrm{dim}}\right)^{n-1}$.

\subsubsection{Influence of pattern statistics on statistics of pattern overlap
matrix}

We want to understand how Gaussian inputs are related to the statistics
of the overlap. For this we make the following definition for patterns
$x_{i\alpha}$

\begin{equation}
x_{i\alpha}:=\mu_{i\alpha}+\ensuremath{\eta_{i\alpha}},\quad i=1...N_{\mathrm{dim}}\,.\label{eq:Appendix_DataModel}
\end{equation}
We assume, that the patterns $x_{i\alpha}$ can be in on of two classes
$c(\alpha)\in\{1,2\}$. We further define that the mean $\mu_{i}^{(\alpha)}$
is class dependent. The noise $\eta_{i\alpha}$ is assumed to be Gaussian
and follows

\begin{align}
\left\langle \ensuremath{\eta_{i\alpha}}\right\rangle _{\eta} & =0\,,\nonumber \\
\left\langle \ensuremath{\eta_{i\alpha}}\ensuremath{\eta_{j\beta}}\right\rangle _{\eta} & =\delta_{\alpha\beta}D_{c(\alpha)ij}\,.\label{eq:Appendix_Pattern_NoiseStatistics}
\end{align}
Here the assumption is, that patterns $x_{i\alpha}$ are independently
and identically drawn and correlations only exist within pixels $i,j$
of a single a pattern and not between patterns. In total we have the
following properties for the input

\begin{align}
\mu_{i\alpha} & =\begin{cases}
\mu_{i1} & c(\alpha)=c_{1}\\
\mu_{i2} & c(\alpha)=c_{2}
\end{cases}\,,\label{eq:Appendix5_Definition_PatternMean}\\
\left\langle \ensuremath{\eta_{i\alpha}}\right\rangle _{\eta} & =0\quad\forall i,\alpha\quad,\\
\left\langle \ensuremath{\eta_{i\alpha}}\ensuremath{\eta_{j\beta}}\right\rangle _{\eta} & =\begin{cases}
\delta_{\alpha\beta}D_{1,ij} & c(\alpha)=c_{1}\\
\delta_{\alpha\beta}D_{2,ij} & c(\beta)=c_{2}
\end{cases}\,.\label{eq:Appendix_Definition_InputStatistics}
\end{align}
From this we define the pattern overlap $K_{\alpha\beta}^{x}$ as

\begin{align}
K_{\alpha\beta}^{x} & =\begin{cases}
a & \alpha=\beta\\
\sum_{i=1}^{N_{\mathrm{dim}}}x_{i\alpha}x_{i\beta} & \alpha\neq\beta
\end{cases}\label{eq:DefOverlap_Appendix_InputDepdendent}
\end{align}
Keep in mind that this enforces normalization of patterns and hence
$K_{\alpha\alpha}^{x}$ is deterministic. For the off-diagonal case
$\alpha\neq\beta$ we hence obtain

\begin{align}
K_{\alpha\beta}^{x} & =\sum_{i=1}^{N_{\mathrm{dim}}}\left(\mu_{i\alpha}+\ensuremath{\eta_{i\alpha}}\right)\left(\mu_{i\beta}+\ensuremath{\eta_{i\beta}}\right)\nonumber \\
 & =\sum_{i=1}^{N_{\mathrm{dim}}}\mu_{i\alpha}\mu_{i\beta}+\ensuremath{\eta_{i\alpha}}\mu_{i\beta}+\mu_{i\alpha}\ensuremath{\eta_{i\beta}}+\ensuremath{\eta_{i\alpha}\ensuremath{\eta_{i\beta}}}\nonumber \\
 & =\mu_{\alpha}^{\top}\mu_{\beta}+\eta_{\alpha}^{\top}\mu_{\beta}+\eta_{\beta}^{\top}\mu_{\alpha}+\eta_{\alpha}^{\top}\eta_{\beta}\,.\label{eq:DefStochasticOverlap_Appendix}
\end{align}
From this we obtain the mean-overlap

\begin{align}
\left\langle K_{\alpha\beta}^{x}\right\rangle _{\eta}=m_{\alpha\beta} & =\left\langle \mu_{\alpha}^{\top}\mu_{\beta}+\eta_{\alpha}^{\top}\mu_{\beta}+\eta_{\beta}^{\top}\mu_{\alpha}+\eta_{\alpha}^{\top}\eta_{\beta}\right\rangle _{\eta}\\
 & =\mu_{\alpha}^{\top}\mu_{\beta}\quad\mathrm{with}\:\alpha\neq\beta,\label{eq:Appendix_Computation_MeanOverlapGaussianInputs}
\end{align}
which is independent of the noise $\eta$ statistics. We now consider
the covariance

\begin{align}
\Sigma_{(\alpha\beta)(\alpha\delta)} & =\left\langle \left(K_{\alpha\beta}^{x}-m_{\alpha\beta}\right)\left(K_{\alpha\delta}^{x}-m_{\alpha\delta}\right)\right\rangle _{\eta}\\
 & =\left\langle K_{\alpha\beta}^{x}K_{\alpha\delta}^{x}\right\rangle -m_{\alpha\beta}m_{\alpha\delta}\,.\label{eq:Appendix_Definition_CovarianceOverlaps}
\end{align}
From this definition we can directly evaluate

\begin{align}
\Sigma_{(\alpha\beta)(\alpha\delta)} & =\left\langle \sum_{i=1}^{N_{\mathrm{dim}}}\left(\mu_{i\alpha}+\ensuremath{\eta_{i\alpha}}\right)\left(\mu_{i\beta}+\ensuremath{\eta_{i\beta}}\right)\sum_{j=1}^{N_{\mathrm{dim}}}\left(\mu_{j\alpha}+\ensuremath{\eta_{j\alpha}}\right)\left(\mu_{j\delta}+\ensuremath{\eta_{j\delta}}\right)\right\rangle _{\eta}-m_{\alpha\beta}m_{\alpha\delta}\\
 & =\left\langle \sum_{i,j=1}^{N_{\mathrm{dim}}}\left(\mu_{i\alpha}\mu_{i\beta}+\ensuremath{\eta_{i\alpha}}\mu_{i\beta}+\ensuremath{\eta_{i\beta}\mu_{i\alpha}+\ensuremath{\eta_{i\alpha}\ensuremath{\eta_{i\beta}}}}\right)\left(\mu_{j\alpha}\mu_{j\delta}+\ensuremath{\eta_{j\alpha}}\mu_{j\delta}+\ensuremath{\eta_{j\delta}\mu_{j\alpha}+\ensuremath{\eta_{j\alpha}\ensuremath{\eta_{j\delta}}}}\right)\right\rangle _{\eta}-m_{\alpha\beta}m_{\alpha\delta}\\
 & =\left\langle \sum_{i,j=1}^{N_{\mathrm{dim}}}\ensuremath{\eta_{i\alpha}}\mu_{i\beta}\ensuremath{\eta_{j\alpha}}\mu_{j\delta}+\eta_{i\beta}\mu_{i\alpha}\eta_{j\delta}\mu_{j\alpha}+\ensuremath{\eta_{i\alpha}\ensuremath{\eta_{i\beta}}\ensuremath{\eta_{j\alpha}\ensuremath{\eta_{j\delta}}}}\right\rangle _{\eta}\\
 & =\mu_{\beta}^{\top}D_{c(\alpha)}\mu_{\delta}+\delta_{\beta,\delta}\left[\mu_{\alpha}^{\top}D_{c(\beta)}\mu_{\alpha}+\mathrm{Tr}\left(D_{c(\alpha)}D_{c(\beta)}^{\top}\right)\right]\,,\label{eq:Appendix_CalculationCovarianceGaussianInputs}
\end{align}
where we used the Gaussianity of $\eta$, its vanishing mean so that
all terms with an odd number of factors $\eta$ vanish, taking into
account that $\alpha\neq\beta$ and $\alpha\neq\delta$, because by
\prettyref{eq:DefOverlap_Appendix_InputDepdendent} only the off-diagonal
elements of $K$ show variability, so $\Sigma_{(\alpha\alpha)(\alpha\delta)}=\Sigma_{(\alpha\beta)(\alpha\alpha)}=0$,
and Wick's theorem to simplify the average $\left\langle \ensuremath{\eta_{i\alpha}\ensuremath{\eta_{i\beta}}\ensuremath{\eta_{j\alpha}\ensuremath{\eta_{j\delta}}}}\right\rangle _{\eta}$.
Hence we can deduce the elements of $\mu_{\alpha\beta}$ and $\Sigma_{(\alpha\beta)(\alpha\delta)}$
using \eqref{eq:Appendix_CalculationCovarianceGaussianInputs} and
the definitions in \eqref{eq:Appendix_Definition_InputStatistics}
and \eqref{eq:Appendix5_Definition_PatternMean} as

\begin{align}
\mu_{\alpha\beta} & =\delta_{\alpha\beta}\,a+(1-\delta_{\alpha\beta})\mu_{c(\alpha)c(\beta)}^{\top}\,,\label{eq:eq:Appendix5_OverlapStatisticsGaussianPattern_Means}\\
\Sigma_{(\alpha\beta)(\alpha\beta)} & =\mu_{c(\beta)}^{\top}D_{c(\alpha)}\mu_{c(\beta)}+\mu_{c(\alpha)}^{\top}D_{c(\beta)}\mu_{c(\alpha)}+\mathrm{Tr}\left(D_{c(\alpha)}D_{c(\beta)}^{\top}\right)\quad\mathrm{with\quad\alpha\neq\beta}\,,\label{eq:Appendix5_OverlapStatisticsGaussianPattern_AB}\\
\Sigma_{(\alpha\beta)(\alpha\delta)} & =\mu_{\beta}^{\top}D_{c(\alpha)}\mu_{c(\delta)}\quad\mathrm{with\quad\alpha\neq}\beta\neq\delta\.\label{eq:Appendix5_OverlapStatisticsForGaussianPatterns_ABD}
\end{align}

Hence we see the link between the Gaussian statistics of the input
and the statistical properties of the overlap matrix. We use these
results to compute the tensor elements of $\mu_{\alpha\beta}$ and
$\Sigma_{(\alpha\beta)(\gamma\delta)}$ for the theory curves in the
main text that check our theory for the Gaussianized MNIST setting.

\subsection{Interpretation of different noise sources in symmetric task setting
\label{supp:Interpretation-of-different-noise-sources}}

Assuming a classification tasks with two classes $c_{1},c_{2}$ we
assume the following structure for the mean $m_{\alpha\beta}$ and
the covariance $C_{(\alpha,\beta)(\gamma\delta)}$ of the overlap
matrix

\begin{align}
m_{\alpha\beta} & =\begin{cases}
a & \alpha=\beta,\\
b & c_{\alpha}=c_{\beta},\\
-b & c_{\alpha}\neq c_{\beta},
\end{cases}\nonumber \\
C_{(\alpha\beta)(\alpha\beta)} & =K,\nonumber \\
C_{(\alpha\beta)(\alpha\delta)} & =\begin{cases}
v & c_{\alpha}=c_{\beta}=c_{\delta};c_{\alpha}\neq c_{\beta}=c_{\delta},\\
-v & c_{\alpha}=c_{\beta}\neq c_{\delta};c_{\alpha}=c_{\delta}\neq c_{\beta}.
\end{cases}\label{eq:Statistical_Properties_SimplePattern_local}
\end{align}
One can construct an ensemble of matrices $K_{\alpha\beta}^{y}$ which
have those statistical properties

\begin{align}
K_{\alpha\beta}^{y} & =b+\sqrt{\frac{K-2v}{2}}\left(\omega_{\alpha\beta}+\omega_{\beta\alpha}\right)+\sqrt{v}\left(\eta_{\alpha}+\eta_{\beta}\right)
\end{align}
for $c_{\alpha}=c_{\beta},\alpha\neq\beta$ and 
\begin{equation}
K_{\alpha\beta}^{y}=-b+\sqrt{\frac{K-2v}{2}}\left(\omega_{\alpha\beta}+\omega_{\beta\alpha}\right)-\sqrt{v}\left(\eta_{\alpha}+\eta_{\beta}\right)
\end{equation}
for $c_{\alpha}\neq c_{\beta}$. In both equations the noise components
are i.i.d drawn from a Gaussian $\eta_{i},\omega_{ij}\sim\mathcal{N}(0,1)$.
The diagonal elements are set to the constant $a$ and hence

\begin{equation}
K_{\alpha\alpha}^{y}=a\,.
\end{equation}
This construction allows for the intuitive understanding of the two
noise components: $\left(\omega_{\alpha\beta}+\omega_{\beta\alpha}\right)$
corresponds to the addition of i.i.d symmetric noise on top of the
mean-block structure, whereas $\left(\eta_{\alpha}+\eta_{\beta}\right)$
controls the presence of ``stripes'' within the overlap matrices
due to the correlation of instances $\alpha,\beta$ in the data set.
Both features can be observed on a qualitative level as well in real
data sets such as MNIST and FashionMNIST. Even though the technique
is powerful one needs to keep in mind that one is not able to create
kernels with arbitrary statistics for $K_{1}\ldots-\overline{K}_{2},v_{1}\ldots\overline{v}_{4}$
within the framework.

\subsection{Deep Non-Linear Kernel expressions}\label{supp:Supplement_NonLinearKernelExpressions}

In the main text \prettyref{subsec:Non-Linear-Kernel}, \prettyref{subsec:Multilayer-NonLinear-Kernel}
we need to propagate the heterogeneity in the input kernel matrix
$K_{\alpha\beta}^{0}$ through the network to the output kernel $K_{\alpha\beta}^{y}$.
We use the setup as in the main text \prettyref{eq:Deep-FFN-Architecture}
and assume a fully connected feed forward neural network with the
activation function $\phi(x)$ for which the NNGP kernel obeys the
set of iterative equations

\begin{align}
K_{\alpha\beta}^{0} & \sim\mathcal{N}(\mu,\Sigma)\\
K_{\alpha\beta}^{l} & =\sigma_{w}^{2}\left\langle \phi\left(h_{\alpha}^{l-1}\right)\phi\left(h_{\beta}^{l-1}\right)\right\rangle _{\left(h_{\alpha}^{l-1},h_{\beta}^{l-1}\right)\sim\mathcal{N}\left(0,K_{\alpha\beta}^{l-1}\right)}\quad\mathrm{for\quad}l=1...L\\
K_{\alpha\beta}^{y} & =\sigma_{u}^{2}\left\langle \phi\left(h_{\alpha}^{L}\right)\phi\left(h_{\beta}^{L}\right)\right\rangle _{\left(h_{\alpha}^{L},h_{\beta}^{L}\right)\sim\mathcal{N}\left(0,K_{\alpha\beta}^{L}\right)}\label{eq:Appendix5_KernelPropagation}
\end{align}
We make the assumption that the kernel matrix in the $l$-th layer
can be decomposed into a mean $m^{l}$ and a noise $\eta^{l}$ with
$K^{l}=m^{l}+\eta^{l}$ and $\eta^{l}\sim\mathcal{N}\left(0,C^{l}\right)$.
With the assumption that the noise in each layer obeys $\eta^{l}\ll m^{l}$
and that only the off-diagonal elements of the kernel fluctuate 
it is sufficient to perform a Taylor expansion in each step for $\alpha\neq\beta$

\begin{align}
K_{\alpha\beta}^{l} & =\sigma_{w}^{2}\left\langle \phi\left(h_{\alpha}^{l-1}\right)\phi\left(h_{\beta}^{l-1}\right)\right\rangle _{\left(h_{\alpha}^{l-1},h_{\beta}^{l-1}\right)\sim\mathcal{N}\left(0,K_{\alpha\beta}^{l-1}\right)}=K_{\alpha\beta}^{l}\left[K_{\alpha\beta}^{l-1}\right]\,,\nonumber \\
K_{\alpha\beta}^{l}\left[K_{\alpha\beta}^{l-1}\right]=K_{\alpha\beta}^{l}\left[m_{\alpha\beta}^{l-1}+\eta_{\alpha\beta}^{l-1}\right] & \approx K_{\alpha\beta}^{l}\left[m_{\alpha\beta}^{l-1}\right]+\frac{\partial K_{\alpha\beta}^{l}}{\partial K_{\alpha\beta}^{l-1}}\bigg\rvert_{K_{\alpha\beta}^{l-1}=m_{\alpha\beta}^{l-1}}\eta_{\alpha\beta}^{l-1}\,,\\
K_{\alpha\beta}^{l} & \approx\sigma_{w}^{2}\left\langle \phi_{\alpha}^{l-1}\phi_{\beta}^{l-1}\right\rangle _{m_{\alpha\beta}^{l-1}}\label{eq:Appendix5_PriceTheoremExpressionLinearApprox}\\
 & +\sigma_{w}^{2}\frac{\partial\left\langle \phi_{\alpha}^{l-1}\phi_{\beta}^{l-1}\right\rangle _{m_{\alpha\beta}^{l-1}}}{\partial K_{\alpha\beta}^{l-1}}\bigg\rvert_{K_{\alpha\beta}^{l-1}=m_{\alpha\beta}^{l-1}}\eta_{\alpha\beta}^{l-1}\,,\\
K_{\alpha\beta}^{l} & \approx\sigma_{w}^{2}\left\langle \phi_{\alpha}^{l-1}\phi_{\beta}^{l-1}\right\rangle _{m_{\alpha\beta}^{l-1}}\label{eq:Appendix5_FExpressionLinearPerturbedKernel}\\
 & +\sigma_{w}^{2}\left\langle \left(\phi^{\prime}\right)_{\alpha}^{l-1}\left(\phi^{\prime}\right)_{\beta}^{l-1}\right\rangle _{m_{\alpha\beta}^{l-1}}\eta_{\alpha\beta}^{l-1}\,,
\end{align}
where we used Price's theorem \citep{Price58_69} from \eqref{eq:Appendix5_PriceTheoremExpressionLinearApprox}
to \eqref{eq:Appendix5_FExpressionLinearPerturbedKernel} (see also
Appendix A in \citep{Schuecker16b_arxiv}) and where we introduced
the shorthand

\begin{equation}
\left\langle f\left(h_{\alpha}^{l-1}\right)g\left(h_{\beta}^{l-1}\right)\right\rangle _{\left(f_{\alpha}^{l-1},g_{\beta}^{l-1}\right)\sim\mathcal{N}\left(0,m_{\alpha\beta}^{l-1}\right)}=\left\langle f_{\alpha}^{l-1}g_{\beta}^{l-1}\right\rangle _{m_{\alpha\beta}^{l-1}}\label{eq:Appendix5_DefinitionShorthand}
\end{equation}
Hence the mean $m_{\alpha\beta}^{l}$ transforms as in the NNGP case
and the noise $\eta_{\alpha\beta}^{l}$, to linear order, as

\begin{align}
m_{\alpha\beta}^{l} & =\sigma_{w}^{2}\left\langle \phi_{\alpha}^{l-1}\phi_{\beta}^{l-1}\right\rangle _{m_{\alpha\beta}^{l-1}}\\
\eta_{\alpha\beta}^{l} & =\sigma_{w}^{2}\left\langle \left(\phi^{\prime}\right)_{\alpha}^{l-1}\left(\phi^{\prime}\right)_{\beta}^{l-1}\right\rangle _{m_{\alpha\beta}^{l-1}}\eta_{\alpha\beta}^{l-1}
\end{align}
Propagating this through the architecture in \eqref{eq:Appendix5_KernelPropagation}
yields the linear response approximation for the mean and covariance
in the final layer

\begin{align}
m_{\alpha\beta} & =\left\langle K_{\alpha\beta}^{y}\right\rangle =\sigma_{u}^{2}\left\langle \phi_{\alpha}^{L}\phi_{\beta}^{L}\right\rangle _{m_{\alpha\beta}^{L}}+\delta_{\alpha\beta}\sigma_{\mathrm{reg}}^{2}\,,\,m_{\alpha\beta}^{0}=\mu\,,\,m_{\alpha\beta}^{l}=\sigma_{w}^{2}\left\langle \phi_{\alpha}^{l-1}\phi_{\beta}^{l-1}\right\rangle _{m_{\alpha\beta}^{l-1}}\,\mathrm{with\:}l=1...L\\
C_{(\alpha\beta)(\gamma\delta)} & =\left\langle \left(K_{\alpha\beta}^{y}-m_{\alpha\beta}\right)\left(K_{\gamma\delta}^{y}-m_{\gamma\delta}\right)\right\rangle =\sigma_{u}^{4}\prod_{l=1}^{L+1}\left[\sigma_{w}^{4}\left\langle \left(\phi^{\prime}\right)_{\alpha}^{l-1}\left(\phi^{\prime}\right)_{\beta}^{l-1}\right\rangle _{m_{\alpha\beta}^{l-1}}\left\langle \left(\phi^{\prime}\right)_{\gamma}^{l-1}\left(\phi^{\prime}\right)_{\delta}^{l-1}\right\rangle _{m_{\gamma\delta}^{l-1}}\right]\underbrace{\left\langle \eta_{\alpha\beta}^{0}\eta_{\gamma\delta}^{0}\right\rangle _{\Sigma}}_{\Sigma_{(\alpha\beta)(\gamma\delta)}}.
\end{align}
\subsection{Matrix elements of inverse block matrices\label{supp:AppendixElementsInverseBlockMatrix}}

As we assumed block-structure in the main text, computing the propagator
$m_{\alpha\beta}^{-1}$ requires the computation of the inverse of
such block matrices. We assume the block structure in our matrix $m\in\mathbb{R}^{D\times D}$
with the $D=D_{1}+D_{2}$ training samples

\begin{align}
m & =\left(\begin{array}{ccccccccc}
a & b & \dots & b & \vline & c & \dots & \dots & c\\
b & a & \ddots & \vdots & \vline & \vdots &  &  & \vdots\\
\vdots & \ddots & \ddots & b & \vline & \vdots &  &  & \vdots\\
b & \dots & b & a & \vline & c & \dots & \dots & c\\
\hline c & \dots & \dots & c & \vline & a & d & \dots & d\\
\vdots &  &  & \vdots & \vline & d & a & \ddots & \vdots\\
\vdots &  &  & \vdots & \vline & \vdots & \ddots & \ddots & d\\
c & \dots & \dots & c & \vline & d & \dots & d & a
\end{array}\right)\label{eq:Appendix_Def_MeanBlockMatrix}
\end{align}
 Similarly we assume block structure for the inverse matrix $m^{-1}\in\mathbb{R}^{D\times D}$
which will be our propagator

\begin{equation}
m^{-1}=\left(\begin{array}{ccccccccc}
q_{1} & q_{2} & \dots & q_{2} & \vline & r & \dots & \dots & r\\
q_{2} & q_{1} & \ddots & \vdots & \vline & \vdots &  &  & \vdots\\
\vdots & \ddots & \ddots & q_{2} & \vline & \vdots &  &  & \vdots\\
q_{2} & \dots & q_{2} & q_{1} & \vline & r & \dots & \dots & r\\
\hline r & \dots & \dots & r & \vline & q_{1}^{\prime} & q_{2}^{\prime} & \dots & q_{2}^{\prime}\\
\vdots &  &  & \vdots & \vline & q_{2}^{\prime} & q_{1}^{\prime} & \ddots & \vdots\\
\vdots &  &  & \vdots & \vline & \vdots & \ddots & \ddots & q_{2}^{\prime}\\
r & \dots & \dots & r & \vline & q_{2}^{\prime} & \dots & q_{2}^{\prime} & q_{1}^{\prime}
\end{array}\right).\label{eq:Appendix_DefMeanInverseMatrix}
\end{equation}
We assume that the blocks have the respective sizes $D_{1}\times D_{1},D_{1}\times D_{2},D_{2}\times D_{1}$
and $D_{2}\times D_{2}$, where $D_{1}$ denotes the inputs of class
$c_{1}$ in the data set and $D_{2}$ denotes the inputs of class
$c_{2}$. As the matrices are inverses of each other we can construct
conditions which have to hold for the relation between the matrix
elements $a,b,c,d$ and the matrix elements of the inverse $q_{1},q_{2},q_{1}^{\prime},q_{2}^{\prime},r$.
In particular the matrices need to fulfill $m^{-1}m=\mathbb{I}$.
From this we can derive the conditions to obtain the matrix elements
of the propagator $m^{-1}:$

\begin{align}
q_{1}a+(D_{1}-1)q_{2}b+D_{2}rc & =1\,,\\
q_{1}b+q_{2}a+(D_{2}-1)q_{2}b+D_{2}rc & =0\,,\\
q_{1}c+(D_{1}-1)q_{2}c+ar+(D_{2}-1)rd & =0\,,\\
q_{1}^{\prime}a+(D_{2}-1)q_{2}^{\prime}d+D_{1}cr & =1\,,\\
q_{1}^{\prime}d+aq_{2}^{\prime}+(D_{2}-2)q_{2}^{\prime}d+D_{1}cr & =0\,,\\
ar+(D_{1}-1)br+q_{1}^{\prime}c+q_{2}^{\prime}c(D_{2}-1) & =0\,.\label{eq:Appendix_Equations_BlockMatrixInversion}
\end{align}
By a straightforward calculation one can obtain the following relations
for the elements of the propagator yield

\selectlanguage{english}%
\begin{eqnarray}
\gamma_{b}^{1}:=(a+b(D_{1}-1) & \qquad & \gamma_{d}^{2}:=(a+d(D_{2}-1))\,,\\
\tilde{\lambda}:=(\gamma_{b}^{1}\gamma_{d}^{2}-D_{1}D_{2}c^{2}) & \qquad & r=-\frac{c}{\tilde{\lambda}}\,,\\
q_{2}=\frac{c^{2}D_{2}-b\gamma_{2}^{d}}{(a-b)\tilde{\lambda}} & \qquad & q_{2}^{\prime}=\frac{c^{2}D_{1}-d\gamma_{1}^{b}}{(a-d)\tilde{\lambda}}\,,\\
q_{1}=\frac{1}{a-b}+q_{2} & \qquad & q_{1}^{\prime}=\frac{1}{a-d}+q_{2}^{\prime}\,.\label{eq:Matrix_Elements_InverseBlockMatrix}
\end{eqnarray}
Considering a symmetric task with \textbf{$b=d,c=-b$} results in
the elements

\begin{align}
\gamma_{b}^{1}:=(a+b(D_{1}-1) & \qquad\gamma_{d}^{2}:=(a+b(D_{2}-1))\,,\\
\tilde{\lambda}:=(a-b)^{2}+b(a-b)D & \qquad r=\frac{b}{\tilde{\lambda}}\,,\\
q_{2}=\frac{-b}{\tilde{\lambda}} & \qquad q_{2}^{\prime}=q_{2}\,,\\
q_{1}=\frac{1}{a-b}+q_{2} & \qquad q_{1}^{\prime}=q_{1}\,.\label{eq:Matrix_Elements_InverseBlock_SimplePattern}
\end{align}
This is also the case for the task setting considered in \foreignlanguage{american}{\prettyref{sec:The-Ising-data-set}}
of the main text. In a similar fashion as for the propagator $m^{-1}$
we assume block structure in the training label vector $y$ as

\begin{equation}
y=\left(\begin{array}{c}
y_{1}\\
\vdots\\
y_{1}\\
y_{2}\\
\vdots\\
y_{2}
\end{array}\right)\rightarrow y_{\alpha}=\begin{cases}
y_{1} & c(\alpha)=c_{1}\\
y_{2} & c(\alpha)=c_{2}
\end{cases}\,.\label{eq:Appendix_DefBlockTrainLabels}
\end{equation}
For computational convenience we always choose $y_{1}=-1,y_{2}=1$.
In addition to the mean training data overlap $m_{\alpha\beta}$ in
\ref{eq:Appendix_Def_MeanBlockMatrix} we need to define the test-train
overlap $m_{*,\beta}$. As we assume i.i.d. distributed data-samples
in the training and test set, the intra- and inter-class statistics
are the same. The difference between $m_{*\beta}$ and $m_{\alpha\beta}$
is that $*\neq\beta,\quad\forall\beta$. Hence we do not get diagonal
elements $a$. The matrix $m_{*\beta}\in\mathbb{R}^{D_{\mathrm{test}}\times D}$
therefore reads

\begin{equation}
m_{*,\beta}=\left(\begin{array}{cccccc}
b & \hdots & b & c & \hdots & c\\
\vdots & \ddots & \vdots & \vdots & \ddots & \vdots\\
b & \hdots & b & c & \hdots & c\\
c & \hdots & c & d & \hdots & d\\
\vdots & \ddots & \vdots & \vdots & \ddots & \vdots\\
c & \hdots & c & d & \hdots & d
\end{array}\right)\,.\label{eq:Appendix_DefMeanTrainTestOverlap}
\end{equation}

\subsubsection{Elements of the $g$ matrix}

We defined the $g$-matrix in the calculations above as it frequently
appears in the calculations

\begin{equation}
g_{*\alpha}=m_{*,\beta}m_{\beta,\alpha}^{-1}\,.\label{eq:Appendix_GTensor_Definition}
\end{equation}
We will now derive its entries. We use the block structure in $m$
and $m^{-1}$ given by \eqref{eq:Appendix_Def_MeanBlockMatrix} and
\eqref{eq:Appendix_DefMeanInverseMatrix} andthe matrix elements of
inverse given by \foreignlanguage{american}{\eqref{eq:Matrix_Elements_InverseBlockMatrix}.
The }mean of the test-train kernel matrix $m_{*,\beta}$ is given
by \foreignlanguage{american}{\eqref{eq:Appendix_DefMeanTrainTestOverlap}.
}From its definition \foreignlanguage{american}{\eqref{eq:Appendix_GTensor_Definition}}
the matrix elements of the $g$-matrix read

\begin{equation}
g_{*1}=\begin{cases}
bq_{1}+(D_{1}-1)q_{2}b+D_{2}rc & c(\alpha=1),c(*)=1\\
cq_{1}+(D_{1}-1)q_{2}c+D_{2}rd & c(\alpha=1),c(*)=2
\end{cases},\qquad g_{*2}=\begin{cases}
brD_{1}+q_{1}^{\prime}c+q_{2}^{\prime}(D_{1}-1)c & c(\alpha=2),c(*)=1\\
crD_{1}+q_{1}^{\prime}d+q_{2}^{\prime}(D_{2}-1)d & c(\alpha=2),c(*)=2
\end{cases}\,.
\end{equation}
\\
Inserting the expressions for the matrix elements of the inverse kernel
we get

\begin{align}
g(c(\alpha) & =1,c(*)=1)=\frac{b\gamma_{2}^{d}-D_{2}c^{2}}{\tilde{\lambda}}=\frac{b(a-d)+D_{2}(bd-c^{2})}{\gamma_{2}^{d}\gamma_{1}^{b}-D_{1}D_{2}c^{2}}\sim\mathcal{O}\left(\frac{1}{D_{1}}\right)\,,\\
g(c(\alpha) & =1,c(*)=2)=\frac{c\gamma_{2}^{d}-D_{2}dc}{\tilde{\lambda}}=\frac{c(a-d)}{\tilde{\lambda}}\sim\mathcal{O}\left(\frac{1}{D_{1}D_{2}}\right)\,,\\
g(c(\alpha) & =2,c(*)=1)=\frac{c\gamma_{1}^{b}-D_{1}bc}{\tilde{\lambda}}=\frac{c(a-b)}{\tilde{\lambda}}\sim\mathcal{O}\left(\frac{1}{n_{1}n_{2}}\right)\,,\\
g(c(\alpha) & =2,c(*)=2)=\frac{d\gamma_{1}^{b}-D_{1}c^{2}}{\tilde{\lambda}}=\frac{d(a-b)+D_{1}(bd-c^{2})}{\gamma_{1}^{b}\gamma_{2}^{d}-D_{1}D_{2}c^{2}}\sim\mathcal{O}\left(\frac{1}{n_{2}}\right)\,.
\end{align}

\subsubsection{Elements of $\hat{y}$}

\selectlanguage{american}%
In subsequent expressions we will also encounter expressions for $\hat{y}$

\begin{equation}
\hat{y}_{\alpha}:=m_{\alpha\beta}^{-1}y_{\beta}\,.\label{eq:hat_y_def}
\end{equation}
We assume block structure in the vector of training labels $y$ which
leads to a block structure of $\hat{y}$ as

\selectlanguage{english}%
\begin{equation}
y:=\left(\begin{array}{c}
y_{1}\\
\vdots\\
y_{1}\\
y_{2}\\
\vdots\\
y_{2}
\end{array}\right)\quad\hat{y}=\left(\begin{array}{c}
\hat{y}_{1}\\
\vdots\\
\vdots\\
\hat{y}_{1}\\
\hat{y}_{2}\\
\vdots\\
\vdots\\
\hat{y}_{2}
\end{array}\right).
\end{equation}
As the choice of labels is arbitrary we set $y_{2}:=-y_{1}$ for computational
convenience. For $\hat{y}$ we therefore obtain from \eqref{eq:hat_y_def}
the values for $\hat{y}_{1},\hat{y}_{2}$ as

\begin{align}
\hat{y}_{1} & =y_{1}\left(q_{1}+(D_{1}-1)q_{2}-rD_{2}\right)=y_{1}\frac{\gamma_{2}^{d}+D_{2}c}{\tilde{\lambda}}=y_{1}\frac{(a-d)+(d+c)D_{2}}{\gamma_{1}^{b}\gamma_{2}^{d}-D_{1}D_{2}c^{2}}\sim\mathcal{O}\left(\frac{1}{D_{1}}\right)\,,\\
\hat{y}_{2} & =-y_{1}\left(q_{1}^{\prime}+(D_{2}-1)q_{2}^{\prime}-rD_{1}\right)=-y_{1}\frac{\gamma_{1}^{b}+D_{1}c}{\tilde{\lambda}}=-y_{1}\frac{(a-b)+(b+c)D_{1}}{\gamma_{1}^{b}\gamma_{2}^{d}-D_{1}D_{2}c^{2}}\sim\mathcal{O}\left(\frac{1}{D_{2}}\right)\,.
\end{align}

\selectlanguage{american}%

\subsection{General expressions for the inference formula in asymmetric block
structured task settings\label{supp:Asymmetric-Overlap-Inference-1}}

\selectlanguage{english}%
We here treat the most general case for the statistics of overlaps
as they occur in real world data sets, such as MNIST. As in the main
text we assume that the elements of $K_{*\alpha}^{y},K_{\alpha\beta}^{y}$
are distributed according to a multivariate Gaussian distribution

\begin{equation}
K_{\alpha\beta}^{y}\sim\mathcal{N}\left(m_{\alpha\beta},C_{(\alpha\beta)(\gamma\delta)}\right)\,.
\end{equation}
where $\alpha,\beta,\gamma,\delta$ can be either train or test points.
The most general choice is to assume that the statistics only depends
on the class membership of $\alpha,\beta,\gamma,\delta$. We assume
a binary classification task with the classes $c_{1},c_{2}$. Hence
the mean of the kernel matrices read

\begin{equation}
m_{\alpha\beta}=\begin{cases}
a & \text{\ensuremath{\alpha=\beta}}\\
b & \alpha\neq\beta,c_{\alpha}=c_{\beta}=c_{1}\\
d & \alpha\neq\beta,c_{\alpha}=c_{\beta}=c_{2}\\
c & c_{\alpha}\neq c_{\beta}
\end{cases}\,.\label{eq:Mean_Blocks-1}
\end{equation}
which is a block matrix \prettyref{eq:Appendix_Def_MeanBlockMatrix}.
Both the variance $C_{(\alpha\beta)(\alpha\beta)}$ and the covariance
$C_{(\alpha\beta)(\gamma\delta)}$ inherit the block structure as
well

\begin{align}
C_{(\alpha\beta)(\alpha\beta)} & =\begin{cases}
K_{1} & c_{\alpha}=c_{\beta}=c_{1}\\
\overline{K}_{1} & c_{\alpha}=c_{\beta}=c_{2}\\
K_{2}=\overline{K}_{2} & c_{\alpha}\neq c_{\beta}
\end{cases}\,,\label{eq:Variance_Blocks-1}\\
C_{(\alpha\beta)(\alpha\delta)} & =\begin{cases}
v_{1} & c_{\alpha}=c_{1},c_{\beta}=c_{1},c_{\delta}=c_{1}\\
v_{2} & c_{\alpha}=c_{1},c_{\beta}=c_{2},c_{\delta}=c_{2}\\
v_{3} & c_{\alpha}=c_{1},c_{\beta}=c_{1},c_{\delta}=c_{2}\\
v_{4}=v_{3} & c_{\alpha}=c_{1},c_{\beta}=c_{2},c_{\delta}=c_{1}\\
\overline{v}_{1} & c_{\alpha}=c_{2},c_{\beta}=c_{2},c_{\delta}=c_{2}\\
\overline{v}_{2} & c_{\alpha}=c_{2},c_{\beta}=c_{1},c_{\delta}=c_{1}\\
\overline{v}_{3} & c_{\alpha}=c_{2},c_{\beta}=c_{2},c_{\delta}=c_{1}\\
\overline{v}_{4}=\overline{v}_{3} & c_{\alpha}=c_{2},c_{\beta}=c_{1},c_{\delta}=c_{2}
\end{cases}\,.\label{eq:Covariance_Blocks-1}
\end{align}
\foreignlanguage{american}{Further the tensor elements $C_{(\alpha\beta)(\gamma\delta)}$
for the case where all indices are different $\alpha\neq\beta\neq\gamma\neq\delta$
yields zero. This follows directly from}

\selectlanguage{american}%
\begin{align}
C_{(\alpha\beta)(\gamma\delta)} & =\left\langle \left(K_{\alpha\beta}^{y}-\left\langle K_{\gamma\delta}^{y}\right\rangle \right)\left(K_{\gamma\delta}^{y}-\left\langle K_{\gamma\delta}^{y}\right\rangle \right)\right\rangle \nonumber \\
 & =\left\langle K_{\alpha\beta}^{y}K_{\gamma\delta}^{y}\right\rangle -\left\langle K_{\alpha\beta}^{y}\right\rangle \left\langle K_{\gamma\delta}^{y}\right\rangle \,,\\
\mathrm{i.i.d\,samples\,}\alpha,\beta,\gamma,\delta: & =\left\langle K_{\alpha\beta}^{y}\right\rangle \left\langle K_{\gamma\delta}^{y}\right\rangle -\left\langle K_{\alpha\beta}^{y}\right\rangle \left\langle K_{\gamma\delta}^{y}\right\rangle =0\,.
\end{align}
\foreignlanguage{english}{Hence the tensor $C_{(\alpha\beta)(\gamma\delta)}$
is sparse by construction due to the assumption of independent and
identically distributed training data samples. One additional assumption
we make in the main text is that the diagonal elements of the kernel
matrix $K_{\alpha\alpha}^{y}$ are deterministic $K_{\alpha\alpha}^{y}=\left\langle K_{\alpha\alpha}^{y}\right\rangle =a$.
From this follows directly}

\begin{align}
C_{(\alpha\alpha)(\beta\gamma)} & =\left\langle \left(K_{\alpha\alpha}^{y}-\left\langle K_{\alpha\alpha}^{y}\right\rangle \right)\left(K_{\beta\gamma}^{y}-\left\langle K_{\beta\gamma}^{y}\right\rangle \right)\right\rangle \nonumber \\
 & =\left\langle \left(a-a\right)\left(K_{\beta\gamma}^{y}-\left\langle K_{\beta\gamma}^{y}\right\rangle \right)\right\rangle =0\,,\\
\rightarrow & C_{(\alpha\alpha)(\alpha\beta)}=C_{(\alpha\alpha)(\alpha\alpha)}=0\,.
\end{align}
\foreignlanguage{english}{As $m_{\alpha\beta}^{-1}$ corresponds to
the propagator in the expressions in our main text, we state the matrix
elements of the inverse of a bipartite block structured matrix \prettyref{eq:Mean_Blocks-1}
in} \prettyref{supp:AppendixElementsInverseBlockMatrix}\foreignlanguage{english}{.Based
on the action }\eqref{eq:Appendix_FullDisorderAveragedAction}\foreignlanguage{english}{
we now want to compute the corrections at first order to the mean-inferred
network output which is given by}

\selectlanguage{english}%
\begin{equation}
\langle y_{*}\rangle_{0+1}=g_{*\alpha}y_{\alpha}+g_{*\alpha}C_{(\alpha\beta)(\gamma\delta)}m_{\beta\gamma}^{-1}m_{\delta\delta^{\prime}}^{-1}y_{\delta^{\prime}}-C_{(*\alpha)(\beta\gamma)}m_{\alpha\beta}^{-1}m_{\gamma\gamma^{\prime}}^{-1}y_{\gamma^{\prime}}\,.\label{eq:Appendix_FullExpressionMeanInferredNetworkOutput-1}
\end{equation}
\foreignlanguage{american}{This can be rewritten as}

\begin{align}
\langle y^{*}\rangle_{0+1} & =g_{*\alpha}y_{\alpha}+V_{4}(*)-V_{3}(*)\,,\\
V_{3}(*) & =\sum_{\alpha,\beta,\gamma}C_{(*\alpha)(\beta\gamma)}m_{\alpha\beta}^{-1}\hat{y}_{\gamma}\,,\\
V_{4}(*) & =\sum_{\alpha,\beta,\gamma}g_{*\alpha}C_{(\alpha\beta)(\gamma\delta)}m_{\beta\gamma}^{-1}\hat{y}_{\delta}\,,\\
g_{*,\alpha}= & m_{*\alpha^{\prime}}m_{\alpha^{\prime}\alpha}^{-1}\,,\\
\hat{y}_{\gamma} & =m_{\gamma\gamma^{\prime}}^{-1}y_{\gamma^{\prime}}\,,
\end{align}
where we introduced the shorthand notation $g_{*\alpha},\hat{y},V_{3}(*),V_{4}(*)$
to simplify subsequent calculations in this part of the appendix.
Evaluating the expression \eqref{eq:Appendix_FullExpressionMeanInferredNetworkOutput-1}
is numerically expensive, because it requires computing contractions
over $D$ training examples in up to five indices and hence scales
as $\mathcal{O}(D^{5})$. However, we know from the definition of
our problem setting that the tensors $C_{(\alpha\beta)(\gamma\delta)},C_{(*\alpha)(\beta\gamma)}$
are sparse and have block structure. We can therefore simplify the
problem and compute the contractions analytically. We will do so by
computing $V_{3}(*)$ and $V_{4}(*)$ individually and combine the
results later. It is important to note that, even though the approach
below is general, it is practically restricted to a binary classification
problem. The reason for this is that in our calculation we exploit
the block structure in the tensors by splitting the contractions over
the indices $\alpha,\beta...$ into two parts, assuming that the data
is presented in an ordered fashion

\begin{equation}
\sum_{\alpha}=\sum_{c_{\alpha}=c_{1}}+\sum_{c_{\beta}=c_{2}}\,.
\end{equation}
\foreignlanguage{american}{In principle, an extension to more classes
is possible in an analogous way, but it requires the inversion of
block matrices with more than two blocks. In the case of $V_{3}(*)$
the contractions over the three indices $\alpha,\beta,\gamma$ hence
produces eight terms}

\begin{align}
\left(\sum_{\alpha,c(\alpha)=1}+\sum_{\beta,c(\alpha)=2}\right)\left(\sum_{\beta,c(\beta)=1}+\sum_{\beta,c(\beta)=2}\right)\left(\sum_{\gamma,c(\gamma)=1}+\sum_{\gamma,c(\gamma)=2}\right) & =\sum_{c(\alpha)=c(\beta)=c(\gamma)=1}\nonumber \\
 & +\sum_{c(\alpha)=c(\beta)=1,c(\gamma)=2}+....\sum_{c(\alpha)=c(\beta)=c(\gamma)=2}
\end{align}
whereas $V_{4}(*)$ requires the evaluation of 16 individual terms.
As the number of terms grows exponentially with the number of classes
we restrict ourselves to binary classification. In the remainder of
this section we denote $D_{1}$ as the number of training samples
in class 1 and $D_{2}$ as the number of training samples in class
2.

\subsubsection{Three Point term $V_{3}(*)$}

We evaluate $V_{3}(*)$ by splitting the contractions in eight terms.
However, due to the structure of $C_{(*\alpha)(\beta,\gamma)}$ two
of those terms vanish\foreignlanguage{american}{. The reason for this
is that these terms contain tensor elements $C_{(*\cdot)(\cdot,\cdot)}$
which vanish by construction. Take}

\selectlanguage{american}%
\begin{equation}
\sum_{c_{\alpha}=1,c_{\beta}=2,c_{\gamma}=2}C_{(*,\alpha)(\beta,\gamma)}m_{\alpha,\beta}^{-1}\hat{y}_{\gamma}
\end{equation}
for example. We know, that $C_{(*\alpha)(\beta\gamma)}=0$ if neither
$\alpha=\beta$ nor $\alpha=\gamma$ and because the test point $*\notin\{\alpha,\beta,\gamma\}$.
However, as $c_{\alpha}=1$ and both $c_{\beta}=c_{\gamma}=2$ and
the set of the two patterns are distinct, there is no index combination
of $\alpha\beta\gamma$ which yields $C_{(*\alpha)(\beta\gamma)}\neq0$.
Hence this term vanishes. Due to symmet\foreignlanguage{english}{ry
the equivalent term with $c_{\alpha}=2,c_{\beta}=c_{\gamma}=1$ vanishes
as well.\newline}

\selectlanguage{english}%

\paragraph{\noindent Case 1: $c_{\alpha}=1,c_{\beta}=1,c_{\gamma}=1$\newline}

\selectlanguage{american}%
\noindent \newline For the first term we will spell out the calculation
steps explicitly. The calculations for the subsequent terms follow
along similar lines. We start by introducing the notation

\begin{equation}
C_{(\alpha\beta)(\gamma\delta)}^{c_{\alpha},c_{\beta},c_{\gamma},c_{\delta}}\,\mathrm{with}\,c_{\alpha},c_{\beta},c_{\gamma},c_{\delta}\in\{1,2\}\,.
\end{equation}
From \prettyref{supp:AppendixElementsInverseBlockMatrix}
we know that the expressions $\hat{y}_{\gamma}$ have block structure
as well and $\hat{y}_{\gamma}=\hat{y}_{1}\forall\gamma\,\mathrm{with\,}c_{\gamma}=1$.
To evaluate the first sum

\selectlanguage{english}%
\begin{equation}
\sum_{\alpha,\beta,\gamma}^{c_{\alpha}=c_{\beta}=c_{\gamma}=c_{1}}C_{(*,\alpha)(\beta,\gamma)}m_{\alpha,\beta}^{-1}\hat{y}_{\gamma}\,.
\end{equation}
We can start by recognizing that the expression $C_{(*\alpha)(\beta\gamma)}$
is only unequal to zero if either $\alpha=\beta$ or $\alpha=\gamma$.
Hence we can split the sum

\begin{equation}
\sum_{\alpha,\beta,\gamma}^{c_{\alpha}=c_{\beta}=c_{\gamma}=c_{1}}C_{(*,\alpha)(\beta,\gamma)}m_{\alpha,\beta}^{-1}\hat{y}_{\gamma}=\sum_{\alpha=\beta,\gamma}^{c_{\alpha}=c_{\beta}=c_{\gamma}=c_{1}}C_{(*,\alpha)(\alpha,\gamma)}m_{\alpha,\alpha}^{-1}\hat{y}_{\gamma}+\sum_{\alpha=\gamma,\beta}^{c_{\alpha}=c_{\beta}=c_{\gamma}=c_{1}}C_{(*,\alpha)(\beta,\alpha)}m_{\alpha,\beta}^{-1}\hat{y}_{\gamma}\,.
\end{equation}
From this, and exploiting the block structure in $C_{(*\alpha)(\beta\gamma)},\hat{y}_{\gamma},m_{\alpha\beta}^{-1}$
we get 

\begin{align}
\sum_{\alpha,\beta,\gamma}^{c_{\alpha}=c_{\beta}=c_{\gamma}=c_{1}}C_{(*,\alpha)(\beta,\gamma)}m_{\alpha,\beta}^{-1}\hat{y}_{\gamma} & =\sum_{\alpha=\beta,\gamma}^{c_{\alpha}=c_{\beta}=c_{\gamma}=c_{1}}C_{(*,\alpha)(\alpha,\gamma)}q_{1}\hat{y}_{\gamma}\nonumber \\
 & +\sum_{\alpha=\gamma,\beta}^{c_{\alpha}=c_{\beta}=c_{\gamma}=c_{1}}C_{(*,\alpha)(\beta,\alpha)}q_{2}\hat{y}_{\gamma}\,,\nonumber \\
 & =C_{(*\alpha)(\alpha,\gamma)}^{*,1,1,1}D_{1}(D_{1}-1)\hat{y}_{1}\left(q_{1}+q_{2}\right)\,.
\end{align}
As the subsequent calculations are analogous we will simply state
the results for the sake of brevity if no additional considerations
need to be taken care of.
\selectlanguage{american}%

\paragraph{\newline \foreignlanguage{english}{\noindent Case 2: $c_{\alpha}=1,c_{\beta}=1,c_{\gamma}=2$\newline}}

\noindent \newline \foreignlanguage{english}{For the second term
we could perform a similar decomposition as in the first term}

\selectlanguage{english}%
\begin{equation}
\sum_{\alpha,\beta,\gamma}^{c_{\alpha}=c_{\beta}=c_{1},c_{\gamma}=c_{2}}C_{(*,\alpha)(\beta,\gamma)}m_{\alpha,\beta}^{-1}\hat{y}_{\gamma}=\sum_{\alpha=\beta,\gamma}^{c_{\alpha}=c_{\beta}=c_{1},c_{\gamma}=c_{2}}C_{(*,\alpha)(\alpha,\gamma)}m_{\alpha,\alpha}^{-1}\hat{y}_{\gamma}+\sum_{\alpha=\gamma,\beta}^{c_{\alpha}=c_{\beta}=c_{1},c_{\gamma}=c_{2}}C_{(*,\alpha)(\beta,\alpha)}m_{\alpha,\beta}^{-1}\hat{y}_{\alpha}\,.
\end{equation}
However, as we know that by construction $c_{\alpha}=1,c_{\gamma}=2$
in this particular term, the second expression where $\alpha=\gamma$
is required, can not appear. Hence the calculation for the second
term reduces to

\begin{align}
\sum_{\alpha=\beta,\gamma}^{c_{\alpha}=c_{\beta}=c_{1},c_{\gamma}=c_{2}}C_{(*,\alpha)(\alpha,\gamma)}m_{\alpha,\alpha}^{-1}\hat{y}_{\gamma} & =\sum_{\alpha=\beta,\gamma}^{c_{\alpha}=c_{\beta}=c_{1},c_{\gamma}=c_{2}}C_{(*,\alpha)(\alpha,\gamma)}q_{1}\hat{y}_{\gamma}\nonumber \\
 & =C_{(*\alpha)(\alpha,\gamma)}^{*,1,1,2}D_{1}D_{2}q_{1}\hat{y}_{2}\,.
\end{align}

\selectlanguage{american}%

\paragraph{\newline \foreignlanguage{english}{\noindent Case 3: $c_{\alpha}=1,c_{\beta}=2,c_{\gamma}=1$\newline}}

\noindent \newline \foreignlanguage{english}{We get:}

\selectlanguage{english}%
\begin{align*}
\sum_{\alpha=\gamma,\beta}^{c_{\alpha}=c_{\gamma}=c_{1},c_{\beta}=c_{2}}C_{(*,\alpha)(\beta,\alpha)}m_{\alpha,\beta}^{-1}\hat{y}_{\alpha} & =\sum_{\alpha=\gamma,\beta}^{c_{\alpha}=c_{\gamma}=c_{1},c_{\beta}=c_{2}}C_{(*,\alpha)(\beta,\alpha)}r\hat{y}_{\alpha}\\
 & =C_{(*\alpha)(\beta,\alpha)}^{*,1,2,1}D_{1}D_{2}\hat{y}_{1}r
\end{align*}

\selectlanguage{american}%

\paragraph{\newline \foreignlanguage{english}{\noindent Case 4: $c_{\alpha}=1,c_{\beta}=2,c_{\gamma}=2$\newline}}

\noindent \newline \foreignlanguage{english}{As previously mentioned
this yields 0. The reason is, that $C_{(*\alpha)(\beta\gamma)}\neq0$
requires, that $\alpha$ is either $\beta$ or $\gamma$. But if $c_{\beta}=c_{\gamma}\neq c_{\alpha}$
and because the test point $*$ is unequal to any of the training
points by construction $*\notin\{\alpha,\beta,\gamma\}$ this is simply
not possible. Hence the contribution yields}

\selectlanguage{english}%
\begin{equation}
\sum_{\alpha,\beta,\gamma}^{c_{\alpha}=c_{1},c_{\beta}=c_{\gamma}=c_{2}}C_{(*,\alpha)(\beta,\gamma)}m_{\alpha,\beta}^{-1}\hat{y}_{\gamma}=0\,.
\end{equation}

\selectlanguage{american}%

\paragraph{\newline \foreignlanguage{english}{\noindent Case 5: $c_{\alpha}=2,c_{\beta}=1,c_{\gamma}=1$\newline}}

\noindent \newline This term is also equal to zero by the same argument
as in the fourth term and yields

\begin{equation}
\sum_{\alpha,\beta,\gamma}^{c_{\alpha}=c_{2},c_{\beta}=c_{\gamma}=c_{1}}C_{(*,\alpha)(\beta,\gamma)}m_{\alpha,\beta}^{-1}\hat{y}_{\gamma}=0\,.
\end{equation}

\paragraph{\newline \foreignlanguage{english}{\noindent Case 6: $c_{\alpha}=2,c_{\beta}=1,c_{\gamma}=2$\newline}}

\noindent \newline \foreignlanguage{english}{We get:}

\selectlanguage{english}%
\begin{align}
\sum_{\alpha=\gamma,\beta}^{c_{\alpha}=c_{\gamma}=c_{2},c_{\beta}=c_{1}}C_{(*,\alpha)(\beta,\alpha)}m_{\alpha,\beta}^{-1}\hat{y}_{\alpha} & =\sum_{\alpha=\gamma,\beta}^{c_{\alpha}=c_{\gamma}=c_{2},c_{\beta}=c_{1}}C_{(*,\alpha)(\beta,\alpha)}r\hat{y}_{2}\nonumber \\
 & =C_{(*\alpha,)(\beta,\alpha)}^{*,2,1,2}D_{1}D_{2}r\hat{y}_{2}\,.
\end{align}

\selectlanguage{american}%

\paragraph{\newline \foreignlanguage{english}{\noindent Case 7: $c_{\alpha}=2,c_{\beta}=2,c_{\gamma}=1$\newline}}

\noindent \newline \foreignlanguage{english}{We get:}

\selectlanguage{english}%
\begin{align}
\sum_{\alpha=\beta,\gamma}^{c_{\alpha}=c_{\beta}=c_{2},c_{\gamma}=c_{1}}C_{(*,\alpha)(\alpha,\gamma)}m_{\alpha,\alpha}^{-1}\hat{y}_{\gamma} & =\sum_{\alpha=\beta,\gamma}^{c_{\alpha}=c_{\beta}=c_{2},c_{\gamma}=c_{1}}C_{(*,\alpha)(\alpha,\gamma)}q_{1}^{\prime}\hat{y}_{\gamma}\nonumber \\
 & =C_{(*\alpha)(\alpha,\gamma)}^{*,2,2,1}N_{1}N_{2}q_{1}^{\prime}\hat{y}_{1}\,.
\end{align}

\selectlanguage{american}%

\paragraph{\newline \foreignlanguage{english}{\noindent Case 8: $c_{\alpha}=2,c_{\beta}=2,c_{\gamma}=2$\newline}}

\noindent \newline \foreignlanguage{english}{We get:}

\selectlanguage{english}%
\begin{align}
\sum_{\alpha=\beta,\gamma}^{c_{\alpha}=c_{\beta}=c_{\gamma}=c_{2}}C_{(*,\alpha)(\alpha,\gamma)}m_{\alpha,\alpha}^{-1}\hat{y}_{\gamma}+\sum_{\alpha=\gamma,\beta}^{c_{\alpha}=c_{\beta}=c_{\gamma}=c_{2}}C_{(*,\alpha)(\beta,\alpha)}m_{\alpha,\beta}^{-1}\hat{y}_{\gamma} & =\sum_{\alpha=\beta,\gamma}^{c_{\alpha}=c_{\beta}=c_{\gamma}=c_{2}}C_{(*,\alpha)(\alpha,\gamma)}q_{1}^{\prime}\nonumber \\
 & \hat{y}_{\gamma}+\sum_{\alpha=\gamma,\beta}^{c_{\alpha}=c_{\beta}=c_{\gamma}=c_{2}}C_{(*,\alpha)(\beta,\alpha)}q_{2}^{\prime}\hat{y}_{\alpha}\nonumber \\
 & =N_{2}(N_{2}-1)q_{1}^{\prime}\hat{y}_{2}C_{(*\alpha)(\alpha,\gamma)}^{*,2,2,2}\\
 & +N_{2}(N_{2}-1)q_{2}^{\prime}\hat{y}_{2}C_{(*,\alpha)(\beta,\alpha)}^{*,2,2,2}\nonumber \\
 & =N_{2}(N_{2}-1)\hat{y}_{2}C_{(*\alpha)(\alpha,\gamma)}^{*,2,2,2}\left(q_{1}^{\prime}+q_{2}^{\prime}\right)\,.
\end{align}

\selectlanguage{american}%

\paragraph{\newline \foreignlanguage{english}{\noindent Total expression for
$V_{3}(*)$\newline}}

\noindent \newline \foreignlanguage{english}{Combining all the results
from the eight terms above we can write down the full expression for
$V_{3}(*)$ as}

\selectlanguage{english}%
\begin{align}
V_{3}(*) & =C_{(*\alpha)(\alpha,\gamma)}^{*,1,1,1}N_{1}(N_{1}-1)\hat{y}_{1}\left(q_{1}+q_{2}\right)\nonumber \\
 & +C_{(*\alpha)(\alpha,\gamma)}^{*,1,1,2}N_{1}N_{2}q_{1}\hat{y}_{2}\nonumber \\
 & +C_{(*\alpha)(\beta,\alpha)}^{*,1,2,1}N_{1}N_{2}\hat{y}_{1}r\nonumber \\
 & +C_{(*\alpha,)(\beta,\alpha)}^{*,2,1,2}N_{1}N_{2}r\hat{y}_{2}\nonumber \\
 & +C_{(*\alpha)(\alpha,\gamma)}^{*,2,2,1}N_{1}N_{2}q_{1}^{\prime}\hat{y}_{1}\nonumber \\
 & +N_{2}(N_{2}-1)\hat{y}_{2}C_{(*\alpha)(\alpha,\gamma)}^{*,2,2,2}\left(q_{1}^{\prime}+q_{2}^{\prime}\right)\,.\label{eq:Appendix_V3_Raw-1}
\end{align}
Considering the fact, that we have exchange symmetry in the tensor

\begin{equation}
C_{(*\alpha)(\beta\gamma)}=C_{(*\alpha)(\gamma\beta)}
\end{equation}
we can make the replacements

\begin{align}
C_{(*\alpha)(\alpha\beta)}^{*,1,1,2} & =C_{(*\alpha)(\beta\alpha)}^{*,1,2,1}\,,\\
C_{(*\alpha)(\beta\alpha)}^{*,2,1,2} & =C_{(*\alpha)(\alpha\beta)}^{*,2,2,1}\,,
\end{align}
and simplify the expression in \eqref{eq:Appendix_V3_Raw-1} to

\begin{align}
V_{3}(*) & =C_{(*\alpha)(\alpha,\gamma)}^{*,1,1,1}N_{1}(N_{1}-1)\left(q_{1}\hat{y}_{1}+q_{2}\hat{y}_{1}\right)\nonumber \\
 & +C_{(*\alpha)(\alpha,\gamma)}^{*,1,1,2}N_{1}N_{2}\left(q_{1}\hat{y}_{2}+\hat{y}_{1}r\right)\nonumber \\
 & +C_{(*\alpha,)(\alpha,\beta)}^{*,2,2,1}N_{1}N_{2}\left(r\hat{y}_{2}+q_{1}^{\prime}\hat{y}_{1}\right)\nonumber \\
 & +C_{(*\alpha)(\alpha,\gamma)}^{*,2,2,2}N_{2}(N_{2}-1)\left(q_{1}^{\prime}\hat{y}_{2}+q_{2}^{\prime}\hat{y}_{2}\right)\,.
\end{align}
Depending on the class membership of $*$ we obtain the following
values from the definition of $C_{(*\alpha)(\beta\gamma)}$ above
in \eqref{eq:Covariance_Blocks-1}:

\begin{align}
*\in c_{1} & \rightarrow\begin{cases}
C_{(*\alpha)(\alpha,\gamma)}^{1,1,1,1}=C_{(\alpha,\beta)(\alpha,\delta)}^{1,1,1,1}=v_{1}\\
C_{(*\alpha)(\alpha,\gamma)}^{1,1,1,2}=C_{(\alpha,\beta)(\alpha,\delta)}^{1,1,1,2}=v_{3}\\
C_{(*\alpha,)(\alpha,\beta)}^{1,2,2,1}=C_{(\alpha,\beta)(\alpha,\delta)}^{2,1,2,1}=\overline{v}_{2}\\
C_{(*\alpha)(\alpha,\gamma)}^{1,2,2,2}=C_{(\alpha,\beta)(\alpha,\delta)}^{2,2,2,1}=\overline{v}_{3}
\end{cases}\,,\\
*\in c_{2} & \rightarrow\begin{cases}
C_{(*\alpha)(\alpha,\gamma)}^{2,1,1,1}=C_{(\alpha,\beta)(\alpha,\delta)}^{1,1,1,2}=v_{3}\\
C_{(*\alpha)(\alpha,\gamma)}^{2,1,1,2}=C_{(\alpha,\beta)(\alpha,\delta)}^{1,2,1,2}=v_{2}\\
C_{(*\alpha,)(\alpha,\beta)}^{2,2,2,1}=C_{(\alpha,\beta)(\alpha,\delta)}^{2,1,2,2}=\overline{v}_{3}\\
C_{(*\alpha)(\alpha,\gamma)}^{2,2,2,2}=C_{(\alpha,\beta)(\alpha,\delta)}^{2,2,2,2}=\overline{v}_{1}
\end{cases}\,.
\end{align}
In total $V_{3}(*)$ reads:

\begin{align}
V_{3}(*\in c_{1}) & =v_{1}N_{1}(N_{1}-1)\left(q_{1}\hat{y}_{1}+q_{2}\hat{y}_{1}\right)\nonumber \\
 & +v_{3}N_{1}N_{2}\left(q_{1}\hat{y}_{2}+\hat{y}_{1}r\right)\nonumber \\
 & +\overline{v}_{2}N_{1}N_{2}\left(r\hat{y}_{2}+q_{1}^{\prime}\hat{y}_{1}\right)\nonumber \\
 & +\overline{v}_{3}N_{2}(N_{2}-1)\left(q_{1}^{\prime}\hat{y}_{2}+q_{2}^{\prime}\hat{y}_{2}\right)\,,\\
V_{3}(*\in c_{2}) & =v_{3}N_{1}(N_{1}-1)\left(q_{1}\hat{y}_{1}+q_{2}\hat{y}_{1}\right)\nonumber \\
 & +v_{2}N_{1}N_{2}\left(q_{1}\hat{y}_{2}+\hat{y}_{1}r\right)\nonumber \\
 & +\overline{v}_{3}N_{1}N_{2}\left(r\hat{y}_{2}+q_{1}^{\prime}\hat{y}_{1}\right)\nonumber \\
 & +\overline{v}_{1}N_{2}(N_{2}-1)\left(q_{1}^{\prime}\hat{y}_{2}+q_{2}^{\prime}\hat{y}_{2}\right)\,.
\end{align}

\subsubsection{Four point expression $V_{4}(*)$}

The calculations in this subsection are analogous to the calculations
for $V_{3}(*)$, the difference being that instead of eight terms
we now have four contractions over the training data indices $\alpha,\beta,\gamma,\delta$
and hence we need to consider sixteen terms.

And we deal with the sixteen terms separately.
\selectlanguage{american}%

\paragraph{\newline \foreignlanguage{english}{\noindent Term 1: $c_{\alpha}=1,c_{\beta}=1,c_{\gamma}=1,c_{\delta}=1$\newline}}

\noindent \newline For the first term we will spell out the calculation
steps explicitly. The calculations for the subsequent terms follow
along similar lines. We use the same notation as in the previous subsection
for tensor elements

\begin{equation}
C_{(\alpha\beta)(\gamma\delta)}^{c_{\alpha},c_{\beta},c_{\gamma},c_{\delta}}\,\mathrm{with\,}c_{\alpha},c_{\beta},c_{\gamma},c_{\delta}\in\{1,2\}\,.
\end{equation}
For the first term we get:

\selectlanguage{english}%
\begin{align}
\sum_{c_{\alpha}=1,c_{\beta}=1,c_{\gamma}=1,c_{\delta}=1}g_{*,\alpha}C_{(\alpha,\beta)(\gamma,\delta)}m_{\beta,\gamma}^{-1}\hat{y}_{\delta} & =g_{*1}\hat{y}_{1}D_{1}(D_{1}-1)C_{(\alpha,\beta)(\alpha,\beta)}^{1,1,1,1}(q_{1}+q_{2})\nonumber \\
 & +g_{*1}\hat{y}_{1}D_{1}(D_{1}-1)(D_{1}-2))C_{(\alpha,\beta)(\alpha,\delta)}^{1,1,1,1}\left(q_{1}+3q_{2}\right)\,.
\end{align}
The occuring terms have different origins. First we can state that
$g_{*\alpha}=g_{*1}$ and $\hat{y}_{\delta}=\hat{y}_{1}$ because
of the block structure. Next we can decompose the sum into relevant
expressions. One of them covers the cases where either $\alpha=\gamma,\beta=\delta$
and another $\alpha=\delta,\beta=\gamma$ which produce the following
sums

\begin{equation}
\sum_{\alpha=\gamma,\beta=\delta}^{c_{\alpha}=c_{\beta}=c_{1}}g_{*,\alpha}C_{(\alpha,\beta)(\alpha,\beta)}m_{\beta,\alpha}^{-1}\hat{y}_{\beta}+\sum_{\alpha=\delta,\beta=\gamma}^{c_{\alpha}=c_{\beta}=c_{1}}g_{*,\alpha}C_{(\alpha,\beta)(\beta,\alpha)}m_{\beta,\beta}^{-1}\hat{y}_{\alpha}\,.\label{eq:Appendix_V4_Term_1_Sums1-1}
\end{equation}
In both sums the tensor element yields $C_{(\alpha\beta)(\alpha\beta)}^{1,1,1,1}$
because of the exchange symmetry in the tensor $C$. Both sums simplify
to

\begin{equation}
g_{*1}C_{(\alpha\beta)(\alpha\beta)}^{1,1,1,1}\hat{y}_{1}\left(\sum_{\alpha=\gamma,\beta=\delta}^{c_{\alpha}=c_{\beta}=c_{1}}m_{\beta,\alpha}^{-1}+\sum_{\alpha=\delta,\beta=\gamma}^{c_{\alpha}=c_{\beta}=c_{1}}m_{\beta,\beta}^{-1}\right)\,.
\end{equation}
Because of the block structure in the propagator
$m_{\alpha\beta}^{-1}$ we can also replace $m_{\alpha\beta}^{-1}=q_{2}$
and $m{{}^-}_{\beta\beta}^{1}=q_{1}$ according to our definitions
in \prettyref{supp:AppendixElementsInverseBlockMatrix}. The sums evaluate to
\begin{align}
\sum_{\alpha=\gamma,\beta=\delta}^{c_{\alpha}=c_{\beta}=c_{1}}m_{\beta,\alpha}^{-1}+\sum_{\alpha=\delta,\beta=\gamma}^{c_{\alpha}=c_{\beta}=c_{1}}m_{\beta,\beta}^{-1} & =\sum_{\alpha=\gamma,\beta=\delta}^{c_{\alpha}=c_{\beta}=c_{1}}q_{2}+\sum_{\alpha=\delta,\beta=\gamma}^{c_{\alpha}=c_{\beta}=c_{1}}q_{1}\nonumber \\
 & =D_{1}\left(D_{1}-1\right)\left(q_{2}+q_{1}\right)\,.
\end{align}
Here we counted the terms in the sums in the following way: We know
that we need to enforce $\alpha\neq\beta$ because otherwise the tensor
would yield $C_{(\alpha\alpha)(\alpha\alpha)}=0$ and hence the terms
would vanish. We can choose any of the $D_{1}$ training examples
for the index $\alpha$. Because $\alpha\neq\beta$ we are left with
$D_{1}-1$ choices for $\beta$ which produces the corresponding prefactor.
In addition to the sums \eqref{eq:Appendix_V4_Term_1_Sums1-1} one
also needs to consider the sub cases that only two indices in $C_{(\alpha,\beta)(\gamma,\delta)}$
are equal. This produces the sums

\begin{align}
\sum_{c_{\alpha}=1,c_{\beta}=1,c_{\gamma}=1,c_{\delta}=1}g_{*,\alpha}C_{(\alpha,\beta)(\gamma,\delta)}m_{\beta,\gamma}^{-1}\hat{y}_{\delta} & =\sum_{c_{\alpha}=1,c_{\beta}=1,c_{\delta}=1}g_{*,\alpha}C_{(\alpha,\beta)(\beta,\delta)}m_{\beta,\beta}^{-1}\hat{y}_{\delta}\nonumber \\
 & +\sum_{c_{\alpha}=1,c_{\beta}=1,c_{\delta}=1}g_{*,\alpha}C_{(\alpha,\beta)(\alpha,\delta)}m_{\beta,\alpha}^{-1}\hat{y}_{\delta}\nonumber \\
 & +\sum_{c_{\alpha}=1,c_{\beta}=1,c_{\gamma}=1}g_{*,\alpha}C_{(\alpha,\beta)(\gamma,\alpha)}m_{\beta,\gamma}^{-1}\hat{y}_{\alpha}\nonumber \\
 & +\sum_{c_{\alpha}=1,c_{\beta}=1,c_{\gamma}=1}g_{*,\alpha}C_{(\alpha,\beta)(\gamma,\beta)}m_{\beta,\gamma}^{-1}\hat{y}_{\beta}\,.
\end{align}
As above we can extract the terms $g_{*1},C,\hat{y}_{1}$ due to the
block structure in the tensors and we can insert the explicit expressions
for the propagator elements $m_{\beta\beta}^{-1},m_{\alpha\beta}^{-1}$:

\begin{align}
\sum_{c_{\alpha}=1,c_{\beta}=1,c_{\gamma}=1,c_{\delta}=1}g_{*,\alpha}C_{(\alpha,\beta)(\gamma,\delta)}m_{\beta,\gamma}^{-1}\hat{y}_{\delta} & =g_{*1}\hat{y}_{1}C_{(\alpha\beta)(\alpha\delta)}^{1,1,1,1}\bigg(\sum_{c_{\alpha}=1,c_{\beta}=1,c_{\delta}=1}q_{1}+\sum_{c_{\alpha}=1,c_{\beta}=1,c_{\delta}=1}q_{2}\nonumber \\
 & +\sum_{c_{\alpha}=1,c_{\beta}=1,c_{\gamma}=1}q_{2}+\sum_{c_{\alpha}=1,c_{\beta}=1,c_{\gamma}=1}q_{2}\bigg)\text{\,.}
\end{align}
Again we need to consider the admissible elements for the sums: We
assume that the index $\alpha$ can take any of the $D_{1}$ values.
The index $\beta$ can hence take $D_{1}-1$ values as we require
$\alpha\neq\beta$. Likewise $\text{\ensuremath{\gamma}}$ can take
$D_{1}-2$ values as we have $\alpha\neq\gamma,\beta\neq\gamma$.
If this were not the case we would over count elements of the sum
as cases where $\gamma=\beta$ are already covered in \eqref{eq:Appendix_V4_Term_1_Sums1-1}.
Combining all of these results we get the first term

\selectlanguage{english}%
\begin{align}
\sum_{c_{\alpha}=1,c_{\beta}=1,c_{\gamma}=1,c_{\delta}=1}g_{*,\alpha}C_{(\alpha,\beta)(\gamma,\delta)}m_{\beta,\gamma}^{-1}\hat{y}_{\delta} & =g_{*1}\hat{y}_{1}D_{1}(D_{1}-1)C_{(\alpha,\beta)(\alpha,\beta)}^{1,1,1,1}(q_{1}+q_{2})\nonumber \\
 & =+g_{*1}\hat{y}_{1}D_{1}(D_{1}-1)(D_{1}-2))C_{(\alpha,\beta)(\alpha,\delta)}^{1,1,1,1}\left(q_{1}+3q_{2}\right)\,.
\end{align}

\selectlanguage{american}%

\paragraph{\newline \foreignlanguage{english}{\noindent Term 2 $c_{\alpha}=1,c_{\beta}=1,c_{\gamma}=1,c_{\delta}=2$\newline}}

\noindent \newline \foreignlanguage{english}{We get:}

\selectlanguage{english}%
\begin{align}
\sum_{c_{\alpha}=1,c_{\beta}=1,c_{\gamma}=1,c_{\delta}=2}g_{*,\alpha}C_{(\alpha,\beta)(\gamma,\delta)}m_{\beta,\gamma}^{-1}\hat{y}_{\delta} & =\sum_{\alpha,\beta,\delta}g_{*\alpha}C_{(\alpha,\beta)(\alpha,\delta)}m_{\beta,\alpha}^{-1}\hat{y}_{\delta}+\sum_{\alpha,\beta,\delta}g_{*\alpha}C_{(\alpha,\beta)(\beta,\delta)}m_{\beta,\beta}^{-1}\hat{y}_{\delta}\nonumber \\
 & =g_{*1}\hat{y}_{2}D_{1}(D_{1}-1)N_{2}(q_{2}+q_{1})C_{(\alpha,\beta)(\alpha,\delta)}^{1,1,1,2}\,.
\end{align}

\selectlanguage{american}%

\paragraph{\newline \foreignlanguage{english}{\noindent Term 3 $c_{\alpha}=1,c_{\beta}=1,c_{\gamma}=2,c_{\delta}=1$\newline}}

\noindent \newline \foreignlanguage{english}{We get:}

\selectlanguage{english}%
\begin{align}
\sum_{c_{\alpha}=1,c_{\beta}=1,c_{\gamma}=2,c_{\delta}=1}g_{*,\alpha}C_{(\alpha,\beta)(\gamma,\delta)}m_{\beta,\gamma}^{-1}\hat{y}_{\delta} & =\sum_{\alpha,\beta,\delta}g_{*\alpha}C_{(\alpha,\beta)(\gamma,\alpha)}m_{\beta,\gamma}^{-1}\hat{y}_{\alpha}+\sum_{\alpha,\beta,\delta}g_{*\alpha}C_{(\alpha,\beta)(\gamma,\beta)}m_{\beta,\gamma}^{-1}\hat{y}_{\beta}\nonumber \\
 & =g_{*1}\hat{y}_{1}D_{1}(D_{1}-1)D_{2}(r+r)C_{(\alpha,\beta)(\gamma,\alpha)}^{1,1,2,1}\,.
\end{align}

\selectlanguage{american}%

\paragraph{\newline \foreignlanguage{english}{\noindent Term 4 $c_{\alpha}=1,c_{\beta}=1,c_{\gamma}=2,c_{\delta}=2$\newline}}

\noindent \newline \foreignlanguage{english}{We get:}

\selectlanguage{english}%
\begin{align}
\sum_{c_{\alpha}=1,c_{\beta}=1,c_{\gamma}=2,c_{\delta}=2}g_{*,\alpha}C_{(\alpha,\beta)(\gamma,\delta)}m_{\beta,\gamma}^{-1}\hat{y}_{\delta} & =0
\end{align}

\selectlanguage{american}%

\paragraph{\newline \foreignlanguage{english}{\noindent Term 5 $c_{\alpha}=1,c_{\beta}=2,c_{\gamma}=1,c_{\delta}=1$\newline}}

\noindent \newline \foreignlanguage{english}{We get:}

\selectlanguage{english}%
\begin{align}
\sum_{c_{\alpha}=1,c_{\beta}=2,c_{\gamma}=1,c_{\delta}=1}g_{*,\alpha}C_{(\alpha,\beta)(\gamma,\delta)}m_{\beta,\gamma}^{-1}\hat{y}_{\delta} & =\sum_{c_{\alpha}=1,c_{\beta}=2,c_{\delta}=1}g_{*,\alpha}C_{(\alpha,\beta)(\alpha,\delta)}m_{\beta,\alpha}^{-1}\hat{y}_{\delta}\nonumber \\
 & +\sum_{c_{\alpha}=1,c_{\beta}=2,c_{\gamma}=1}g_{*,\alpha}C_{(\alpha,\beta)(\gamma,\alpha)}m_{\beta,\gamma}^{-1}\hat{y}_{\alpha}\\
 & =g_{*1}\hat{y}_{1}D_{1}D_{2}(D_{1}-1)(r+r)C_{(\alpha,\beta)(\alpha,\delta)}^{1,2,1,1}\,.
\end{align}

\selectlanguage{american}%

\paragraph{\newline \foreignlanguage{english}{\noindent Term 6 $c_{\alpha}=1,c_{\beta}=2,c_{\gamma}=1,c_{\delta}=2$\newline}}

\noindent \newline \foreignlanguage{english}{We get:}

\selectlanguage{english}%
\begin{align}
\sum_{c_{\alpha}=1,c_{\beta}=2,c_{\gamma}=1,c_{\delta}=2}g_{*,\alpha}C_{(\alpha,\beta)(\gamma,\delta)}m_{\beta,\gamma}^{-1}\hat{y}_{\delta} & =\sum_{c_{\alpha}=1,c_{\beta}=2}g_{*,\alpha}C_{(\alpha,\beta)(\alpha,\beta)}m_{\beta,\alpha}^{-1}\hat{y}_{\beta}\nonumber \\
 & +\sum_{c_{\alpha}=1,c_{\beta}=2,c_{\delta}=2,\beta\neq\delta}g_{*,\alpha}C_{(\alpha,\beta)(\alpha,\delta)}m_{\beta,\alpha}^{-1}\hat{y}_{\delta}\\
 & +\sum_{c_{\alpha}=1,c_{\beta}=2,c_{\gamma}=1,\alpha\neq\gamma}g_{*,\alpha}C_{(\alpha,\beta)(\gamma,\beta)}m_{\beta,\gamma}^{-1}\hat{y}_{\beta}\\
 & =g_{*1}\hat{y}_{2}D_{1}D_{2}r\left(K_{2}+(D_{2}-1)C_{(\alpha,\beta)(\alpha,\delta)}^{1,2,1,2}+(D_{1}-1)C_{(\alpha,\beta)(\gamma,\beta)}^{1,2,1,2}\right)\,.
\end{align}

\selectlanguage{american}%

\paragraph{\newline \foreignlanguage{english}{\noindent Term 7 $c_{\alpha}=1,c_{\beta}=2,c_{\gamma}=2,c_{\delta}=1$\newline}}

\noindent \newline \foreignlanguage{english}{We get:}

\selectlanguage{english}%
\begin{align}
\sum_{c_{\alpha}=1,c_{\beta}=2,c_{\gamma}=2,c_{\delta}=1}g_{*,\alpha}C_{(\alpha,\beta)(\gamma,\delta)}m_{\beta,\gamma}^{-1}\hat{y}_{\delta} & =\sum_{c_{\alpha}=1,c_{\beta}=2}g_{*,\alpha}C_{(\alpha,\beta)(\beta,\alpha)}m_{\beta,\beta}^{-1}\hat{y}_{\alpha}\nonumber \\
 & +\sum_{c_{\alpha}=1,c_{\beta}=2,c_{\gamma}=2,\beta\neq\gamma}g_{*,\alpha}C_{(\alpha,\beta)(\gamma,\alpha)}m_{\beta,\gamma}^{-1}\hat{y}_{\alpha}\\
 & +\sum_{c_{\alpha}=1,c_{\beta}=2,c_{\delta}=1,\alpha\neq\delta}g_{*,\alpha}C_{(\alpha,\beta)(\beta,\delta)}m_{\beta,\beta}^{-1}\hat{y}_{\delta}\\
 & =g_{*1}\hat{y}_{1}D_{1}D_{2}\left(C_{(\alpha,\beta)(\alpha,\beta)}^{1,2,1,2}q_{1}^{\prime}+(D_{2}-1)C_{(\alpha,\beta)(\gamma,\alpha)}^{1,2,2,1}q_{2}^{\prime}\right)\\
 & +g_{*1}\hat{y}_{1}D_{1}D_{2}(D_{1}-1)C_{(\alpha,\beta)(\beta,\delta)}^{1,2,2,1}q_{1}^{\prime}\,.
\end{align}

\selectlanguage{american}%

\paragraph{\newline \foreignlanguage{english}{\noindent Term 8 $c_{\alpha}=1,c_{\beta}=2,c_{\gamma}=2,c_{\delta}=2$\newline}}

\noindent \newline \foreignlanguage{english}{We get:}

\selectlanguage{english}%
\begin{align}
\sum_{c_{\alpha}=1,c_{\beta}=2,c_{\gamma}=2,c_{\delta}=2}g_{*,\alpha}C_{(\alpha,\beta)(\gamma,\delta)}m_{\beta,\gamma}^{-1}\hat{y}_{\delta} & =\sum_{c_{\alpha}=1,c_{\beta}=2,c_{\delta}=2,\beta\neq\delta}g_{*,\alpha}C_{(\alpha,\beta)(\beta,\delta)}m_{\beta,\beta}^{-1}\hat{y}_{\delta}\nonumber \\
 & +\sum_{c_{\alpha}=1,c_{\beta}=2,c_{\gamma}=2,\beta\neq\gamma}g_{*,\alpha}C_{(\alpha,\beta)(\gamma,\beta)}m_{\beta,\gamma}^{-1}\hat{y}_{\beta}\\
 & =g_{*1}\hat{y}_{2}*D_{1}D_{2}(D_{2}-1)(C_{(\alpha,\beta)(\beta,\delta)}^{1,2,2,2}q_{1}^{\prime}+C_{(\alpha,\beta)(\gamma,\beta)}^{1,2,2,2}q_{2}^{\prime})\,.
\end{align}
\_
\selectlanguage{american}%

\paragraph{\newline \foreignlanguage{english}{\noindent Term 9 $c_{\alpha}=2,c_{\beta}=1,c_{\gamma}=1,c_{\delta}=1$\newline}}

\noindent \newline \foreignlanguage{english}{We get:}

\selectlanguage{english}%
\begin{align}
\sum_{c_{\alpha}=2,c_{\beta}=1,c_{\gamma}=1,c_{\delta}=1}g_{*,\alpha}C_{(\alpha,\beta)(\gamma,\delta)}m_{\beta,\gamma}^{-1}\hat{y}_{\delta} & =\sum_{c_{\alpha}=2,c_{\beta}=1,c_{\delta}=1,\beta\neq\delta}g_{*,\alpha}C_{(\alpha,\beta)(\beta,\delta)}m_{\beta,\beta}^{-1}\hat{y}_{\delta}\nonumber \\
 & +\sum_{c_{\alpha}=2,c_{\beta}=1,c_{\gamma}=1,\beta\neq\gamma}g_{*,\alpha}C_{(\alpha,\beta)(\gamma,\beta)}m_{\beta,\gamma}^{-1}\hat{y}_{\beta}\\
 & =g_{*2}\hat{y}_{1}D_{2}D_{1}(D_{1}-1)(C_{(\alpha,\beta)(\beta,\delta)}^{2,1,1,1}q_{1}+C_{(\alpha,\beta)(\gamma,\beta)}^{2,1,1,1}q_{2})\,.
\end{align}

\selectlanguage{american}%

\paragraph{\newline \foreignlanguage{english}{\noindent Term 10 $c_{\alpha}=2,c_{\beta}=1,c_{\gamma}=1,c_{\delta}=2$\newline}}

\noindent \newline \foreignlanguage{english}{We get:}

\selectlanguage{english}%
\begin{align}
\sum_{c_{\alpha}=2,c_{\beta}=1,c_{\gamma}=1,c_{\delta}=2}g_{*,\alpha}C_{(\alpha,\beta)(\gamma,\delta)}m_{\beta,\gamma}^{-1}\hat{y}_{\delta} & =\sum_{c_{\alpha}=2,c_{\beta}=1}g_{*,\alpha}C_{(\alpha,\beta)(\beta,\alpha)}m_{\beta,\beta}^{-1}\hat{y}_{\alpha}\nonumber \\
 & +\sum_{c_{\alpha}=2,c_{\beta}=1,c_{\gamma}=1,\beta\neq\gamma}g_{*,\alpha}C_{(\alpha,\beta)(\gamma,\alpha)}m_{\beta,\gamma}^{-1}\hat{y}_{\alpha}\\
 & +\sum_{c_{\alpha}=2,c_{\beta}=1,c_{\delta}=2,\alpha\neq\delta}g_{*,\alpha}C_{(\alpha,\beta)(\beta,\delta)}m_{\beta,\beta}^{-1}\hat{y}_{\delta}\\
 & =g_{*2}\hat{y}_{2}D_{2}D_{1}\left(C_{(\alpha,\beta)(\alpha,\beta)}^{2,1,2,1}q_{1}+(D_{1}-1)C_{(\alpha,\beta)(\gamma,\alpha)}^{2,1,1,2}q_{2}\right)\\
 & +g_{*2}\hat{y}_{2}D_{2}D_{1}(D_{2}-1)C_{(\alpha,\beta)(\beta,\delta)}^{2,1,1,2}q_{1}\,.
\end{align}

\selectlanguage{american}%

\paragraph{\newline \foreignlanguage{english}{\noindent Term 11 $c_{\alpha}=2,c_{\beta}=1,c_{\gamma}=2,c_{\delta}=1$\newline}}

\noindent \newline \foreignlanguage{english}{We get:}

\selectlanguage{english}%
\begin{align}
\sum_{c_{\alpha}=2,c_{\beta}=1,c_{\gamma}=2,c_{\delta}=1}g_{*,\alpha}C_{(\alpha,\beta)(\gamma,\delta)}m_{\beta,\gamma}^{-1}\hat{y}_{\delta} & =\sum_{c_{\alpha}=2,c_{\beta}=1}g_{*,\alpha}C_{(\alpha,\beta)(\alpha,\beta)}m_{\beta,\alpha}^{-1}\hat{y}_{\beta}\nonumber \\
 & +\sum_{c_{\alpha}=2,c_{\beta}=1,c_{\delta}=1,\beta\neq\delta}g_{*,\alpha}C_{(\alpha,\beta)(\alpha,\delta)}m_{\beta,\alpha}^{-1}\hat{y}_{\delta}\\
 & +\sum_{c_{\alpha}=2,c_{\beta}=1,c_{\gamma}=2,\alpha\neq\gamma}g_{*,\alpha}C_{(\alpha,\beta)(\gamma,\beta)}m_{\beta,\gamma}^{-1}\hat{y}_{\beta}\\
 & =g_{*2}\hat{y}_{1}D_{1}D_{2}r\left(C_{(\alpha,\beta)(\alpha,\beta)}^{2,1,2,1}+(D_{1}-1)C_{(\alpha,\beta)(\alpha,\delta)}^{2,1,2,1}\right)\\
 & +g_{*2}\hat{y}_{1}D_{1}D_{2}r(D_{2}-1)C_{(\alpha,\beta)(\gamma,\beta)}^{2,1,2,1}\,.
\end{align}

\selectlanguage{american}%

\paragraph{\newline \foreignlanguage{english}{\noindent Term 12 $c_{\alpha}=2,c_{\beta}=1,c_{\gamma}=2,c_{\delta}=2$\newline}}

\noindent \newline \foreignlanguage{english}{We get:}

\selectlanguage{english}%
\begin{align}
\sum_{c_{\alpha}=2,c_{\beta}=1,c_{\gamma}=2,c_{\delta}=2}g_{*,\alpha}C_{(\alpha,\beta)(\gamma,\delta)}m_{\beta,\gamma}^{-1}\hat{y}_{\delta} & =\sum_{c_{\alpha}=2,c_{\beta}=1,c_{\delta}=2,\alpha\neq\delta}g_{*,\alpha}C_{(\alpha,\beta)(\alpha,\delta)}m_{\beta,\alpha}^{-1}\hat{y}_{\delta}\nonumber \\
 & +\sum_{c_{\alpha}=2,c_{\beta}=1,c_{\gamma}=2,\alpha\neq\gamma}g_{*,\alpha}C_{(\alpha,\beta)(\gamma,\alpha)}m_{\beta,\gamma}^{-1}\hat{y}_{\alpha}\\
 & =g_{*2}\hat{y}_{2}D_{2}D_{1}(D_{2}-1)(r+r)C_{(\alpha,\beta)(\alpha,\delta)}^{2,1,2,2}\,.
\end{align}

\selectlanguage{american}%

\paragraph{\newline \foreignlanguage{english}{\noindent Term 13 $c_{\alpha}=2,c_{\beta}=2,c_{\gamma}=1,c_{\delta}=1$\newline}}

\noindent \newline \foreignlanguage{english}{We get:}

\selectlanguage{english}%
\begin{align}
\sum_{c_{\alpha}=2,c_{\beta}=2,c_{\gamma}=1,c_{\delta}=1}g_{*,\alpha}C_{(\alpha,\beta)(\gamma,\delta)}m_{\beta,\gamma}^{-1}\hat{y}_{\delta} & =0\,.
\end{align}

\selectlanguage{american}%

\paragraph{\newline \foreignlanguage{english}{\noindent Term 14 $c_{\alpha}=2,c_{\beta}=2,c_{\gamma}=1,c_{\delta}=2$\newline}}

\noindent \newline \foreignlanguage{english}{We get:}

\selectlanguage{english}%
\begin{align}
\sum_{c_{\alpha}=2,c_{\beta}=2,c_{\gamma}=1,c_{\delta}=2}g_{*,\alpha}C_{(\alpha,\beta)(\gamma,\delta)}m_{\beta,\gamma}^{-1}\hat{y}_{\delta} & =\sum_{c_{\alpha}=2,c_{\beta}=2,c_{\gamma}=1,\alpha\neq\beta}g_{*,\alpha}C_{(\alpha,\beta)(\gamma,\alpha)}m_{\beta,\gamma}^{-1}\hat{y}_{\alpha}\nonumber \\
 & +\sum_{c_{\alpha}=2,c_{\beta}=2,c_{\gamma}=1,\alpha\neq\beta}g_{*,\alpha}C_{(\alpha,\beta)(\gamma,\beta)}m_{\beta,\gamma}^{-1}\hat{y}_{\beta}\\
 & =g_{*2}\hat{y}_{2}D_{2}(D_{2}-1)D_{1}(r+r)C_{(\alpha,\beta)(\gamma,\alpha)}^{2,2,1,2}\,.
\end{align}

\selectlanguage{american}%

\paragraph{\newline \foreignlanguage{english}{\noindent Term 15 $c_{\alpha}=2,c_{\beta}=2,c_{\gamma}=2,c_{\delta}=1$\newline}}

\noindent \newline \foreignlanguage{english}{We get:}

\selectlanguage{english}%
\begin{align}
\sum_{c_{\alpha}=1,c_{\beta}=2,c_{\gamma}=2,c_{\delta}=2}g_{*,\alpha}C_{(\alpha,\beta)(\gamma,\delta)}m_{\beta,\gamma}^{-1}\hat{y}_{\delta} & =\sum_{c_{\alpha}=1,c_{\beta}=2,c_{\gamma}=2,\beta\neq\gamma}g_{*,\alpha}C_{(\alpha,\beta)(\gamma,\beta)}m_{\beta,\gamma}^{-1}\hat{y}_{\beta}\nonumber \\
 & +\sum_{c_{\alpha}=1,c_{\beta}=2,c_{\delta}=2,\beta\neq\delta}g_{*,\alpha}C_{(\alpha,\beta)(\beta,\delta)}m_{\beta,\beta}^{-1}\hat{y}_{\delta}\\
 & =g_{*2}\hat{y}_{1}D_{2}(D_{2}-1)D_{1}(q_{2}^{\prime}+q_{1}^{\prime})C_{(\alpha,\beta)(\alpha,\delta)}^{2,2,2,1}\,.
\end{align}

\selectlanguage{american}%

\paragraph{\newline \foreignlanguage{english}{\noindent Term 16 $c_{\alpha}=2,c_{\beta}=2,c_{\gamma}=2,c_{\delta}=2$\newline}}

\noindent \newline \foreignlanguage{english}{We get:}

\selectlanguage{english}%
\begin{align}
\sum_{c_{\alpha}=2,c_{\beta}=2,c_{\gamma}=2,c_{\delta}=2}g_{*,\alpha}C_{(\alpha,\beta)(\gamma,\delta)}m_{\beta,\gamma}^{-1}\hat{y}_{\delta} & =\sum_{c_{\alpha}=2,c_{\beta}=2}g_{*,\alpha}C_{(\alpha,\beta)(\beta,\alpha)}m_{\beta,\beta}^{-1}\hat{y}_{\alpha}\nonumber \\
 & +\sum_{c_{\alpha}=2,c_{\beta}=2}g_{*,\alpha}C_{(\alpha,\beta)(\alpha,\beta)}m_{\beta,\alpha}^{-1}\hat{y}_{\beta}\\
 & +\sum_{c_{\alpha}=2,c_{\beta}=2,c_{\delta}=2}g_{*,\alpha}C_{(\alpha,\beta)(\beta,\delta)}m_{\beta,\beta}^{-1}\hat{y}_{\delta}\\
 & +\sum_{c_{\alpha}=2,c_{\beta}=2,c_{\delta}=2}g_{*,\alpha}C_{(\alpha,\beta)(\beta,\delta)}m_{\beta,\beta}^{-1}\hat{y}_{\delta}\\
 & +\sum_{c_{\alpha}=2,c_{\beta}=2,c_{\gamma}=2}g_{*,\alpha}C_{(\alpha,\beta)(\gamma,\beta)}m_{\beta,\gamma}^{-1}\hat{y}_{\beta}\\
 & +\sum_{c_{\alpha}=2,c_{\beta}=2,c_{\gamma}=2}g_{*,\alpha}C_{(\alpha,\beta)(\gamma,\alpha)}m_{\beta,\gamma}^{-1}\hat{y}_{\alpha}\\
 & +\sum_{c_{\alpha}=2,c_{\beta}=2,c_{\gamma}=2}g_{*,\alpha}C_{(\alpha,\beta)(\alpha,\delta)}m_{\beta,\alpha}^{-1}\hat{y}_{\delta\text{\textasciiacute}}\\
 & =g_{*2}\hat{y}_{2}D_{2}(D_{2}-1)C_{(\alpha,\beta)(\alpha,\beta)}^{2,2,2,2}(q_{1}^{\prime}+q_{2}^{\prime})\nonumber \\
 & +g_{*2}\hat{y}_{2}D_{2}(D_{2}-1)(D_{2}-2)C_{(\alpha,\beta)(\alpha,\delta)}^{2,2,2,2}\left(q_{1}^{\prime}+3q_{2}^{\prime}\right)\,.
\end{align}

\selectlanguage{american}%

\paragraph{\newline \foreignlanguage{english}{\noindent} Total expression
for $V_{4}(*)$\foreignlanguage{english}{\newline}}

\noindent \newline \foreignlanguage{english}{The total expression
for $V_{4}(*)$ hence reads:}

\selectlanguage{english}%
\begin{align*}
V_{4}(*) & =g_{*1}\hat{y}_{1}D_{1}(D_{1}-1)*\left(\underbrace{C_{(\alpha,\beta)(\alpha,\beta)}^{1,1,1,1}}_{K_{1}}(q_{1}+q_{2})+\underbrace{C_{(\alpha,\beta)(\alpha,\delta)}^{1,1,1,1}}_{v_{1}}(D_{1}-2)\left(q_{1}+3q_{2}\right)\right)\\
2\vert & +g_{*1}\hat{y}_{2}D_{1}(D_{1}-1)D_{2}(q_{2}+q_{1})\underbrace{C_{(\alpha,\beta)(\alpha,\delta)}^{1,1,1,2}}_{v_{3}}\\
3\vert & +g_{*1}\hat{y}_{1}D_{1}(D_{1}-1)D_{2}(r+r)\underbrace{C_{(\alpha,\beta)(\gamma,\alpha)}^{1,1,2,1}}_{v_{3}}
\end{align*}
\\
\begin{align*}
4\vert & +0\\
5\vert & +g_{*1}\hat{y}_{1}D_{1}D_{2}(D_{1}-1)(r+r)\underbrace{C_{(\alpha,\beta)(\alpha,\delta)}^{1,2,1,1}}_{v_{3}}\\
6\vert & +g_{*1}\hat{y}_{2}D_{1}D_{2}r\left(\underbrace{C_{(\alpha,\beta)(\alpha,\beta)}^{1,2,1,2}}_{K_{2}}+(D_{2}-1)\underbrace{C_{(\alpha,\beta)(\alpha,\delta)}^{1,2,1,2}}_{v_{2}}+(D_{1}-1)\underbrace{)C_{(\alpha,\beta)(\gamma,\beta)}^{1,2,1,2}}_{\overline{v}_{2}}\right)\\
7\vert & +g_{*1}\hat{y}_{1}D_{1}D_{2}\left(\underbrace{C_{(\alpha,\beta)(\alpha,\beta)}^{1,2,1,2}}_{K_{2}}q_{1}^{\prime}+(D_{2}-1)\underbrace{C_{(\alpha,\beta)(\gamma,\alpha)}^{1,2,2,1}}_{v_{2}}q_{2}^{\prime}+(D_{1}-1)\underbrace{C_{(\alpha,\beta)(\beta,\delta)}^{1,2,2,1}}_{\overline{v}_{2}}q_{1}^{\prime}\right)\\
8\vert & +g_{*1}\hat{y}_{2}D_{1}D_{2}(D_{2}-1)(\underbrace{C_{(\alpha,\beta)(\beta,\delta)}^{1,2,2,2}}_{\overline{v}_{3}}q_{1}^{\prime}+\underbrace{C_{(\alpha,\beta)(\gamma,\beta)}^{1,2,2,2}}_{\overline{v}_{3}}q_{2}^{\prime})
\end{align*}

\selectlanguage{american}%
\begin{align*}
9\vert & +g_{*2}\hat{y}_{1}D_{2}D_{1}(D_{1}-1)(\underbrace{C_{(\alpha,\beta)(\beta,\delta)}^{2,1,1,1}}_{v_{3}}q_{1}+\underbrace{C_{(\alpha,\beta)(\gamma,\beta)}^{2,1,1,1}}_{v_{3}}q_{2})\\
10\vert & +g_{*2}\hat{y}_{2}D_{2}D_{1}\left(\underbrace{C_{(\alpha,\beta)(\alpha,\beta)}^{1,2,1,2}}_{K_{2}}q_{1}+(D_{1}-1)\underbrace{C_{(\alpha,\beta)(\gamma,\alpha)}^{2,1,1,2}}_{\overline{v}_{2}}q_{2}+(D_{2}-1)\underbrace{C_{(\alpha,\beta)(\beta,\delta)}^{2,1,1,2}}_{v_{2}}q_{1}\right)\\
11\vert & +g_{*2}\hat{y}_{1}D_{1}D_{2}r\left(\underbrace{C_{(\alpha,\beta)(\alpha,\beta)}^{2,1,2,1}}_{\overline{K}_{2}}+(D_{1}-1)\underbrace{C_{(\alpha,\beta)(\alpha,\delta)}^{2,1,2,1}}_{\overline{v}_{2}}+(D_{2}-1)\underbrace{C_{(\alpha,\beta)(\gamma,\beta)}^{2,1,2,1}}_{v_{2}}\right)\\
12\vert & +g_{*2}\hat{y}_{2}D_{2}D_{1}(D_{2}-1)(r+r)\underbrace{C_{(\alpha,\beta)(\alpha,\delta)}^{2,1,2,2}}_{\overline{v}_{3}}\\
13\vert & +0\\
14 & +g_{*2}\hat{y}_{2}D_{2}(D_{2}-1)D_{1}(r+r)\underbrace{C_{(\alpha,\beta)(\gamma,\alpha)}^{2,2,1,2}}_{\overline{v}_{3}}\\
15\vert & +g_{*2}\hat{y}_{1}D_{2}(D_{2}-1)D_{1}(q_{2}^{\prime}+q_{1}^{\prime})\underbrace{C_{(\alpha,\beta)(\alpha,\delta)}^{2,2,2,1}}_{\overline{v}_{3}}\\
16\vert & +g_{*2}\hat{y}_{2}D_{2}(D_{2}-1)*\left(\underbrace{C_{(\alpha,\beta)(\alpha,\beta)}^{2,2,2,2}}_{\overline{K}_{1}}(q_{1}^{\prime}+q_{2}^{\prime})+\underbrace{C_{(\alpha,\beta)(\alpha,\delta)}^{2,2,2,2}}_{\overline{v}_{1}}(D_{2}-2)\left(q_{1}^{\prime}+3q_{2}^{\prime}\right)\right)
\end{align*}
\foreignlanguage{english}{We can summarize the term by using the definitions
from above for $C_{(...)(...)}$ and exploiting the symmetry under
exchange of indices.}

\selectlanguage{english}%
\begin{align*}
V_{4}(*) & =\\
1\vert & g_{*1}\hat{y}_{1}D_{1}(D_{1}-1)*\left(K_{1}(q_{1}+q_{2})+v_{1}(D_{1}-2)\left(q_{1}+3q_{2}\right)\right)\\
16\vert & +g_{*2}\hat{y}_{2}D_{2}(D_{2}-1)*\left(\overline{K}_{1}(q_{1}^{\prime}+q_{2}^{\prime})+\overline{v}_{1}(D_{2}-2)\left(q_{1}^{\prime}+3q_{2}^{\prime}\right)\right)\\
2+3+4+5+9\vert & +D_{1}D_{2}(D_{1}-1)(g_{*1}(q_{1}+q_{2})\hat{y}_{2}+g_{*1}4r\hat{y}_{1}+g_{*2}\hat{y}_{1}(q_{1}+q_{2}))v_{3}\\
12+13+14+15+8\vert & +D_{2}(D_{2}-1)D_{1}(g_{*2}(q_{2}^{\prime}+q_{1}^{\prime})\hat{y}_{1}+g_{*2}4r\hat{y}_{2}+g_{*1}\hat{y}_{2}(q_{1}^{\prime}+q_{2}^{\prime}))\overline{v}_{3}\\
\mathrm{Mix\,of\,6,7,10}\vert & +\overline{v}_{2}D_{1}D_{2}(D_{1}-1)\left(r*g_{*1}\hat{y}_{2}+g_{*1}\hat{y}_{1}q_{1}^{\prime}+g_{*2}\hat{y}_{2}q_{2}+g_{*2}\hat{y}_{1}r\right)\\
\mathrm{Mix\,of\,6,7,10}\vert & +v_{2}D_{1}D_{2}(D_{2}-1)\left(g_{*1}\hat{y}_{2}r+g_{*1}\hat{y}_{1}q_{2}^{\prime}+g_{*2}\hat{y}_{2}q_{1}+g_{*2}\hat{y}_{1}r\right)\\
\mathrm{Mix\,of\,6,7,10}\vert & +K_{2}D_{1}D_{2}\left(g_{*1}\hat{y}_{2}r+g_{*2}\hat{y}_{1}r+g_{*1}\hat{y}_{1}q_{1}^{\prime}+g_{*2}\hat{y}_{2}q_{1}\right)
\end{align*}

\selectlanguage{american}%
Further we inserted the explicit tensor elements for $C_{(\alpha\beta)(\gamma\delta)}$,
\selectlanguage{english}%

\subsubsection{Full result}

Taking the full results from $V_{3}(*)$ and $V_{4}(*)$ we get:

\begin{align}
\left\langle y_{*}\right\rangle _{\text{0+1}} & =\underbrace{g_{*1}D_{1}y_{1}+g_{*2}D_{2}y_{2}}_{y_{0}}\nonumber \\
 & +g_{*1}\hat{y}_{1}D_{1}(D_{1}-1)*\left(K_{1}(q_{1}+q_{2})+v_{1}(D_{1}-2)\left(q_{1}+3q_{2}\right)\right)\nonumber \\
 & +g_{*2}\hat{y}_{2}D_{2}(D_{2}-1)*\left(\overline{K}_{1}(q_{1}^{\prime}+q_{2}^{\prime})+\overline{v}_{1}(D_{2}-2)\left(q_{1}^{\prime}+3q_{2}^{\prime}\right)\right)\nonumber \\
 & +D_{1}D_{2}(D_{1}-1)(g_{*1}(q_{1}+q_{2})\hat{y}_{2}+g_{*1}4r\hat{y}_{1}+g_{*2}\hat{y}_{1}(q_{1}+q_{2}))v_{3}\nonumber \\
 & +D_{2}(D_{2}-1)D_{1}(g_{*2}(q_{2}^{\prime}+q_{1}^{\prime})\hat{y}_{1}+g_{*2}4r\hat{y}_{2}+g_{*1}\hat{y}_{2}(q_{1}^{\prime}+q_{2}^{\prime}))\overline{v}_{3}\nonumber \\
 & +\overline{v}_{2}D_{1}D_{2}(D_{1}-1)\left(r*g_{*1}\hat{y}_{2}+g_{*1}\hat{y}_{1}q_{1}^{\prime}+g_{*2}\hat{y}_{2}q_{2}+g_{*2}\hat{y}_{1}r\right)\nonumber \\
 & +v_{2}D_{1}D_{2}(D_{2}-1)\left(g_{*1}\hat{y}_{2}r+g_{*1}\hat{y}_{1}q_{2}^{\prime}+g_{*2}\hat{y}_{2}q_{1}+g_{*2}\hat{y}_{1}r\right)\nonumber \\
 & +K_{2}D_{1}D_{2}\left(g_{*1}\hat{y}_{2}r+g_{*2}\hat{y}_{1}r+g_{*1}\hat{y}_{1}q_{1}^{\prime}+g_{*2}\hat{y}_{2}q_{1}\right)\nonumber \\
 & -v_{1}D_{1}(D_{1}-1)\left(q_{1}\hat{y}_{1}+q_{2}\hat{y}_{1}\right)\nonumber \\
 & -v_{3}D_{1}D_{2}\left(q_{1}\hat{y}_{2}+\hat{y}_{1}r\right)\nonumber \\
 & -\overline{v}_{2}D_{1}D_{2}\left(r\hat{y}_{2}+q_{1}^{\prime}\hat{y}_{1}\right)\nonumber \\
 & -\overline{v}_{3}D_{2}(D_{2}-1)\left(q_{1}^{\prime}\hat{y}_{2}+q_{2}^{\prime}\hat{y}_{2}\right)\,,\\
 & =y_{\text{0}}\nonumber \\
 & +K_{1}(q_{1}+q_{2})g_{*1}\hat{y}_{1}D_{1}(D_{1}-1)+\overline{K}_{1}(q_{1}^{\prime}+q_{2}^{\prime})g_{*2}\hat{y}_{2}D_{2}(D_{2}-1)\nonumber \\
 & +v_{1}\hat{y}_{1}D_{1}(D_{1}-1)\left[(D_{1}-2)\left(q_{1}+3q_{2}\right)*g_{*1}-\left(q_{1}+q_{2}\right)\right]\nonumber \\
 & +\overline{v}_{1}(D_{2}-2)\left(q_{1}^{\prime}+3q_{2}^{\prime}\right)*g_{*2}\hat{y}_{2}D_{2}(D_{2}-1)\nonumber \\
 & +v_{3}D_{1}D_{2}\left[(D_{1}-1)(g_{*1}(q_{1}+q_{2})\hat{y}_{2}+g_{*1}4r\hat{y}_{1}+g_{*2}\hat{y}_{1}(q_{1}+q_{2}))-\left(q_{1}\hat{y}_{2}+\hat{y}_{1}r\right)\right]\nonumber \\
 & +\overline{v}_{3}D_{2}(D_{2}-1)\left[D_{1}(g_{*2}(q_{2}^{\prime}+q_{1}^{\prime})\hat{y}_{1}+g_{*2}4r\hat{y}_{2}+g_{*1}\hat{y}_{2}(q_{1}^{\prime}+q_{2}^{\prime}))-\left(q_{1}^{\prime}\hat{y}_{2}+q_{2}^{\prime}\hat{y}_{2}\right)\right]\nonumber \\
 & +\overline{v}_{2}D_{1}D_{2}\left[(D_{1}-1)\left(r*g_{*1}\hat{y}_{2}+g_{*1}\hat{y}_{1}q_{1}^{\prime}+g_{*2}\hat{y}_{2}q_{2}+g_{*2}\hat{y}_{1}r\right)-\left(r\hat{y}_{2}+q_{1}^{\prime}\hat{y}_{1}\right)\right]\nonumber \\
 & +v_{2}D_{1}D_{2}(D_{2}-1)\left(g_{*1}\hat{y}_{2}r+g_{*1}\hat{y}_{1}q_{2}^{\prime}+g_{*2}\hat{y}_{2}q_{1}+g_{*2}\hat{y}_{1}r\right)\nonumber \\
 & +K_{2}D_{1}D_{2}\left(g_{*1}\hat{y}_{2}r+g_{*2}\hat{y}_{1}r+g_{*1}\hat{y}_{1}q_{1}^{\prime}+g_{*2}\hat{y}_{2}q_{1}\right)\,.\label{eq:Appendix_FullAsymmetricExpression-1}
\end{align}

\subsubsection{Symmetric Case}

The expression in \eqref{eq:Appendix_FullAsymmetricExpression-1}
can be greatly simplified if one considers a symmetric task setting.
This is the case in the main text for the our artificial Ising task.
Hence we can already make the following substitutions

\begin{align}
K_{1},\overline{K}_{1},K_{2} & \rightarrow K\,,\\
v_{1},v_{2},\overline{v}_{1},\overline{v}_{2} & \rightarrow v\,,\\
v_{3},v_{4},\overline{v}_{3},\overline{v}_{4} & \rightarrow-v\,,
\end{align}
which yields

\begin{align}
\left\langle y_{*}\right\rangle _{0+1} & =g_{*1}D_{1}y_{1}+g_{*2}D_{2}y_{2}\nonumber \\
 & +K(q_{1}+q_{2})g_{*1}\hat{y}_{1}D_{1}(D_{1}-1)+K(q_{1}+q_{2})g_{*2}\hat{y}_{2}D_{2}(D_{2}-1)\nonumber \\
 & +v\hat{y}_{1}D_{1}(D_{1}-1)\left[(D_{1}-2)\left(q_{1}+3q_{2}\right)*g_{*1}-\left(q_{1}+q_{2}\right)\right]\nonumber \\
 & +v(D_{2}-2)\left(q_{1}+3q_{2}\right)*g_{*2}\hat{y}_{2}D_{2}(D_{2}-1)\nonumber \\
 & -vD_{1}D_{2}\left[(D_{1}-1)(g_{*1}(q_{1}+q_{2})\hat{y}_{2}+g_{*1}4r\hat{y}_{1}+g_{*2}\hat{y}_{1}(q_{1}+q_{2}))-\left(q_{1}\hat{y}_{2}+\hat{y}_{1}r\right)\right]\nonumber \\
 & -vD_{2}(D_{2}-1)\left[D_{1}(g_{*2}(q_{2}+q_{1})\hat{y}_{1}+g_{*2}4r\hat{y}_{2}+g_{*1}\hat{y}_{2}(q_{1}+q_{2}))-\left(q_{1}\hat{y}_{2}+q_{2}\hat{y}_{2}\right)\right]\nonumber \\
 & +vD_{1}D_{2}\left[(D_{1}-1)\left(r*g_{*1}\hat{y}_{2}+g_{*1}\hat{y}_{1}q_{1}+g_{*2}\hat{y}_{2}q_{2}+g_{*2}\hat{y}_{1}r\right)-\left(r\hat{y}_{2}+q_{1}\hat{y}_{1}\right)\right]\nonumber \\
 & +vD_{1}D_{2}(D_{2}-1)\left(g_{*1}\hat{y}_{2}r+g_{*1}\hat{y}_{1}q_{2}+g_{*2}\hat{y}_{2}q_{1}+g_{*2}\hat{y}_{1}r\right)\nonumber \\
 & +KD_{1}D_{2}\left(g_{*1}\hat{y}_{2}r+g_{*2}\hat{y}_{1}r+g_{*1}\hat{y}_{1}q_{1}+g_{*2}\hat{y}_{2}q_{1}\right)\,,
\end{align}
where we used that from the setting for the mean $m_{\alpha,\beta}$
we can also get symmetries in the matrix elements of the propagator
$m_{\alpha\beta}^{-1}$

\begin{align}
m_{\beta\beta}^{-1}: & q_{1}^{\prime}=q_{1}\,,\\
m_{\alpha\beta}^{-1}: & q_{2}^{\prime}=q_{2}\,.
\end{align}
Further we derive some symmetries in $g_{*1},g_{*2},r,q_{1},q_{2},\hat{y}_{1},\hat{y}_{2}$
from the expressions in \foreignlanguage{american}{\eqref{eq:Matrix_Elements_InverseBlockMatrix}}

\begin{align}
g(c(\alpha) & =1,c(*)=1)=\frac{b\gamma_{2}^{d}-D_{2}c^{2}}{\tilde{\lambda}}=\frac{b(a-d)+D_{2}(bd-c^{2})}{\gamma_{2}^{d}\gamma_{1}^{b}-D_{1}D_{2}c^{2}}=\frac{b(a-b)}{(a-b)^{2}+b(a-b)D}=\frac{b}{(a-b)+bD}\,,\\
g(c(\alpha) & =2,c(*)=1)=\frac{c\gamma_{1}^{b}-D_{1}bc}{\tilde{\lambda}}=\frac{c(a-b)}{\tilde{\lambda}}=\frac{-b(a-b)}{(a-b)^{2}+b(a-b)D}=-g_{*2}\,,\\
q_{2} & =\frac{c^{2}D_{2}-b\gamma_{2}^{b}}{(a-b)\tilde{\lambda}}=\frac{-b(a-b)}{(a-b)((a-b)^{2}+b(a-b)D)}=\frac{-b}{(a-b)^{2}+b(a-b)D}\,,\\
q_{1} & =\frac{1}{a-b}+q_{2}\,,\\
r & =-\frac{-b}{(a-b)^{2}+b(a-b)D}=\frac{b}{(a-b)^{2}+b(a-b)D}=-q_{2}\,,\\
\hat{y}_{1} & =y_{1}\frac{(a-d)+(d+c)D_{2}}{\gamma_{1}^{b}\gamma_{2}^{d}-D_{1}D_{2}c^{2}}=y_{1}\frac{(a-b)}{(a-b)^{2}+(a-b)bD}\,,\\
\hat{y}_{2} & =-y_{1}\frac{(a-b)+(b+c)D_{1}}{\gamma_{1}^{b}\gamma_{2}^{d}-D_{1}D_{2}c^{2}}=-y_{1}\frac{(a-b)}{\gamma_{1}^{b}\gamma_{2}^{d}-D_{1}D_{2}c^{2}}=-\hat{y}_{1}\,.
\end{align}
Using those replacements we further simplify

\begin{align}
\left\langle y_{*}\right\rangle _{0+1} & =g_{*1}D_{1}\hat{y}_{1}-g_{*1}D_{2}(-y_{1})\nonumber \\
 & +K(q_{1}+q_{2})g_{*1}\hat{y}_{1}D_{1}(D_{1}-1)-K(q_{1}+q_{2})g_{*1}(-\hat{y}_{1})D_{2}(D_{2}-1)\nonumber \\
 & +v\hat{y}_{1}D_{1}(D_{1}-1)\left[(D_{1}-2)\left(q_{1}+3q_{2}\right)*g_{*1}-\left(q_{1}+q_{2}\right)\right]\nonumber \\
 & -v(D_{2}-2)\left(q_{1}+3q_{2}\right)*g_{*1}(-\hat{y}_{1})D_{2}(D_{2}-1)\nonumber \\
 & -vD_{1}D_{2}\left[(D_{1}-1)(-g_{*1}(q_{1}+q_{2})\hat{y}_{1}+g_{*1}4r\hat{y}_{1}-g_{*1}\hat{y}_{1}(q_{1}+q_{2}))-\left(-q_{1}\hat{y}_{1}-q_{2}\hat{y}_{1}\right)\right]\nonumber \\
 & -vD_{2}(D_{2}-1)\left[D_{1}(-g_{*1}(q_{2}+q_{1})\hat{y}_{1}-g_{*1}4r(-\hat{y}_{1})-g_{*1}(\hat{y}_{1})(q_{1}+q_{2}))+\left(q_{1}\hat{y}_{1}+q_{2}\hat{y}_{1}\right)\right]\nonumber \\
 & +vD_{1}D_{2}\left[(D_{1}-1)\left(-r*g_{*1}\hat{y}_{1}+g_{*1}\hat{y}_{1}q_{1}+g_{*1}\hat{y}_{1}q_{2}-g_{*1}\hat{y}_{1}r\right)-\left(+q_{2}\hat{y}_{1}+q_{1}\hat{y}_{1}\right)\right]\nonumber \\
 & +vD_{1}D_{2}(D_{2}-1)\left(-g_{*1}\hat{y}_{1}r+g_{*1}\hat{y}_{1}q_{2}+g_{*1}\hat{y}_{1}q_{1}-g_{*1}\hat{y}_{1}r\right)\nonumber \\
 & +KD_{1}D_{2}\left(-g_{*1}\hat{y}_{1}r-g_{*1}\hat{y}_{1}r+g_{*1}\hat{y}_{1}q_{1}+g_{*1}\hat{y}_{1}q_{1}\right)\,.
\end{align}
Simplifying further we get:

\begin{align}
\left\langle y_{*}\right\rangle _{0+1} & =g_{*1}D\hat{y}_{1}\nonumber \\
 & +\underbrace{K(q_{1}+q_{2})g_{*1}\hat{y}_{1}\left(D_{1}(D_{1}-1)+D_{2}(D_{2}-1)\right)+2KD_{1}D_{2}g_{*1}\hat{y}_{1}\left(q_{1}+q_{2}\right)}_{K(q_{1}+q_{2})g_{*1}\hat{y}_{1}\left(D^{2}-D\right)}\nonumber \\
 & v\hat{y}_{1}g_{*1}\left(q_{1}+3q_{2}\right)\left(D_{1}(D_{1}-1)(D_{1}-2)+D_{2}(D_{2}-1)(D_{2}-2)+2D_{1}D_{2}(D_{1}-1)+2D_{2}(D_{2}-1)D_{1}\right)\nonumber \\
 & -v\hat{y}_{1}D_{1}(D_{1}-1)\left[\left(q_{1}+q_{2}\right)\right]\nonumber \\
 & -v\hat{y}_{1}D_{1}D_{2}\left[\left(q_{1}+q_{2}\right)\right]\nonumber \\
 & -v\hat{y}_{1}D_{2}(D_{2}-1)\left[\left(q_{1}+q_{2}\right)\right]\nonumber \\
 & -v\hat{y}_{1}D_{1}D_{2}\left(q_{1}+q_{2}\right)\nonumber \\
 & +g_{*1}\hat{y}_{1}vD_{1}D_{2}(D_{1}-1)\left(q_{1}+3q_{2}\right)\nonumber \\
 & +g_{*1}\hat{y}_{1}vD_{1}D_{2}(D_{2}-1)\left(q_{1}+3q_{2}\right)\,.
\end{align}
We can consolidate some of the terms by ordering expressions by the
occurrence of the tensor elements $K,v$

\begin{align}
\left\langle y_{*}\right\rangle _{0+1} & =g_{*1}D\hat{y}_{1}\nonumber \\
 & +K(q_{1}+q_{2})g_{*1}\hat{y}_{1}\left(D^{2}-D\right)\nonumber \\
 & +v\hat{y}_{1}g_{*1}\left(q_{1}+3q_{2}\right)\left(D_{1}(D_{1}-1)(D_{1}-2)+D_{2}(D_{2}-1)(D_{2}-2)+2D_{1}D_{2}(D_{1}-1)+2D_{2}(D_{2}-1)D_{1}\right)\nonumber \\
 & -v\hat{y}_{1}\left(q_{1}+q_{2}\right)\left(D_{1}(D_{1}-1)+2D_{1}D_{2}+D_{2}(D_{2}-1)\right)\nonumber \\
 & +g_{*1}\hat{y}_{1}v\left(q_{1}+3q_{2}\right)\left(D_{1}D_{2}(D_{1}-1)+D_{1}D_{2}(D_{2}-1)\right)\,.
\end{align}
This yields

\begin{align*}
\left\langle y_{*}\right\rangle _{0+1} & =g_{*1}D\hat{y}_{1}\\
 & +K(q_{1}+q_{2})g_{*1}\hat{y}_{1}\left(D^{2}-D\right)\\
 & +v\hat{y}_{1}g_{*1}\left(q_{1}+3q_{2}\right)\bigg(D_{1}(D_{1}-1)(D_{1}-2)+D_{2}(D_{2}-1)(D_{2}-2)+2D_{1}D_{2}(D_{1}-1)+2D_{2}(D_{2}-1)D_{1}\\
 & +D_{1}D_{2}(D_{1}-1)+D_{1}D_{2}(D_{2}-1)\bigg)\\
 & -v\hat{y}_{1}\left(q_{1}+q_{2}\right)\left(D^{2}-D\right)
\end{align*}
Where we can use binomial formula to simplify the third line to 

\begin{align}
 & v\hat{y}_{1}g_{*1}\left(q_{1}+3q_{2}\right)\left(D_{1}(D_{1}-1)(D_{1}-2)+D_{2}(D_{2}-1)(D_{2}-2)+2D_{1}D_{2}(D_{1}-1)+2D_{2}(D_{2}-1)D_{1}\right)\,,\\
= & v\hat{y}_{1}g_{*1}\left(q_{1}+3q_{2}\right)\bigg(D^{3}-3D^{2}+2D\bigg)\,.
\end{align}
which corresponds to the result \prettyref{eq:mean_SimplePatternStatistics_MFT}
and \prettyref{eq:mean_SimplePatternStatistics_FirstOrder} stated
in the main text 
\selectlanguage{american}%

\subsection{Limiting value for $D\rightarrow\infty$ of $\left\langle y_{*}\right\rangle _{0+1}$
in the symmetric setting}\label{supp:LimitingValue}

\selectlanguage{english}%
We can now compute expansions of the mean of the predictive distribution
in $1/D\ll1$ in order to obtain the leading order expressions for
large number of training samples and hence get the asymptotic value
for $\lim_{D\rightarrow\infty}\left\langle y_{*}\right\rangle _{0+1}$.
Starting from the full expression \prettyref{eq:mean_SimplePatternStatistics_MFT}
and \prettyref{eq:mean_SimplePatternStatistics_FirstOrder} we compute
the limiting value for $D\rightarrow\infty$. We introduce the notation

\begin{align}
k & =k^{(0)}+k^{(1)}+...\quad\mathrm{with\quad}k^{(i)}\sim\mathcal{O}\left(\frac{1}{D^{i}}\right)\,
\end{align}
for the quantities $g_{*},\hat{y}_{*},q_{1},q_{2}$ in order to make
the $D$ dependence explicit:
\begin{align}
g_{*} & =g_{*}^{(1)}+g_{*}^{(2)}\,,\\
q_{1} & =\frac{1}{a-b}+q_{2}=q_{1}^{(0)}+q_{2}^{(1)}+q_{2}^{(2)}\,,\\
\hat{y}_{1} & =\hat{y}_{1}^{(1)}+\hat{y}_{1}^{(2)}\,.
\end{align}
with

\begin{align}
g_{*} & \approx1*\frac{1}{D^{1}}-\frac{a-b}{b}\frac{1}{D^{2}}\,,\\
q_{1} & \approx\frac{1}{a-b}\frac{1}{D^{\text{0}}}-\frac{1}{a-b}\frac{1}{D^{1}}+\frac{1}{b}\frac{1}{D^{2}}\,,\\
\hat{y}_{1} & \approx\frac{y_{1}}{b}\frac{1}{D^{1}}-y_{1}\frac{a-b}{b^{2}}\frac{1}{D^{2}}\,.
\end{align}
which follows directly from the results in \prettyref{supp:AppendixElementsInverseBlockMatrix}.
We can hence insert the expression:

\begin{align*}
\left\langle y^{*}\right\rangle _{0+1} & =(g_{*}^{(1)}+g_{*}^{(2)})Dy_{1}\\
 & +K(q_{1}^{(0)}+2q_{2}^{(1)}+2q_{2}^{(2)})(g_{*}^{(1)}+g_{*}^{(2)})(\hat{y}_{1}^{(1)}+\hat{y}_{1}^{(2)})(D^{2}-D)\\
 & +v\left(q_{1}^{(0)}+4q_{2}^{(1)}+4q_{2}^{(2)}\right)(g_{*}^{(1)}+g_{*}^{(2)})(\hat{y}_{1}^{(1)}+\hat{y}_{1}^{(2)})(D^{3}-3D^{2}+2D)\\
 & -v\left(q_{1}^{(0)}+2q_{2}^{(1)}+2q_{2}^{(2)}\right)(\hat{y}_{1}^{(1)}+\hat{y}_{1}^{(2)})(D^{2}-D)
\end{align*}
We now order in powers of $D$

\begin{align}
\mathcal{O}(D) & :vq_{1}^{(0)}g_{*}^{(1)}\hat{y}_{1}^{(1)}D^{3}-vq_{1}^{(0)}\hat{y}_{1}^{(1)}D^{2}\,,\\
\mathcal{O}(1) & :g_{*}^{(1)}Dy_{1}\nonumber \\
 & +Kq_{1}^{(0)}g_{*}^{(1)}\hat{y}_{1}^{(1)}D^{2}\nonumber \\
 & +vq_{1}^{(0)}g_{*}^{(1)}\hat{y}_{1}^{(1)}(-3D^{2})+v4q_{2}^{(1)}g_{*}^{(1)}\hat{y}_{1}^{(1)}D^{3}+vq_{1}^{(0)}g_{*}^{(2)}\hat{y}_{1}^{(1)}D^{3}+vq_{1}^{(0)}g_{*}^{(1)}\hat{y}_{1}^{(2)}D^{3}\nonumber \\
 & +vq_{1}^{(0)}\hat{y}_{1}^{(1)}D-vq_{1}^{(0)}\hat{y}_{1}^{(2)}D^{2}-2vq_{2}^{(1)}\hat{y}_{1}^{(1)}D^{2}\,.
\end{align}
This simplifies to:

\begin{align}
\mathcal{O}(D) & :vq_{1}^{(0)}\frac{1}{D}\hat{y}_{1}^{(1)}D^{3}-vq_{1}^{(0)}\hat{y}_{1}^{(1)}D^{2}=0\,,\\
\mathcal{O}(1) & :y_{1}\nonumber \\
 & +K\frac{1}{a-b}\frac{y_{1}}{b}\nonumber \\
 & -\frac{3v}{a-b}\frac{y_{1}}{b}-\frac{4v}{a-b}\frac{y_{1}}{b}-\frac{v}{a-b}\frac{a-b}{b}\frac{y_{1}}{b}-\frac{v}{a-b}\frac{a-b}{b^{2}}y_{1}\nonumber \\
 & +\frac{v}{a-b}\frac{y_{1}}{b}+v\frac{1}{a-b}y_{1}\frac{a-b}{b^{2}}+\frac{2v}{a-b}\frac{y_{1}}{b}\,.
\end{align}
which can be further simplified to

\begin{align}
\mathcal{O}(D) & :0\,,\\
\mathcal{O}(1) & :y_{1}\nonumber \\
 & +K\frac{1}{a-b}\frac{y_{1}}{b}\nonumber \\
 & -\frac{3v}{a-b}\frac{y_{1}}{b}-\frac{4v}{a-b}\frac{y_{1}}{b}-2v\frac{y_{1}}{b^{2}}\nonumber \\
 & +\frac{v}{a-b}\frac{y_{1}}{b}+\frac{v}{b^{2}}y_{1}+\frac{2v}{a-b}\frac{y_{1}}{b}\,.
\end{align}
Combining both results we obtain limiting value from the main text

\begin{equation}
\langle y\rangle_{0+1}(D\rightarrow\infty)=y_{1}+\frac{y_{1}}{b}\left(\frac{1}{a-b}(K-4v)-\frac{v}{b}\right)\,.
\end{equation}

\selectlanguage{american}%

\end{document}